\documentclass[11pt,a4paper]{article}
\usepackage{jheppub}

\usepackage{cleveref}

\newcommand{\refeq}[1]{(\ref{#1})}
\newcommand{\nn}{\nonumber \\}
\newcommand{\beq}{\begin{equation}}
\newcommand{\eeq}{\end{equation}}

\renewcommand{\d}{\partial}

\newcommand{\Ocal}{\mathcal{O}}

\newcommand{\Mcal}{\mathcal{M}}
\newcommand{\eps}{\epsilon}

\newcommand{\im}{\operatorname{Im}}

\newcommand{\ket}[1]{| #1 \rangle}
\newcommand{\bra}[1]{\langle #1 |}
\newcommand{\braket}[2]{\langle #1| #2 \rangle}

\title{Undecay}

\author[a]{Eugenio Meg\'ias,}
\author[b]{Manuel P\'erez-Victoria}
\author[c]{and Mariano Quir\'os}

\affiliation[a]{Departamento de F\'isica At\'omica, Molecular y Nuclear and Instituto Carlos I de F\'isica Te\'orica y Computacional, 
Universidad de Granada, Campus de Fuentenueva, E-18071 Granada, Spain}
\affiliation[b]{CAFPE and Departamento de F\'{\i}sica Te\'orica y del Cosmos,
Universidad de Granada, Campus de Fuentenueva, E-18071 Granada, Spain}
\affiliation[c]{Institut de F\'isica d'Altes Energies (IFAE) and The Barcelona Institute of Science and Technology (BIST), 
Campus UAB, 08193 Bellaterra, Barcelona, Spain}

\emailAdd{emegias@ugr.es}
\emailAdd{mpv@ugr.es}
\emailAdd{quiros@ifae.es}

\abstract{Unstable particles decay sooner or later, so they are not described by asymptotic one-particle states and they should not be included as independent states in unitarity relations such as the optical theorem. The same applies to any countable collection of unstable particles. We show that the behaviour of unparticle stuff, that is, a continuous collection of particles with different masses and common decay channels, is pretty different: it has a non-vanishing probability of surviving for ever and the corresponding asymptotic states must be taken into account to comply with unitarity. We also discuss compressed spectra and the transition from the discrete to the continuous case.}

\begin{document}

\maketitle

%%%%%%%%%%%%%%%%%%%

\section{Introduction}
\label{s:intro}

New physics beyond the Standard Model (SM) typically involve extra particles. They are often heavier than the known particles and are often unstable.
Besides their possible indirect effects, the new particles can in principle be discovered as resonances in the scattering of the SM particles. In particular, the models that incorporate an extra sector formed by a confining gauge theory predict a rich spectrum of resonances, which could potentially be observed at present or future colliders.\footnote{In certain scenarios, some SM particles are actually the lightest of these resonances. This is the case of a pseudo-Goldstone composite Higgs model, for instance~\cite{Kaplan:1983fs}. In the following, when we speak of SM fields and particles we will be referring to the ones that do not belong to the extra sector, unless otherwise indicated.} Similar features are shared by models in extra compact dimensions, which in some geometries can actually be understood as holographic duals of strongly-coupled theories~\cite{Maldacena:1997re,Arkani-Hamed:2000ijo,Rattazzi:2000hs,Perez-Victoria:2001lex,Gherghetta:2006ha}. 

New physics without new particles is nevertheless possible, at least in principle. Such a scenario was proposed by Howard Georgi in~\cite{Georgi:2007ek}. It involves a hidden sector with an infrared (IR) conformal fixed point, weakly coupled to the SM. Scale invariance forbids particle states with definite non-zero mass, so an extra non-trivial sector of this sort will have a gapless continuous spectrum of {\em unparticle} states. One can also envisage a similar type of model in which a controlled breaking of conformal invariance gives rise to a mass gap with a continuum on top~\cite{Cacciapaglia:2007jq}, or directly consider an extra sector with a gapped continuous spectrum that is not necessarily scale-invariant~\cite{vanderBij:2007um}.\footnote{Partially continuous mass spectra are actually a universal feature of quantum field theories. Indeed, the multi-particle states of the stable  particles associated to the discrete part of the spectrum have continuous invariant-mass distributions.}  At any rate, the mass gap suppresses the effects of the extra sector at lower energies and allows for viable models with relatively large couplings to the SM particles. Extra-dimensional models with such spectra have been constructed and studied in Refs.~\cite{Cacciapaglia:2008ns,Falkowski:2008fz,Falkowski:2008yr,Falkowski:2009uy,Csaki:2018kxb,Megias:2019vdb,Megias:2021mgj}. They involve non-compact extra dimensions and require a critical value of the parameters determining the geometry. 

The phenomenology of extra sectors with continuous spectra is very different from the usual one involving a discrete set of particles. In particular, a continuous invariant-mass distribution can evade or relax the limits from standard direct searches of new physics, which look for resonant peaks. On the other hand, a discrete set of particles can mimic a continuum when either the energy-momentum resolution or the width of the resonances is larger than the mass spacings~\cite{Perez-Victoria:2008pbb}. Such compressed spectra are expected generically when conformal symmetry is explicitly or spontaneously broken (within the hidden sector or by its interactions with the SM sector) and in extra-dimensional models with small departures (in the right direction) from the critical parameters mentioned above. For example, a spectrum with a mass gap followed by narrowly-spaced masses is actually a characteristic feature of the linear dilaton geometry in five dimensions~\cite{Antoniadis:2001sw} and of clockwork models~\cite{Kaplan:2015fuy,Giudice:2016yja}, which can be seen as a deconstruction of the former. In realistic generic models, most of the extra particles will actually be unstable and decay into the lightest ones, which will be stable except for their possible decay into SM particles. This is akin to the QCD sector of the SM, with the lightest hadrons decaying only by electroweak interactions. This and other features that are usually present in hidden-sector gauge theories imply that unparticle models with conformal breaking and a sufficiently large mass gap typically have {\em hidden valley\/} phenomenology~\cite{Strassler:2006im,Strassler:2008bv}, with striking potential signals such us displaced vertices and abundant high-multiplicity final states at high energies.

In all cases, it is possible in principle to construct an effective field theory for the SM fields by integrating out the degrees of freedom of the weakly- or strongly- interacting extra sector. This effective theory will have non-local operators, with form factors given by the different correlation functions of the operators of the extra sector that couple to the SM fields. Only at energy scales smaller than the mass gap, if it exists, can a truncated derivative expansion be used to find a local effective theory that approximates well the physics of the exact theory. But here we are interested in physics at larger energies. The non-local effective theory, if calculated exactly, provides a complete description of any process involving SM particles in the initial and final states. Unitarity may not be apparent in the effective theory, as the unitarity relations involve not only the SM particles but also the asymptotic states (of particle or unparticle nature) associated to the extra sector, which we will call hereafter {\em hidden states}. But the unitarity cuts of the form factors contain information about these hidden states and this information can be used to calculate inclusive cross sections of processes with both external SM particles and external hidden states. So, the missing contributions to unitarity relations are trivially provided by the cuts of the form factors and the effective theory is self-consistent if they are taken into account. The crucial point here is that all the relevant hidden states are (collectively) created by interpolating operators formed by the SM fields. 

In this work we study the states of the extra sector from the point of view of the effective theory. For this, we use a toy model with one complex scalar field $\varphi$ (playing the role of the SM fields) which couples in a cubic interaction to a single operator $\mathcal{O}$ of the extra sector, with strength $g$. We will refer to $\varphi$ as the {\em elementary} field, and to the particles associated with it as the elementary particles. For further simplicity, we will focus on the form factor given by the two-point function of $\mathcal{O}$, that is, we will neglect higher-point correlators. Even if the latter can have a strong impact on the phenomenology of this type of model, as emphasized in~\cite{Strassler:2008bv}, the two-point function already has important information about the fundamental states that can be produced by the scattering of SM particles and about their evolution. Moreover, we can consider the extreme case of gauge theories in the large $N$ limit, in which the higher-point correlation functions vanish. In the same limit, which is dual to small coupling in extra-dimensional models, the resonances of that sector become infinitely narrow (when the coupling to the elementary field vanishes). In section~\ref{s:discussion} we will briefly comment on the possible impact of higher-order functions in our results.

The K\"all\'en-Lehmann representation of the two-point function $\langle \Ocal(x_1) \Ocal(x_2) \rangle$ for $g=0$ provides the spectral density of hidden states. We will consider only the cases of a purely discrete or purely continuum spectrum, both with a mass gap larger than twice the mass of the field $\varphi$. Correspondingly, the two-point function in momentum space has a series of simple poles in the discrete case and a branch cut in the continuous one. The two-point function also encodes the propagation of the state that is created by the operator $\Ocal$, and can be used as a free propagator for the perturbative expansion in powers of $g$ of the Green functions in the effective theory. 

The singularities of the free propagator are associated to stable states of the extra sector. However, these states need not be stable in the complete theory when $g$ does not vanish. Because they are heavier than the production threshold, they can decay into elementary particles. In the discrete case, all the hidden states decay, although, as we will see, not following in general an exponential law. In the continuous case, on the other hand, the optical theorem implies, as shown below in detail, that there are stable hidden states also in the presence of interactions with light elementary particles. This fact, which has been ignored in most of the unparticle literature, has significant phenomenological implications. For example, in~\cite{Falkowski:2009uy} it was shown that the visible decay of a hypothetical Higgs unparticle~\cite{Stancato:2008mp} into $b\bar{b}$ would be very suppressed due to the existence of an invisible decay mode, which helped in evading LEP limits. From the effective-theory perspective, the existence of these stable hidden states is not completely obvious and raises at least a couple of questions: What is the decay law that allows some unparticle states survive for an arbitrarily long time? Can this behaviour be approximated by a discrete spectrum with arbitrarily small mass spacings? In this article we answer these questions and shed some light on the dynamics of continuous and compressed spectra. We do this by analyzing in detail the time evolution of the state created by the operator $\Ocal$ in the toy model described above, in both the discrete and the continuous cases. We will see that the state of the system undergoes simultaneously decay into elementary particles and oscillations into different states of the extra sector. It turns out that the latter are irreversible in the continuum limit and give rise to an alternative decay mode into hidden states. 

Unparticle decay into SM particles has been studied before in a few occasions. Let us briefly comment on them. In~\cite{Stephanov:2007ry}, unparticle stuff was described as the continuum limit of a model with a discrete spectrum. This provided an intuitive understanding of several peculiar properties of continuous spectra. In particular, it was argued that, because the couplings of each discrete mode become smaller with smaller spacings and approach zero in the continuum limit, ``in a certain sense a true unparticle, once produced, never decays''. 
One might think that this accounts for the stability of the hidden states. But then, how is it possible that visible final states are also observed? The flaw in the argument is that the individual unparticle modes with well defined mass cannot form normalizable states on their own, so it makes no sense to treat them separately. Actually, their vanishing couplings in the continuum limit are nothing but a reflection of the infinite norm of generalized states in the continuum. Normalizable physical states are continuous linear combinations of these generalized states and actual physical processes always involve an uncountable number of modes. The interference of their different contributions to the corresponding amplitude is unavoidable and must be taken into account. As we will see, this interference is crucial and completely changes the simple but ill-defined picture with isolated modes. In particular, the stable hidden states are a consequence of interference. Similar considerations apply to the case of compressed discrete spectra. In~\cite{Rajaraman:2008bc}, it was correctly observed that the collective effect of the many modes compensates the infinitesimal couplings and gives a finite decay rate. But the cross section was written as an integrated form of the narrow width approximation, which neglects the interference among different modes. This is not a good approximation in the continuum nor for sufficiently compressed spectra, and it precludes the existence of invisible decay channels. Unparticle decay was also studied in~\cite{Delgado:2008gj}. In that reference, it was shown (in a toy model essentially equal to ours) that when the threshold of elementary particle production is smaller than the continuum mass gap, the dressed unparticle propagator (with resummed self-energy diagrams) develops a complex pole on the second Riemann sheet. This pole was associated, in the standard way, to the decay into elementary particles. Here we will see that the characteristic branch cut of the continuum also plays an important role in the time evolution in the system. In fact, a single complex pole without other singularities would imply an exponential decay law, which is not observed.\footnote{The case of production threshold larger than the mass gap, which we do not consider here, was also studied in~\cite{Delgado:2008gj}. It was found that the dressed propagator has in that case a real pole, corresponding to a stable one-particle state. The existence of a stable state below the production threshold is not completely surprising, but the optical theorem shows that there are other independent hidden states, which cannot be understood as multi-particle states of the real-pole particle, much like the ones that we study here.} Finally, the visible decay of the particles in the extra sector has been described in explicit ultraviolet (UV) completions in~\cite{Strassler:2008bv}, together with other observable hidden-valley effects in theories of this type. 

The structure of the article is the following. In section~\ref{s:setup} we present the toy model to be studied, write the general expressions of the free and dressed propagator and describe the associated spectral densities. We also give an alternative formulation of the same model given by a tower of fields with standard quadratic terms. In section~\ref{s:unitarity} we show that the optical theorem requires the presence of hidden stable states in the continuous case. In section~\ref{s:eigenstates} we identify a basis of generalized energy eigenstates for both the free and the interacting Hamiltonian. They will be useful, as usual in quantum mechanics, in the study of time evolution. In section~\ref{s:holography} we introduce a simple model in five dimensions, which is the holographic dual of our toy model. Holography will provide a very intuitive picture of the time evolution of the system and we actually use the holographic model in the explicit computations. Time evolution is first studied in section~\ref{s:freetime} for an isolated extra sector, that is, when the coupling $g$ to the elementary fields vanishes. We finally study time evolution in the presence of interactions between both sectors in section~\ref{s:interactiontime}. We study the survival probability of the initial state and distinguish decay into elementary particles from oscillations into different states of the extra sector. This allows us to identify the origin of the hidden states in the continuum as the asymptotic oscillation into a subspace that is not connected to the elementary fields. In section~\ref{s:processes} we discuss how these somewhat formal results are relevant to, and can be observed in, the scattering of elementary particles. We emphasize in this regard the decisive role of energy-momentum uncertainties. We summarize our results in section~\ref{s:summary}. Finally, in section~\ref{s:discussion} we discuss variations of the basic setup and briefly comment on phenomenological implications. 

%%%%%%%%%%%%%%%%%%

\section{Setup}
\label{s:setup}
We will work with a simple toy model that captures essential features of the scenario we have described in the introduction. It is a four-dimensional effective theory of a real scalar field $A$ coupled locally to a massless scalar field $\varphi$, which we call the elementary field, with Lagrangian 
\beq
\mathcal{L} = -\frac{1}{2} A \Pi(-\d^2) A + \d_\mu \varphi^\dagger \d^\mu \varphi + g A \varphi^\dagger \varphi , \label{Lag}
\eeq
where $\Pi$ is an arbitrary form factor, up to the restrictions enforced by the axioms of quantum field theory. As discussed above, this effective theory can arise from a more conventional one with some extra sector that has been integrated out (without any derivative expansion). The field $A$ is to be understood as a local operator of the extra sector that has not been integrated out or as a mediator field that couples linearly to such an operator (more formally, $A$ can be the Legendre dual of the operator, treated as a dynamical field as in~\cite{Perez-Victoria:2001lex}). For simplicity we do not include  $A$ self-interactions in the effective theory, which correspond to three-point and higher-point correlation functions in the extra sector. Even if they will be generically present, they can be suppressed in certain regimes, such as the large $N$ limit of an $SU(N)$ gauge theory. We will comment on the impact of $A$ self-interactions in the conclusions. The local coupling to the elementary field $\varphi$ arises if it only interacts with the extra sector via $A$. Choosing the mass term of $\varphi$ to vanish requires fine tuning and is not essential, but we make this choice for maximal simplicity. In section~\ref{s:holography} we give a five-dimensional realization of a UV completion of this model, inspired by holographic dualities. 

We will work throughout the paper in the approximation with tree-level $\varphi$ propagator,  dressed $A$ propagator---obtained by the Dyson resummation of all the contributions with an arbitrary number of $A$ self-energies, which we calculate at one loop---and no vertex or higher-point one-particle irreducible loop corrections. We will call this the {\em A1 approximation}. The resummation for the $A$ propagator is essential to describe correctly the behaviour of the unstable (un)particles associated to the field $A$. 
To calculate the self-energy we use dimensional regularization and the $\overline{\mathrm{MS}}$ scheme.\footnote{We also impose as a renormalization condition that the one-point function of $A$ vanishes. For a masless $\varphi$ this is automatic in $\overline{\mathrm{MS}}$ at one loop for the Lagrangian~\refeq{Lag}.} 
The renormalized one-loop contribution to the $A$ self-energy is then
\beq
\Sigma(p^2) = -\frac{g^2}{16\pi^2} \log \frac{-p^2}{M^2},
\eeq
with $M$ a renormalization scale. Changes in $M$ can be absorbed into $\Pi(0)$. The logarithm above is defined as usual with a branch cut along the negative real axis, so $\Sigma$ has a branch cut along the positive real axis. This is relevant because we will often make use of amplitudes and Green functions evaluated at complex momenta. We will need the imaginary part of $\Sigma$ right above the real axis, which corresponds to Feynman boundary conditions and is equal to half the discontinuity across the branch cut: 
\beq
\im \Sigma(p^2+i 0^+) = \frac{g^2}{16 \pi} \theta(p^2) . \label{imSigma}
\eeq
Note that it is positive. 
In this approximation, the quadratic part in $A$ of the quantum effective action reads
\beq
\Gamma^{(2)} = - \int d^4 x \frac{1}{2} A \big[ \Pi(-\d^2) + \Sigma(-\d^2) \big] A  .
\eeq
The dressed propagator for the field $A$ is just the inverse of the operator in the square bracket above:
\begin{align}
i G(p^2) &= \frac{i}{\Pi(p^2) + \Sigma(p^2)}  \nn
&= \frac{i}{\Pi(p^2)} \left[ 1 - \frac{\Sigma(p^2)}{\Pi(p^2)} + \dots   \right] . \label{G}
\end{align}
The expansion in the second line is a geometric series in powers of $g$. When $g=0$ all the fields in~\refeq{Lag} are free and the propagator reduces to
\beq
i G^{(0)}(p^2) = \frac{i}{\Pi(p^2)} .
\eeq
The K\"all\'en-Lehmann spectral representation of the propagator is
\beq
i G(p^2) = \int_0^\infty d \mu^2 \sigma(\mu^2) \frac{i}{p^2-\mu^2} , \label{KL}
\eeq
with the spectral density $\sigma$ given by
\begin{align}
\sigma(\mu^2) & = -\frac{1}{\pi} \im G(\mu^2+ i 0^+) \nn
                    & = \frac{1}{\pi} \frac{\im \left(\Pi(\mu^2+i 0^+)+\Sigma(\mu^2+i0^+) \right)}{\left|\Pi(\mu^2+i 0^+) + \Sigma(\mu^2+i 0^+)\right|^2} , \label{sigma}
\end{align}
with real and positive $\mu^2$. In the free theory, we obviously have
\begin{align}
\sigma^{(0)}(\mu^2) & = -\frac{1}{\pi} \im G^{(0)}(\mu^2+ i 0^+) \nn
                    & = \frac{1}{\pi} \frac{\im \Pi(\mu^2+i 0^+)}{\left|\Pi(\mu^2+i 0^+)\right|^2} . \label{sigma0}
\end{align}
In the interacting theory the spectral density in Eq.~\refeq{sigma} can be written as $\sigma=\sigma_1 + \sigma_{2}$, where
\begin{align}
\sigma_1(\mu^2) & = \frac{1}{\pi}\frac{\mathrm{Im}\, \Pi(\mu^2+i 0^+)}{|\Pi(\mu^2+i 0^+)+\Sigma(\mu^2+i 0^+)|^2}  \nn
& = \sigma^{(0)}(\mu^2) + O(g^2)  \label{sigma1}
\end{align}
and
\begin{align}
\sigma_2(\mu^2) & = \frac{1}{\pi}\frac{\mathrm{Im}\, \Sigma(\mu^2+i 0^+)}{|\Pi(\mu^2+i 0^+)+\Sigma(\mu^2+i 0^+)|^2}  \nn
& =  \frac{1}{\pi}\frac{\mathrm{Im}\, \Sigma(\mu^2+i 0^+)}{|\Pi(\mu^2+i 0^+)|^2}+ O(g^4).  \label{sigma2}
\end{align}
Using the spectral representation, we can also decompose $G=G_1+G_2$ with
\beq
G_i(p^2) = \int_0^\infty d\mu^2 \frac{\sigma_i(\mu^2)}{p^2-\mu^2},~~ i=1,2. \label{spectralrep12}
\eeq
Let $P_\mu$ be the momentum operator, with Hamiltonian $P_0=H$ and $P^{(0)}_\mu$ be the momentum operator in the free theory ($g=0$), with free Hamiltonian $P^{(0)}_0=H_0 = H|_{g=0}$. Up to a point at $\mu^2=0$ associated to the massless particles created by $\varphi$, the spectra of $P^\mu P_\mu$ and ${P^{(0)}}^\mu P^{(0)}_\mu$ are given by the support of $\sigma$ and $\sigma^{(0)}$, respectively. We shall consider two cases, depending on the type of spectrum of the free theory: in the {\em discrete case},
\beq
\sigma^{(0)}(\mu^2) = \sum_n F_n \delta(\mu^2-m_n^2) ,
\eeq
while in the {\em continuous case} $\sigma^{(0)}$ is a non-singular function. Mixed situations with both discrete and continuous spectrum are also possible, but we will not study them in this paper. We assume that in both cases the free theory for the field $A$ has a mass gap $\mu_0$, that is, $\sigma^{(0)}(\mu^2)=0$ for $\mu^2<\mu_0^2$. %For finite $g$, the branch cut in $\Sigma$ (in our case with branch point at $p^2=0$) moves the poles away from the real line and $\sigma$ becomes a smooth function with peaks near $\mu^2=m_n^2$. Physically, this means that in the interacting theory $A$ creates no stable one-particle states. 
On the other hand, in the interacting theory $\sigma$ is always a non-singular function.\footnote{\label{tachyon}In the examples we have considered $\sigma$ actually contains one discrete delta function $F_T \delta(\mu^2-m_T^2)$ with $m_T^2<0$ and $|m_T^2|\ll M^2$. This tachyonic mode signals a failure of the $A1$ approximation, due to IR divergences in the presence of the massless elementary field $\varphi$. The tachyon would not appear in this approximation if the elementary field were massive and in the massless case it could in principle be avoided by a Sudakov-like resummation. Nevertheless, in this paper we insist in working with a massless elementary field in the $A1$ approximation, for simplicity, and in the following we just ignore the tachyon in our formulas and discussions. This is possible  because the value of $F_T$ is exponentially small for perturbative values of the coupling $g$, so all the effects of the tachyon on the physics we want to describe are negligible. Because we ignore the tachyon, we will write the integrals on $\mu^2$ with range from 0 to $\infty$ (rather than from $-\infty$ to $\infty$), as we have already done in~\refeq{KL}. } The contribution $\sigma_1$ has support included in $[\mu_0^2,\infty)$, while $\sigma_2$ is non vanishing in all $\mathbb{R}^+$. Moreover, $\sigma_1$ identically vanishes in the discrete case for $g\neq 0$.

\begin{figure}
  \centering
  \includegraphics[width=0.4\textwidth]{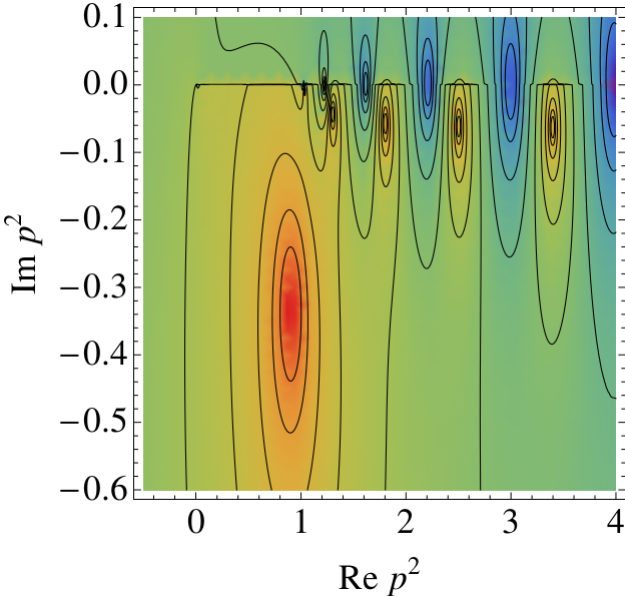}%{analytic_discrete.pdf}
 \hspace{1cm}
  \includegraphics[width=0.4\textwidth]{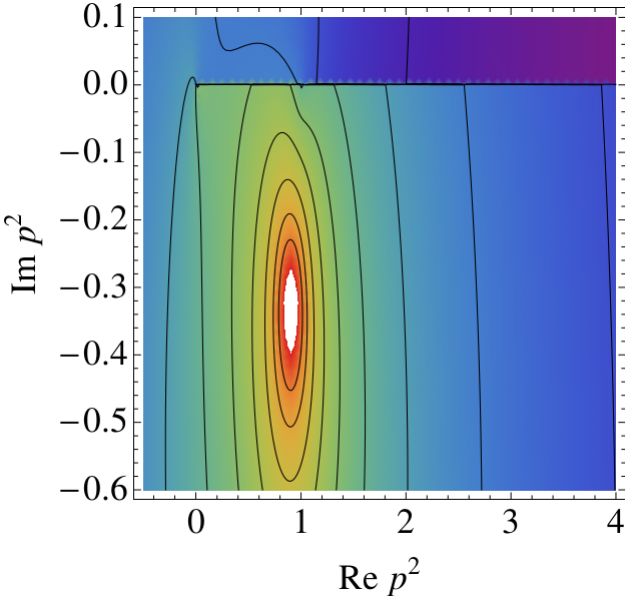}%{analytic_continuous.pdf}
  \caption{Analytic structure of the propagator on the second Riemann sheet for complex $s$ in a particular model, to be introduced in section~\ref{s:holography} . We plot $|G^{\mathrm{II}}(s)|$ in the discrete case for $L= 10$ (left) and in the continuous case (right), with $g= 4$, $\mu_0 = 1$ and $M = 10$. Here, $L$ is approximately the inverse of the mass spacing. }
  \label{fig:analytic}
\end{figure}

All this is directly related to the analytical structure of $G$. In the free theory,  the spectral representation~\refeq{KL} implies that $G^{(0)}(p^2)$ has simple poles at the real values $m_n^2$ in the discrete case, 
\beq
G^{(0)}(p^2) = \sum_n F_n \frac{1}{p^2-m_n^2} ,
\eeq
while it has a branch cut along the real interval $[\mu_0,\infty)$ in the continuous case. This cut can be intuitively understood in the continuum limit of the discrete case as the collective effect of closer and closer poles with smaller and smaller residua. In the interacting theory, the $G_2$ part of the propagator $G$ has a branch cut along $[0,\infty)$ in both cases. In the discrete case, when the interaction is turned on the poles move away from the real axis into the fourth quadrant of the second Riemann sheet and $G_1=0$. 
In the continuous case with interaction, $G_1$ does not vanish and has a branch cut along the real interval $[\mu_0,\infty)$. Therefore, in this case $G$ has two branch cuts, partially superimposed.\footnote{Further cuts,  with branch points at higher values of $\mu^2$, occur beyond the A1 approximation.} In the continuum limit, the complex poles of the interacting discrete case move closer and closer to the real axis, which agrees with the fact that the extra branch cut in the continuum case is located on the real axis. At least one complex pole can also be present in the interacting continuous case, as shown and emphasized in~\cite{Delgado:2008gj}. The analytic structure of $G^{\mathrm{II}}$, the propagator analytically continued into the second Riemann sheet across the real interval $(0,\mu_0)$, is illustrated in figure~\ref{fig:analytic} for the model introduced in section~\ref{s:holography}.

To finish this section, we give an alternative form of theory~\refeq{Lag}, which is obtained by {\em integrating in\/} a discrete or continuous tower of real scalar fields $B_\mu$ (here and in several quantities below we use the subindex $\mu:=\sqrt{\mu^2}$, not to be confused with a Lorentz index) with standard kinetic terms:
\begin{align}
\mathcal{L}^\prime = & \d_\nu \varphi^\dagger \d^\nu \varphi   \nn
&  \mbox{} + \int_0^\infty d\mu^2 \sigma^{(0)}(\mu^2) \Big( \frac{1}{2} \d_\nu B_\mu \d^\nu B_\mu - \frac{\mu^2}{2} B_\mu^2 
 + g B_\mu \varphi^\dagger \varphi \Big) . \label{Lagprime}
\end{align} 
The Lagrangians $\mathcal{L}$ and $\mathcal{L}^\prime$ represent equivalent theories, as can be shown by integrating out the fields $B_\mu$, subject to the constraint
\beq
A = \int_0^\infty d\mu^2 \sigma^{(0)}(\mu^2)  B_\mu .  \label{constraint}
\eeq
Using a Lagrange multiplier for the constraint~\refeq{constraint}, this is a Gaussian functional integral, which straightfowardly gives~\refeq{Lag} as a result.

\section{Unitarity}
\label{s:unitarity}

\begin{figure}
  \centering
  \includegraphics[width=0.7\textwidth]{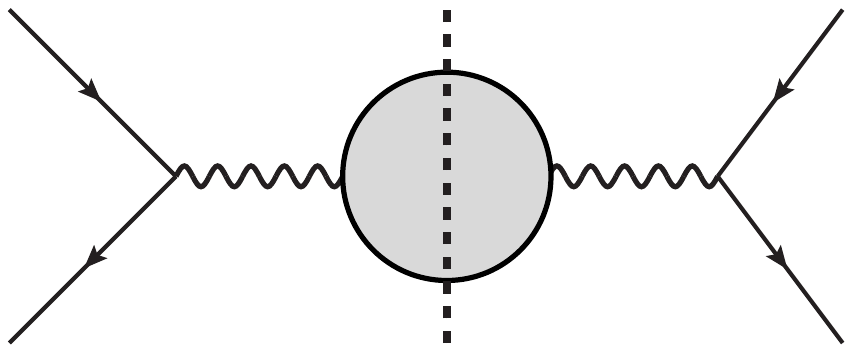}
  \vspace{-9cm}
  \caption{Diagram contributing to $\Mcal_\varphi$ with dressed $A$ propagator in the $s$ channel represented as a wavy line with a blob; the cut extracts the discontinuity across the real axis. The external lines correspond to $\varphi$ fields.}
  \label{fig:dressedcut}
\end{figure}

We assume that the effective theory described by~\refeq{Lag} arises from a unitary theory.  In this section, we use unitarity to learn about possible hidden states, for both discrete and continuous spectrum, without knowledge of the UV completion. Specifically, we consider the optical theorem for $\varphi \bar{\varphi} \to \varphi \bar{\varphi}$ scattering, where $\varphi$ ($\bar{\varphi}$) denotes the (anti) particle created by the field $\varphi$. In our model, this process is mediated by the $A$ field. The Feynman diagrams for this process have the same topology as the ones for Bhabha scattering in QED. The optical theorem relates the cross section and the imaginary part of the forward amplitude. Moreover, in perturbation theorem there are also {\em partial} optical theorems satisfied separately for the imaginary part of the sum of certain subsets of Feynman diagrams and certain contributions (not necessarily of the same diagrams) to the cross section. This has been shown in~\cite{Veltman:1963th} as part of a proof of unitarity with resummed propagators, assuming as we do the spectral representation~\refeq{KL} (for the free theory). We make this observation because, for the amplitude we are considering, the optical theorem involves not only selfenergies, but also higher-point loop subdiagrams. To simplify the discussion, we will consider instead the partial optical theorem that involves only the contribution $\mathcal{M}_\varphi^{(s)}$ to the amplitude, given in this approximation by the s-channel diagram displayed in figure~\ref{fig:dressedcut}. 
%and restricting the number of particles in the final state.

Let $\Mcal_\varphi(s,t) =\Mcal\left(\bar{\varphi} (p_1) \varphi (p_2) \to \bar{\varphi}(p_3) \varphi(p_4)\right)$, with $s=(p_1+p_2)^2$ and $t=(p_1-p_3)^2$ the Mandelstam variables, be the on-shell amplitude of the process. The total cross section with  initial state $\bar{\varphi}(p_1)\varphi(p_2)$ is proportional to 
 \beq
 \tilde{\sigma}_{\varphi}(p_1,p_2) =   \sum_X \int d \Pi_X(2\pi)^4 \delta^{(4)}(p_1+p_2-p_X) |\Mcal(\bar{\varphi}(p_1)\varphi(p_2) \to X)|^2 , \label{totalcrosssection}
 \eeq
with $X$ any possible final state and $d\Pi_X$ the corresponding Lorentz invariant phase space measure. 
The optical theorem for this amplitude is the statement
\beq
\im \mathcal{M}_\varphi (s,0) = \frac{1}{2} \tilde{\sigma}_\varphi(p_1,p_2) . \label{optical}
\eeq
We concentrate on the partial optical theorem
\beq
\im \mathcal{M}^{(s)}_\varphi (s,0) = \frac{1}{2} \tilde{\sigma}^{(s)}_\varphi(p_1,p_2) , \label{partialoptical}
\eeq
where $\tilde{\sigma}^{(s)}$ refers to contributions to the total cross section mediated by the dressed $A$ propagator in the $s$ channel. When $A$ creates only one particle in the free theory, this particle is unstable in the interacting theory. Unstable particles do not form independent states of the Hilbert space, as they do not survive at asymptotic times. In fact, Veltman has shown in \cite{Veltman:1963th} that, as long as dressed propagators are used, the optical theorem is satisfied without including unstable particles as final states in the total cross section. In that familiar particular case, it is easy to see that~\refeq{partialoptical} holds in the A1 approximation if we only include the final states $X=\ket{\bar{\varphi}(q_1) \varphi(q_2)}$, corresponding to elastic scattering with final momenta $q_i$. Then $\tilde{\sigma}^{(s)}$ is proportional to $| \mathcal{M}^{(s)}_\varphi|^2$. In general, the s-channel amplitude is
\beq
\mathcal{M}^{(s)}_\varphi (s,t) = - g^2 G(s+i0^+).
\eeq
This gives the quadratic s-channel contribution of the $\bar{\varphi}\varphi$ final state to the total cross section
\beq
\tilde{\sigma}^{(s)\bar{\varphi}\varphi}_\varphi(p_1,p_2) = \frac{g^4}{8\pi}  \frac{1}{|\Pi(s+i0^+)+\Sigma(s+i 0^+)|^2}   \label{stotalcrosssection}
\eeq
and
\begin{align}
\im \Mcal^{(s)}_\varphi(s,0) & = g^2 \pi \sigma(s) \nn
& = g^2 \left(\im \Pi(s+i 0^+)+\frac{g^2}{16\pi} \right) \frac{1}{|\Pi(s+i 0^+)+\Sigma(s+i 0^+)|^2 } .
\end{align}
We see that the partial optical theorem~\refeq{partialoptical} is satisfied, including only the $\bar{\varphi}\varphi$ final state, if and only if $\mathrm{Im} \,\Pi(s+i 0^+) =0$. Therefore, it is satisfied  in the discrete case, just as for the case of a single unstable particle,\footnote{When $g=0$ the infinitesimal imaginary parts in the discrete case give rise to the delta functions in $\sigma^{(0)}$, but this does not happen when $g\neq 0$ because then the denominator does not vanish for any real values of $s$.} but not in the continuum case. For the latter, there is an excess in the imaginary part:
\beq
\Delta \im \Mcal^{(s)}_\varphi(s,0) = \frac{g^2 \im \Pi(s+i0^+)}{|\Pi(s+i0^+)+\Sigma(s+i 0^+)|^2 } > 0. \label{excess}
\eeq
In a unitary theory, this excess must correspond to other possible final states in this process. So it is actually a deficit in the total cross section. One could think of final states formed by two or more $\varphi\bar{\varphi}$ pairs, but the phase space of these processes does not match the form of the excess~\refeq{excess}. They actually correspond to cuts in multi-loop diagrams beyond the A1 approximation.\footnote{For instance, the final state $\varphi\bar{\varphi}\varphi\bar{\varphi}$ enters a partial optical theorem that involves the imaginary part of a box contribution to  $\mathcal{M}_\varphi (s,0)$ with two dressed $A$ propagators carrying timelike momentum.}
This means that in the continuous case of the interacting theory, even if the states created by $A$ can decay into $\bar{\varphi}\varphi$, there exist asymptotic states that are not superpositions of multi-particle states of $\varphi$ and $\bar{\varphi}$. This differs drastically from the familiar situation of an unstable particle, and, more generally, from the discrete case.\footnote{In fact, in strict perturbation theory to order $g^2$, the optical theorem is satisfied if we include the particles or unparticle stuff created by $A$ as a final state and calculate the cross section with Georgi's formula (which works both in the discrete and continuous case at the tree level):
\begin{align}
\tilde{\sigma}(\bar{\varphi}\varphi \to A)(p_1,p_2) &= \int \frac{d^4 p_A}{(2\pi)^4} \theta(E_A) 2\pi \sigma^{(0)}(p_A^2) (2\pi)^4 \delta^{(4)}(p_1+p_2-p_A)  g^2 + O(g^4) \nn
& = 2 g^2 \pi \sigma^{(0)}(s) +O(g^4).
\end{align} 
This is not surprising because the particles or unparticle stuff in the final state of this process are stable to order $g^2$, and thus have asymptotic states in both the discrete and the continuous cases. But all this is an artifact of strict perturbation theory.}  Note that Veltman's proof of unitarity without extra asymptotic states assumes in the final step a particular form of the free propagator, without any branch cut, so it does not apply to the continuous case.

Our main purpose in this paper is identifying the extra asymptotic states within the effective-theory description and understanding their production mechanism. We also want to understand the discontinuity in the continuum limit of~\refeq{excess}: $\Delta \im \Mcal^{(s)}_\varphi(s,0)$ vanishes for arbitrarily small values of the spacings $\Delta m_n ^2 = m_{n+1}^2-m_{n}^2$, but not in the continuum. We will throw light on these issues by studying in detail the time evolution of the state created by the field $A$.

%%%%%%%%%%%%%%%%%%%

\section{Energy eigenstates}
\label{s:eigenstates}

In order to study the time evolution of the states in the theory, we first identify a basis of generalized energy eigenvectors for both the free and the complete Hamiltonians, using the corresponding spectral functions.

\subsection{Free Hamiltonian}

The Fock vacuum $\ket{0}$ is the (unique) eigenstate of $H_0$ with vanishing eigenvalue. It is also an eigenstate of $\vec{P}$ with vanishing eigenvalues. The Hilbert space of the theory is spanned by the Fock vacuum and the free multi-particle states, which are tensor products of the free one-particle states created by the fields $\varphi$ and $A$. Let us concentrate on the latter. We define the {\em free $A$ one-particle eigenstates\/} $\ket{\mu,\vec{p}}_0$ as the eigenstates of $\vec{P}$ and $H_0$ with eigenvalues $\vec{p}$ and $\omega_{\mu,p}:=\sqrt{\vec{p}^2+\mu^2}$, respectively, which are created by the action of $A$ on the Fock vacuum:
\beq
{}_0\bra{\mu,\vec{p}} A(0) \ket{0} =  Z_{0\mu}^{\frac{1}{2}} \neq 0 . \label{creation}
\eeq
The possible values of $\mu^2$ correspond to the points in the support of $\sigma^{(0)}$. We can assume that these eigenvectors are unique for each $\mu$ and $\vec{p}$, as any degeneracy would be unobservable in the theory~\refeq{Lag}. We choose the phases such that $Z_{0\mu}^{1/2}$ is real. The free eigenvectors fulfil the orthogonality relation
\beq
{}_0\braket{\mu,\vec{p}}{\nu,\vec{q}}_0 = (2\pi)^3 \frac{2 \omega_{\mu,p}}{H_\mu}  \delta_{\mu,\nu} \delta^{3}(\vec{p}-\vec{q}), \label{orthodisc}
\eeq
in the discrete case, and
\beq
{}_0\braket{\mu,\vec{p}}{\nu,\vec{q}}_0 = (2\pi)^3 \frac{2 \omega_{\mu,p}}{H(\mu^2)} \delta(\mu^2-\nu^2) \delta^{3}(\vec{p}-\vec{q}), \label{orthocont}
\eeq
in the continuous case. We have kept an arbitrary normalization, parametrized by $H_\mu$ or $H(\mu^2)$. Note that the products $Z_{0\mu} H_\mu$ and $Z_{0\mu} H(\mu^2)$ do not depend on the normalization of the eigenstates. The free $A$ one-particle states are by definition the normalized states belonging to the Hilbert subspace $\mathcal{H}^{(0)}_{1A}$ spanned by the set of $\ket{\mu,\vec{p}}_0$. The corresponding completeness relation reads
\beq
1^{(0)}_{1A} = \int_0^\infty d \mu^2 \rho(\mu^2)   \int \frac{d^3 p}{(2\pi)^3} \frac{1}{2 \omega_{\mu,p}} \ket{\mu,\vec{p}}_0{}_0\bra{\mu,\vec{p}}, \label{identityfree}
\eeq
where $1^{(0)}_{1A} $ is the identity in $\mathcal{H}^{(0)}_{1A}$ and
\begin{align}
\rho(\mu^2) & := \sum_{\nu^2 \in \mathcal{K}_0} H_\nu \, \delta(\mu^2-\nu^2)~~\mbox{in the discrete case}, \\
\rho(\mu^2) & := H(\mu^2) \, \mathbf{1}_{\mathcal{K}_0}(\mu^2)~~\mbox{in the continuous case}, \label{rho}
\end{align}
with $\mathcal{K}_0$ the support of $\sigma^{(0)}$ and $\mathbf{1}_C(x)$ the indicator function, giving 1 if $x\in C$ and 0 otherwise.
Using~\refeq{identityfree} in the two-point function of the free theory, the spectral representation of the free propagator follows in the textbook way, with
\beq
\theta(q^0) \sigma^{(0)}(q^2) =   \int_0^\infty d \mu^2 \rho(\mu^2)   \int \frac{d^3 p}{2 \omega_{\mu,p}} \delta^4(q-p_{(\mu)}) |{}_0\bra{\mu,\vec{p}} A(0) \ket{0}|^2 , \label{freesigma}
\eeq
where $p_{(\mu)}^i = \vec{p}_i$ and $p_{(\mu)}^0 = \omega_{\mu,p}$.
Furthermore, using~\refeq{creation} in \refeq{freesigma}, 
\begin{align}
\theta(q^0) \sigma^{(0)}(q^2) & =   \int_0^\infty d \mu^2 \rho(\mu^2)  Z_{0\mu}  \int \frac{d^3 p}{2 \omega_{\mu,p}} \delta^4(q-p_{(\mu)}) \nn
& =  \int_0^\infty d \mu^2 \rho(\mu^2)  Z_{0\mu} \frac{1}{2 \omega_{\mu,q}} \delta(q^0-w_{\mu,q}) \nn
& =  \int_0^\infty d \mu^2 \rho(\mu^2)  Z_{0\mu}  \theta(q^0) \delta(q^2-\mu^2) \nn
& = \theta(q^0) \rho(q^2) Z_{0q} ,
\end{align}
so we identify $\sigma^{(0)}(\mu^2) = Z_{0\mu} \rho(\mu^2)$ and hence, $F_n=Z_{0 m_n} H_{m_n}$. The free $A$ one-particle eigenstates are in one-to-one correspondence with the fields $B_\mu$ in the Lagrangian $\mathcal{L}^\prime$ in~\refeq{Lagprime}:
\begin{align}
{}_0\bra{\mu,\vec{p}} B_\nu(0) \ket{0} & = \delta_{\mu\nu} \frac{1}{Z_{0\mu}^{1/2} H_\mu} ~~\mbox{in the discrete case}, \\
{}_0\bra{\mu,\vec{p}} B_\nu(0) \ket{0} & = \delta(\mu^2-\nu^2) \frac{1}{Z_{0\mu}^{1/2} H(\mu^2)} ~~\mbox{in the continuous case}.
\end{align}

\subsection{Complete Hamiltonian}
The basis of eigenstates of the free Hamiltonian can also be used when $g\neq 0$, that is, in the presence of interactions. But in this case it is often more convenient to use a basis of generalized eigenstates of $\vec{P}$ and the complete Hamiltonian $H$. This basis is formed by the physical vacuum $\ket{\Omega}$, which is a proper eigenstate of $H$ and has the smallest eigenenergy, and tensor products of asymptotic one-particle states. As usual in scattering theory, the latter are characterized by the appearance of their linear combinations to observers at very early or very late times. However, the interaction cannot be neglected at asymptotic times due to the effects of the self-energy, and the spectra of $H$ and $H_0$ are actually different if $g\neq 0$. Hence, we cannot use the free one-particle states to label the asymptotic one-particle states. The asymptotic one-particle eigenstates are, by definition (see~\cite{Weinberg:1995mt}), fully described by the eigenvalues of $P^2$ and $\vec{P}$ plus possible discrete labels, such as the species identification. In our theory, they can be written as $\ket{\mu,\vec{p}}$, $\ket{\vec{p}}_\varphi$, and $\ket{\vec{p}}_{\bar{\varphi}}$, for the particle states created by $A$, particle states created by $\varphi$ and antiparticle states created by $\varphi$, respectively, when acting on the physical vacuum. For the first ones,
\beq
\bra{\mu,\vec{p}} A(0) \ket{\Omega} =  Z_{\mu}^{\frac{1}{2}} \neq 0 . \label{exactcreation}
\eeq
This is only relevant for the continuous case, as it is clear from the discussion of unitarity that $A$ does not create any asymptotic one-particle states in the discrete case when $g\neq 0$.  In the continuous case, we normalize these states similarly to the free $A$ eigenstates:
\beq
\braket{\mu,\vec{p}}{\nu,\vec{q}} = (2\pi)^3 \frac{2 \omega_{\mu,p}}{H_1(\mu^2)} \delta(\mu^2-\nu^2) \delta^{3}(\vec{p}-\vec{q}), \label{orthoint}
\eeq
and we define
\beq
\rho_1(\mu^2) := H_1(\mu^2) \, \mathbf{1}_{\mathcal{K}_0}(\mu^2) .
\eeq
To unify the equations below, we also define $\rho_1:=0$ in the discrete case. We choose the phases of the eigenstates such that $Z_{\mu}^{1/2}$ is real. Of course, in the interacting theory $A$ can create also multi-particle states, which have additional continuous labels. In the A1 approximation, the relevant multi-particle states for our analysis are the $\varphi \bar{\varphi}$ asymptotic two-particle states $\ket{X}=\ket{\vec{p}_1,\vec{p}_2}_{\varphi\bar{\varphi}}$, where $X$ is a collective label and $\vec{p}_i$ is the momentum of each particle. Their total momentum is $\vec{p}_X=\vec{p}_1+\vec{p}_2$ and their total energy, $E_X=|\vec{p}_1| + |\vec{p}_2|$, as the particles are massless. We keep using the adjective ``asymptotic" to emphasize that these states only look like a $\varphi \bar{\varphi}$ pair of particles to observers at asymptotic times (when they form wave packets). We normalize them by
\beq
\braket{X}{Y} = \delta(X-Y),
\eeq
where the delta function is defined by $\int d\Pi_X \delta(X-Y) f(Y) = f(X)$, for any test function $f$, with $d\Pi_X$ the Lorentz-invariant phase factor accounting for all the possible momentum configurations:
\beq
d\Pi_X = \Pi_{j=1}^2 \frac{d^3p_j}{(2\pi)^3}\frac{1}{2\omega_{0,p_j}}.
\eeq

Let $\mathcal{H}_1$ be the Hilbert subspace spanned by the asymptotic one-particle states $\ket{\mu,\vec{p}}$ and $\mathcal{H}_2$ be the Hilbert subspace spanned by the asymptotic two-particle states $\ket{\vec{p},\vec{q}}_{\varphi\bar{\varphi}}$. By definition, these two subspaces are orthogonal. The relevant Hilbert subspace for our purposes is then $\mathcal{H}_A = \{ \ket{\Omega}\} \oplus \mathcal{H}_1 \oplus \mathcal{H}_2$. In the discrete case, $\mathcal{H}_1$ is trivial. The completeness relation within $\mathcal{H}_A$ reads
\beq
1= \ket{\Omega}\bra{\Omega} + \int_0^\infty d \mu^2 \rho_1(\mu^2)   \int \frac{d^3 p}{(2\pi)^3} \frac{1}{2 \omega_{\mu,p}} \ket{\mu,\vec{p}}\bra{\mu,\vec{p}}
% + \int d \Pi_{\{\vec{p}\}} \ket{\{\vec{p}\}}\bra{\{\vec{p}\}} .
+ \int d \Pi_X \ket{X}\bra{X}.  \label{complete}
\eeq
The states in both $\mathcal{H}_1$ and $\mathcal{H}_2$ contribute to the spectral density $\sigma$ in the A1 approximation, while $\ket{\Omega}$ does not because $\bra{\Omega} A \ket{\Omega} =0$ in our renormalization scheme. The discontinuities in $\sigma$ or its derivatives signal, via unitarity, the thresholds of new physical processes. In our theory, the thresholds are at $\mu_0^2$ for the $A$ one-particle states and $0$ for the two-particle states $\ket{X}$. Hence, the states in $\mathcal{H}_1$ and $\mathcal{H}_2$ contribute to $\sigma_1$ and $\sigma_2$, respectively and we have
\begin{align}
& \theta(q^0) \sigma_1(q^2) =   \int_0^\infty d \mu^2 \rho_1(\mu^2)   \int \frac{d^3 p}{2 \omega_{\mu,p}} \delta^4(q-p_{(\mu)}) |\bra{\mu,\vec{p}} A(0) \ket{\Omega}|^2 ,  \label{sigma1states} \\
 & \theta(q^0) \sigma_2(q^2) = \int d \Pi_{X}  (2\pi)^3 \delta^4(q-p_{X}) |\bra{X} A(0) \ket{\Omega}|^2  \label{sigma2states} .
\end{align}
Using~\refeq{exactcreation} we see that $\sigma_1(\mu^2) = Z_\mu \rho_1(\mu^2)$. We can also understand these features of the spectrum in terms of Feynman diagrams by expanding $G$ as a geometric series in the selfenergy. Then, $\sigma_2$ is associated to cuts in the selfenergy blobs, while $\sigma_1$ is associated to cuts in the free $A$ propagators. 

%%%%%%%%%%%%%%%%%%%%%

\section{A holographic model}
\label{s:holography}

In this section, we introduce a specific holographic model, which is dual to a theory with an extra spatial dimension. This will serve to illustrate our general results and will also provide an intuitive picture of time evolution in our model. Even if holographic dualities for renormalizable theories are only well-defined in asymptotically-AdS geometries, it will be sufficient for our purposes to work with a model in flat space. The very same qualitative results can be obtained in more involved geometries, such us the ones employed in~\cite{Falkowski:2008fz}.\footnote{As an example, the momentum-space calculations in asymptotically-AdS spaces with gapped continuous spectrum performed in~\cite{Falkowski:2008yr,Falkowski:2009uy} agree qualitatively with the results to be given in section~\ref{s:processes}.}  The model has at least one boundary (which in AdS would correspond to a UV cutoff~\cite{Perez-Victoria:2001lex}.) We will consider both a compact and a non-compact extra spatial dimension, corresponding to the discrete and continuum cases, respectively.
So, the space-time geometry is $M_4\times I$, with $M_4$ the four-dimensional Minkowski space and $I$ a finite or semi-infinite interval. Locally, the geometry is five-dimensional Minkowski space. The interval $I$ can be parametrized by a coordinate $z \in [0,L]$ with the non-compact model corresponding to $L= \infty$, that is, $z\in [0,\infty)$. We call the boundary at $z=0$ the UV boundary and the one at $L$ (for finite $L$), the IR boundary. The model contains a real scalar field $B$, which propagates in all five dimensions and interacts with a four-dimensional complex scalar field $\varphi$ confined to the UV boundary. The action is
\begin{align}
S[B,\varphi,\varphi^\dagger] = \int d^4x & \left[  \int_0^L dz  \frac{1}{2} \left( \d^M B(x,z) \d_M B(x,z) - \mu_0^2 B(x,z)^2 \right) \right. \nn
&\left. \mbox{} +  \d^\mu \varphi^\dagger(x) \d_\mu \varphi(x) 
%+ \frac{1}{2} m^2 B(x,0)^2 
+ g B(x,0) \varphi^\dagger(x) \varphi(x)  \right] , \label{5Daction}
\end{align}
where the indices  run over all five coordinates while the indices $\mu$ run over the coordinates of $M_4$. We impose the Neumann boundary conditions
\begin{align}
& \left. \d_z B(x,z) \right|_{z=0} = 0 , \nn
& \left. \d_z B(x,z) \right|_{z=L}=0 ~~~\mbox{if $L < \infty$}.
\end{align}
A four-dimensional holographic action is obtained by integrating out the bulk ($z>0$) degrees of freedom of the field $B$. Because there are no bulk interactions, integrating out at the classical level gives an exact result, which can be used at the quantum level. To do this, we work in four-dimensional momentum space and write the on-shell $B$ field as
\beq
\bar{B}(p,z) = A(p) \mathcal{K}(p^2,z) , 
\eeq
where the bulk-to-boundary propagator $\mathcal{K}$ obeys the bulk equation of motion
\beq
(-\d_z^2 + \mu_0^2-p^2) \mathcal{K}(p^2,z) = 0 \label{eom}
\eeq
and the boundary conditions
\begin{align}
& \mathcal{K}(p^2,0)= 1 ,  \label{bc1} \\
& \left. \d_z \mathcal{K}(p^2,z) \right|_{z=L}=0 ~~~\mbox{if $L < \infty$}, \label{bc2finite} \\
& \lim_{z\to \infty} \mathcal{K}(p^2+i \epsilon,z) = 0,~\epsilon>0,  ~~~\mbox{if $L = \infty$}. \label{bc2infinite}
\end{align}
The last condition, to be used in the non-compact case, ensures that the bulk degrees of freedom to be integrated out are only excited from their ground state (which vanishes at Euclidean infinity) by their interaction with the boundary fields. Substituting $B$ by $\bar{B}$ in~\refeq{5Daction}, integrating by parts and using~\refeq{eom}, we find the holographic action 
\begin{align}
S_h[A,\varphi,\varphi^\dagger] & = S[\bar{B}(A),\varphi,\varphi^\dagger] \nn 
& =  \int d^4x  \left[ - \frac{1}{2} A(x) \Pi(-\d^2) A(x) +  \d^\mu \varphi^\dagger(x) \d_\mu \varphi(x) + g A(x) \varphi^\dagger(x) \varphi(x)    \right] \nn
& = \int d^4 x \mathcal{L}(x),  \label{holoaction}
\end{align}
where the Lagrangian $\mathcal{L}$ is exactly of the form~\refeq{Lag} and
\beq
\Pi(p^2) = \left. \d_z \mathcal{K}(p^2,z) \right|_{z=0}.
\eeq
In the holographic interpretation, the boundary field $A$ is an elementary field that couples locally to an operator of the strongly-coupled sector~\cite{Perez-Victoria:2001lex}. The propagator for $A$ and the corresponding spectral density $\sigma$ follow as in section~\ref{s:unitarity}. Let us comment in passing on the relation of these quantities with the Kaluza-Klein decomposition of the five-dimensional field $B$.
The corresponding profiles $f_\mu$ are the solutions of the eigenvalue problem
\beq
(-\d_z^2 + \mu_0^2) f_\mu(z) = \mu^2 f_\mu(z) 
\eeq
with the same boundary conditions as $B$. When $L=\infty$, they have infinite norm, that is, they do not belong to $\mathrm{L}^2([0,\infty))$. Nevertheless, choosing the $L$ and $\mu$-independent normalization $f_\mu(0)=1$, they obey the orthogonality relation
\begin{align}
& \int_0^L dz f_\mu(z) f_\nu(z) = \frac{1}{Z_{0\mu}H_\mu} \delta_{\mu,\nu},~~L<\infty ,\label{ortho1}\\ 
& \int_0^L dz f_\mu(z) f_\nu(z) = \frac{1}{Z_{0\mu} H(\mu^2)} \delta(\mu^2-\nu^2),~~L=\infty , \label{ortho2}
\end{align}
which gives an alternative way of identifying $\sigma^{(0)}$,
and the completeness relation (in the space of functions satisfying the same boundary conditions)
\beq
\int_0^\infty d\mu^2 \sigma^{(0)}(\mu^2) f_\mu(z) f_\mu(z^\prime) = \delta(z-z^\prime) .\label{completeprofile}
\eeq
The latter relation implies that we can write
\beq
B(x,z) = \int_0^\infty d\mu^2 \sigma^{(0)}(\mu^2) f_\mu(z) B_\mu(x).  \label{KKexpansion}
\eeq
Inserting this in the five-dimensional action~\refeq{5Daction}, we get, for both the compact and non-compact cases
\begin{align}
S[B,\varphi,\varphi^\dagger] = & \int d^4 x  \Big[ \d^\nu \varphi^\dagger(x) \d_\nu \varphi(x)   \nn
&  \mbox{} + \int d\mu^2 \sigma^{(0)}(\mu^2) \Big( \frac{1}{2} \d^\nu B_\mu(x) \d_\nu B_\mu(x) - \frac{\mu^2}{2} B_\mu(x)^2 
 + g B_\mu(x) \varphi^\dagger(x) \varphi(x) \Big) 
\Big]  \nn
= & \int d^4 x \mathcal{L}^\prime(x).
 \label{KKaction}
\end{align} 
So, we recover the form~\refeq{Lagprime} of the four-dimensional theory. Coming back to the holographic method, in our specific model the solution to~\refeq{eom} with the boundary conditions~\refeq{bc1} and~\refeq{bc2finite} or \refeq{bc2infinite} is
\begin{align}
& \mathcal{K}(p^2,z) = \frac{\cos\left(\sqrt{p^2-\mu_0^2}(L-z)\right)}{\cos\left( \sqrt{p^2-\mu_0^2} L \right)} ,~~ L<\infty, \\
& \mathcal{K}(p^2,z) = e^{i \sqrt{p^2-\mu_0^2} z}, ~~L=\infty ,
\end{align}
where the square root is defined as the analytic continuation of the real square root with a branch cut on the negative real axis. 
Then, the kinetic form factor is\footnote{The very same results are obtained for a model with a free scalar field propagating in AdS space with squared bulk mass $-15/4 k^2$, where $k$ is the AdS inverse radius, and subject to a bulk potential $\mu_0^2/z^3$.}
\begin{align}
& \Pi(p^2)  = \sqrt{p^2-\mu_0^2} \tan\left( \sqrt{p^2-\mu_0^2}\, L \right), ~~ L<\infty , \label{kineticdiscrete}\\
& \Pi(p^2) = -\sqrt{\mu_0^2-p^2}, ~~ L=\infty. \label{kineticcontinuous}
\end{align}
From the corresponding free propagator we find the free spectral function
\begin{align}
& \sigma^{(0)}(\mu^2) = \frac{1}{L} \left[ \delta(\mu^2-m_0^2) + 2 \sum_{n=1}^\infty \delta(\mu^2-m_n^2) \right] , ~~ L<\infty  \label{sigma0holodisc} \\ 
& \sigma^{(0)}(\mu^2) = \frac{1}{\pi \sqrt{\mu^2-\mu_0^2}} \theta(\mu^2-\mu_0^2), ~~ L = \infty, \label{sigma0holocont}
\end{align}
with $m_n^2 = \mu_0^2 + \pi^2 n^2/L^2$ the Kaluza-Klein masses in the compact case. Note that the mass spacing for large $n$ is $\Delta m_n := m_{n+1}-m_n = \pi/L$. The Kaluza-Klein profiles for the modes are in both cases
\beq
f_\mu(z) = \cos\left(\sqrt{\mu^2-\mu_0^2}\, z \right) ,
\eeq
with $\mu^2 \in \{m_n^2, n=0,1,\dots \}$ in the compact case. The orthogonality and completeness relations~(\ref{ortho1}-\ref{completeprofile}) can be checked explicitly. 
The partial spectral functions in the interacting theory are\footnote{For our explicit kinetic function $\Pi$ in \refeq{kineticcontinuous}, the tachyonic mode mentioned in \cref{tachyon} has squared mass $\mu_T^2 \simeq - M^2 \mathrm{exp}\left(-\frac{16\pi^2 \mu_0}{g^2} \right)$ and residue $F_T \simeq \frac{16 \pi^2}{g^2} \mu_T^2$. We see the residue, and hence the coupling of the tachyon, is a non-perturbative effect. For instance, for the values $g=1$, $g=2$ and $g=3$  (with $\mu_0=1$, $M=10$), which we use in some of our examples, we have $F_T \simeq 4.1 \times 10^{-65}$,  $F_T\simeq 2.8 \times 10^{-14}$ and $F_T \simeq 4.2\times 10^{-5}$, respectively. Hence, we can safely ignore it in all our considerations, as anticipated.}
\begin{align}
& \sigma_1(\mu^2) = 0,~~ L<\infty, \\
& \sigma_1 (\mu^2) = \frac{1}{\pi} \frac{\sqrt{\mu^2-\mu_0^2}} {\left|\sqrt{\mu_0^2-\mu^2-i0^+} + \Sigma(\mu^2+i 0^+) \right|^2} \theta(\mu^2-\mu_0^2) ,~~L=\infty,
\end{align}
and 
\begin{align}
&  \sigma_2(\mu^2) = \frac{1}{\pi} \frac{\mathrm{Im} \,\Sigma(\mu^2+i 0^+)} {\left|  \sqrt{\mu^2+i0^+ -\mu_0^2} \tan\left( \sqrt{\mu^2+i 0^+ -\mu_0^2}\, L \right) + 
\Sigma(\mu^2+i 0^+) \right|^2} ,~~ L<\infty, \\
& \sigma_2(\mu^2) = \frac{1}{\pi} \frac{\mathrm{Im} \, \Sigma(\mu^2+i 0^+)} {\left|\sqrt{\mu_0^2-\mu^2-i0^+} + \Sigma(\mu^2+i 0^+) \right|^2} ,~~ L=\infty.
\end{align}

It will also be interesting to study the propagation of the field from the boundary to an arbitrary point in the bulk. The corresponding resummed 5D propagator can be easily computed in our model, thanks to the fact that the self-energy of the field $B(z)$ is localized on the UV boundary and can be incorporated as a modified boundary condition. The resummed propagator $G(p^2;z,z^\prime)$ is then a Green's function of the equation of motion,
\beq
(-\d_z^2 + \mu_0^2-p^2) G(p^2;z,z^\prime) = \delta(z-z^\prime), \label{Green}
\eeq
satisfying the same IR boundary condition as $\mathcal{K}$ and the UV boundary condition
\beq
\left. \d_z G(p^2;z,z^\prime) \right|_{z=0} = -\Sigma(p^2) G(p^2,0,z^\prime) .
\eeq
For the particular case with one point on the UV boundary, we find
\beq
G(p^2;0,z) = \frac{\mathcal{K}(p^2,z)}{\Pi(p^2)+ \Sigma(p^2)},
\eeq
which generalizes our previous result for $z=z^\prime=0$. It can be readily checked that the free 5D propagator has the spectral representation
\beq
G^{(0)}(p^2;z,z^\prime) = \int_0^\infty d\mu^2 \sigma^{(0)}(\mu^2) \frac{f_\mu(z)f_\mu(z^\prime)}{p^2-\mu^2} .
\eeq

We will use the functions $\Pi$ in eqs.~\refeq{kineticdiscrete} and~\refeq{kineticcontinuous} in our explicit calculations and plots below.

%%%%%%%%%%%%%%%%%%%%

\section{Time evolution for a free field}
\label{s:freetime}

We have now all the tools we need to study the time evolution of states associated to the field $A$. We will derive general results and will illustrate them using the holographic theory presented in section~\ref{s:holography}. In this section we consider the free theory for the field $A$, that is, we set $g=0$ in~\refeq{Lag} and neglect $\varphi$. As we will see, already in this case the time evolution is non-trivial. The two-point function in this free theory is 
\beq
\bra{0} T A(x) A(0) \ket{0} = \int \frac{d^4 p}{2\pi^4} e^{-i x p} i G^{(0)}(p^2+i 0^+).
\eeq
At time $t=0$ the field $A(0,\vec{x})$ creates from the vacuum a one-particle state, which can be expanded in the $\ket{\mu,\vec{p}}_0$ basis:
\begin{align}
A(0,\vec{x}) \ket{0}  & = \int_0^\infty d \mu^2 \rho(\mu^2) \int  \frac{d^3 p}{(2\pi)^3}\frac{1}{2 \omega_{\mu,p}} {}_0\bra{\mu,\vec{p}} A(0,\vec{x}) \ket{0}  \ket{\mu,\vec{p}}_0 \nn
& =  \int_0^\infty d \mu^2 \rho(\mu^2) Z_{0\mu}^{\frac{1}{2}} \int  \frac{d^3 p}{(2\pi)^3}\frac{1}{2 \omega_{\mu,p}}  e^{-i \vec{x} \cdot \vec{p}}  \ket{\mu,\vec{p}}_0 \nn
& = \int \frac{d^3 p}{(2\pi)^3} e^ {-i \vec{x} \cdot \vec{p}} \int_0^\infty d \mu^2 \rho(\mu^2) Z_{0\mu}^{\frac{1}{2}} \frac{1}{2 \omega_{\mu,p}} \ket{\mu,\vec{p}}_0 \nn
& =:  \int \frac{d^3 p}{(2\pi)^3} e^ {-i \vec{x} \cdot \vec{p}} \ket{\mathcal{A}^0_{\vec{p}}} .  \label{creationfromvacuum}
\end{align}
The state $\ket{\mathcal{A}^0_{\vec{p}}}$ has well-defined spatial momentum, but not well-defined energy. Therefore, it evolves non-trivially in time:
\begin{align}
\ket{\mathcal{A}^0_{\vec{p}},t} & = e^{-i t H_0}  \ket{\mathcal{A}^0_{\vec{p}}} \nn
& = \int_0^\infty d \mu^2 \rho(\mu^2) Z_{0\mu}^{\frac{1}{2}} \frac{1}{2 \omega_{\mu,p}} e^{-i t \omega_{\mu,p}} \ket{\mu,\vec{p}} .
\end{align}
The overlap with the initial state is given by
\begin{align}
\braket{\mathcal{A}^0_{\vec{p}}}{\mathcal{A}^0_{\vec{q}},t}  & =  (2\pi)^3 \delta^3(\vec{p}-\vec{q})
 \int_0^\infty d\mu^2 e^{-i t \omega_{\mu,p}}  \frac{\rho(\mu^2) Z_{0\mu}}{2\omega_{\mu,p}}  \nn
  &= (2\pi)^3 \delta^3(\vec{p}-\vec{q})
 \int_0^\infty d\mu^2 e^{-i t \omega_{\mu,p}}  \frac{\sigma^{(0)}(\mu^2) }{2\omega_{\mu,p}},\label{hola}
\end{align}
where we have used the orthogonality relation~\refeq{orthodisc} or~\refeq{orthocont} and the definition~\refeq{rho}.
Note that $\braket{\mathcal{A}^0_{\vec{p}}}{\mathcal{A}^0_{\vec{q}},-t} = \braket{\mathcal{A}^0_{\vec{p}}}{\mathcal{A}^0_{\vec{q}},t} ^*$. 
This overlap is nothing but the two-point function in the time-momentum representation: for $t\geq0$,
\begin{align}
\braket{\mathcal{A}^0_{\vec{p}}}{\mathcal{A}^0_{\vec{q}},t}  & = \int d^3 x  \int d^3 y \, e^{-i \vec{x}\cdot \vec{p}} e^{i \vec{y}\cdot \vec{q}}  \bra{0} A(0,\vec{x}) e^{-i t H_0}  A(0,\vec{y}) \ket{0} \nn
&=  \int d^3 x \int d^3 y \, e^{-i \vec{x}\cdot \vec{p}} e^{i \vec{y}\cdot \vec{q}}  \bra{0} A(t,\vec{x})  A(0,\vec{y}) \ket{0} \nn
&=  \int d^3 x  \int d^3 y \, e^{-i \vec{x}\cdot \vec{p}} e^{i \vec{y}\cdot \vec{q}}  \bra{0} T A(t,\vec{x})  A(0,\vec{y}) \ket{0} \nn
& =  \int d^3 x  \int d^3 y \, e^{-i \vec{x}\cdot \vec{p}} e^{i \vec{y}\cdot \vec{q}}  \bra{0} T A(t,\vec{x}-\vec{y})  A(0,\vec{0}) \ket{0} \nn
& =  (2\pi)^3  \delta^3(\vec{p}-\vec{q}) i \tilde{G}^{(0)}(t,\vec{p}) , \label{overlapprop}
\end{align}
where we have used the invariance of the vacuum under time translations and symmetry of the theory under spatial translations. In the last line we have defined (for arbitrary $t$)
\beq
\tilde{G}^{(0)}(t,\vec{p}) : = \int_{-\infty}^\infty \frac{dE}{2\pi} e^{-i E t} G^{(0)}(E,\vec{p}) , \label{energytime}
\eeq
with $G^{(0)}(p_0,\vec{p}) :=G^{(0)}(p^2+i 0^+)$.
The $t<0$ version of~\refeq{overlapprop}  is obtained by complex conjugating its right-hand side.
In terms of $\tilde{G}^{(0)}$, Eq.~\refeq{hola} reads
\beq
i \tilde{G}^{(0)}(t,\vec{p}) =  \int_0^\infty d\mu^2 e^{-i |t| \omega_{\mu,p}} \frac{\sigma^{(0)}(\mu^2)}{2\omega_{\mu,p}} ,   \label{spectralGrhoZ}
\eeq
which is just another form of the spectral representation. 

As the function $\tilde{G}^{(0)}$ gives, up to a momentum-conservation delta, the overlap of initial and final states with well-defined momentum, its modulus square should give the survival probability of the initial state after a time $t$ has elapsed. This entails a technical problem, however: for this interpretation, we need to normalize the initial state\footnote{The norm is preserved  under the unitary time evolution.}, but the state is not normalizable for an arbitrary kinetic operator $\Pi$. Indeed, the norm is given by Eq.~\refeq{overlapprop} with $t=0$, and $\tilde{G}^{(0)}(0,\vec{p})$ 
diverges unless $\sigma^{(0)}(\mu^2)/\omega_{\mu,p} \lesssim C \mu^{-2-\delta}$ at large $\mu^2$ for some positive $C$ and $\delta$. We will actually find this divergence in our explicit examples, as $\delta=0$ for the spectral function~\refeq{sigma0holocont}. This is just an instance of a generic feature of quantum field theory: the correlation functions of fields have singularities at coincident points (even at the tree level) and are actually not functions but tempered distributions.\footnote{In an interacting theory, renormalization is required at the loop level because the product of tempered distributions is not a tempered distribution in general.} Correspondingly, the local fields are also to be understood as operator-valued distributions, to be smeared with Schwartz test functions~\cite{Streater:1989vi}\footnote{Schwartz test functions are $C^\infty$ functions of fast decrease (decreasing faster than any power of the coordinates when they approach infinity).}:
\beq
A(x) \to A_f = \int d^4x A(x) f(x) .  \label{smearA}
\eeq
As discussed in section~\ref{s:processes}, in actual scattering processes time and spatial smearing arises from the wave functions of initial and final states. Thus far we have only smeared the operators in the spatial coordinates, when performing the three-dimensional Fourier transform\footnote{Because $e^{-i \vec{x}\cdot \vec{p}}$ is not a Schwartz function, the Fourier transform actually gives rise to another distribution. This is taken into account in~\refeq{overlapprop} by the Dirac delta of the momentum difference.}. When $\tilde{G}^{(0)}(0,\vec{p})$ is divergent, we need to smear the operators in time as well.  When acting on the vacuum, such smeared operators will create normalizable states $A_f \ket{0}$.  
Most of the time we will choose a Gaussian distribution
\beq
f_\tau(t) = \frac{1}{\sqrt{2\pi} \, \tau} e^{-\frac{1}{2} t^2/\tau^2},
\eeq
with $\tau>0$ playing the role of a time uncertainty. 
We will also use occasionally the distribution
\beq
f_\tau(t) = \frac{1}{\pi \tau} K_0\left(\frac{|t|}{\tau}\right) \label{CauchyF},
\eeq
with $K$ the modified Bessel function of the second kind. 
We will refer to the smearing in Eq.~\refeq{CauchyF} as a Cauchy smearing, because the square of its Fourier transform is a Cauchy distribution (that is, a Breit-Wigner).
For any $f_\tau$, we define the time-smeared field
\beq
A_\tau(t,\vec{x}) = \int_{-\infty}^\infty d t^\prime f_\tau(t-t^\prime) A(t^\prime,\vec{x}) .
\eeq
It can be checked that $A_\tau(t,\vec{x}) = e^{i t H_0} A_\tau(0,\vec{x}) e^{-it H_0}$. This smeared field creates at $t=0$ a state
\begin{align}
A_\tau(0,\vec{x}) \ket{0}& = \int \frac{d^3 p}{(2\pi)^3} e^ {-i \vec{x} \cdot \vec{p}} \int_0^\infty d \mu^2 \rho(\mu^2) Z_\mu^{\frac{1}{2}} \hat{f}_\tau(\omega_{\mu,p}) \frac{1}{2 \omega_{\mu,p}} \ket{\mu,\vec{p}}_0 \nn
& =: \int \frac{d^3 p}{(2\pi)^3} e^ {-i \vec{x} \cdot \vec{p}} \ket{ \mathcal{A}^{0\tau}_{\vec{p}} },
\end{align}
with $\hat{f}_\tau$ the Fourier transform of $f_\tau$.
The transition amplitude is then
\beq
\braket{\mathcal{A}^{0\tau}_{\vec{p}}}{\mathcal{A}^{0\tau}_{\vec{q}},t}  =  (2\pi)^3 \delta^3(\vec{p}-\vec{q})
 \int_0^\infty d\mu^2 e^{-i t \omega_{\mu,p}} \frac{\sigma_\tau^{(0)}(\mu^2,\vec{p}^2)}{2\omega_{\mu,p}} , \label{overlapsuave}  
\eeq
where
\beq
\sigma_\tau^{(0)}  (\mu^2,\vec{p}^2) = \tilde{f}_\tau(\omega_{\mu,p}) \sigma^{(0)}(\mu^2),
\eeq
with $\tilde{f}_\tau = {(\hat{f}_\tau)}^2$. For Gaussian smearing,
\beq
\tilde{f}_\tau(E) =  e^{- E^2 \tau^2}
\eeq
while for Cauchy smearing,
\beq
\tilde{f}_\tau(E) = \frac{1}{1+E^2 \tau^2 } . 
\eeq
The generalized spectral function $\sigma_\tau^{(0)}$ is not Lorentz invariant, as a consequence of our non-covariant smearing. Furthermore, we can write
\beq
\braket{\mathcal{A}^{0\tau}_{\vec{p}}}{\mathcal{A}^{0\tau}_{\vec{q}},t}  =  (2\pi)^3 \delta^3(\vec{p}-\vec{q}) i \tilde{G}^{(0)}_\tau(t,\vec{p}),
\eeq
where for Gaussian smearing we have
\begin{align}
i \tilde{G}_\tau^{(0)}(t,\vec{p}) &  =  
\int_{-\infty}^\infty \frac{dE}{2\pi} e^{-E^2 \tau^2} \left[F_\tau(t,E) \mathrm{Re} (i G^{(0)}(E,\vec{p})) +i H_\tau(t,E) \mathrm{Im}(i G^{(0)}(E,\vec{p}))\right],
 \label{gaussiansmear}
\end{align}
where
\begin{align}
& F_\tau(t,E) = \cos(E t) , \\
& H_\tau(t,E)=\frac{1}{2} \left[ e^{-i E t} \mathrm{Erf}\left(\frac{t}{2\tau}-i E \tau\right)  + e^{i E t}   \mathrm{Erf}\left(\frac{t}{2\tau}+i E \tau\right)     \right],
\end{align}
with Erf the error function. When $t \gg \tau$ an excellent approximation to \refeq{gaussiansmear} is given by the time smearing of the propagator,
\beq
i \tilde{G}_\tau^{(0)}(t,\vec{p})   \simeq i
\int_{-\infty}^\infty \frac{dE}{2\pi} e^{-i E t -E^2 \tau^2}  G^{(0)}(E,\vec{p}) ,~~~  t \gg \tau  . \label{approxgaussiansmear}
\eeq
The reason is that in this case the region with opposite time ordering has negligible contribution to the smearing. Using the symmetry properties of this function it is easy to check that $H_\tau$, like $F_\tau$, is real. Moreover, $H_\tau(0,E)=0$, so
\beq
i \tilde{G}_\tau^{(0)}(0,\vec{p})   =   \int_{-\infty}^\infty \frac{dE}{2\pi} e^{-E^2 \tau^2}  \mathrm{Re} (i G^{(0)}(E,\vec{p})) ,
\eeq
which is real and positive.\footnote{This is not true with the approximation~\refeq{approxgaussiansmear}, which is not good at $t=0$.} This property is already apparent in Eq.~\refeq{overlapsuave} for arbitrary $f_\tau$.
We can thus define a normalized inital state 
\begin{align}
\ket{\mathcal{A}_h^{0\tau}} & = \int \frac{d^3p}{(2\pi)^3} \, h(\vec{p}) \frac{1}{\sqrt{i\tilde{G}^{(0)}_\tau(0,\vec{p})}} \ket{\mathcal{A}^{0\tau}_{\vec{p}}} \nn 
& = :   \int \frac{d^3p}{(2\pi)^3}  \, h(\vec{p})  \ket{\bar{\mathcal{A}}^{0\tau}_{\vec{p}}} ,  \label{hintegral}
\end{align}
where $h$ is a  wave function with
\beq
\int \frac{d^3p}{(2\pi)^3}  |h(\vec{p})|^2 = 1.
\eeq
The survival probability of this state after a time $t$ has elapsed is given by
\begin{align}
\mathcal{P}_{\mathrm{sur}}(t) & = |\braket{\mathcal{A}_h^{0\tau}}{\mathcal{A}_h^{0\tau},t}|^2 \nn
& = \left| \int \frac{d^3p}{(2\pi)^3}  |h(\vec{p})|^2 \frac{\tilde{G}^{(0)}_\tau(t,\vec{p})}{\tilde{G}^{(0)}_\tau(0,\vec{p})} \right|^2 .
\end{align}
In particular, if we consider $h$ strongly peaked at $\vec{p}_0$, we can approximate $|h(\vec{p})|^2 \simeq (2\pi)^3 \delta^3(\vec{p}-\vec{p}_0)$, and hence
\begin{align}
\mathcal{P}_{\mathrm{sur}}(t)  & \simeq \left| \frac{\tilde{G}^{(0)}_\tau(t,\vec{p}_0)}{\tilde{G}^{(0)}_\tau(0,\vec{p}_0)}   \right|^2 \nn
& = \mathcal{N}^{-2} |\tilde{G}^{(0)}_\tau(t,\vec{p}_0)|^2 . \label{survivalprob}
\end{align}
For simplicity we consider such a wave function $h$ in the following (dropping the subindex in $\vec{p}_0$), unless otherwise stated. Furthermore, although we will keep $\vec{p}_0$ arbitrary in the equations, in all the plots we will work in the reference frame with $\vec{p}_0=0$.

%%%%%%%%%%%%%%%%%

Let us next discuss some basic features of the time dependence of the survival probability in the free theory. We consider $t \geq 0$ in this discussion. At short times we can expand the integrand in the right-hand side of Eq.~\refeq{gaussiansmear} in powers of $t$, taking advantage of the exponential damping at large energies. Then the linear term in $t$ cancels out and we find in all cases
\beq
\mathcal{P}(t) = 1 - \frac{t^2}{t_Z^2}  + O(t^4) ,  \label{shorttime}
\eeq
with 
\beq
t_Z = \left[\bra{\mathcal{A}_h^{0\tau}} (H_0)^2 \ket{\mathcal{A}_h^{0\tau}} -  \bra{\mathcal{A}_h^{0\tau}} H_0 \ket{\mathcal{A}_h^{0\tau}}^2  \right]^{-\frac{1}{2}}  \label{Zeno}
\eeq
the so-called Zeno time, which is the inverse of the energy uncertainty in the initial state. In particular, the rate $R(t)=d\mathcal{P}_{\mathrm{sur}}(t)/dt$ vanishes at $t=0$. This non-exponential behaviour at short times is a very general behaviour of quantum systems. In our effective field theory the smearing is crucial for the series expansion at $t=0$ to be valid. 
The evolution at later times depends crucially on the nature of the spectrum. 

\subsection{Discrete case}
In the discrete case,
\beq
i \tilde{G}^{(0)}_\tau(t,\vec{p}_0) = \sum_\nu \alpha_{\nu,p} e^{-i t \omega_{\nu,p}}  , \label{oscillation}
\eeq
with $\alpha_{\nu,p} = \frac{1}{2\omega_\nu} Z_\nu H_\nu \tilde{f}_\tau(\omega_{\nu,p})$.  In the familiar case in which $A$ creates only a single mode, there is only one term in the sum and the survival amplitude is a pure phase. Indeed, in this case the initial state is an eigenstate of the free Hamiltonian and thus stationary, so $\mathcal{P}_{\mathrm{sur}}(t)=1$ at all times. (In this case, \refeq{shorttime} holds with infinite $t_Z$.) In general, Eq.~\refeq{oscillation} corresponds to an oscillation of the state of the system into time-dependent linear combinations. For just two modes of squared masses $m_1^2$ and $m_2^2$, the system undergoes Rabi oscillations of frequency $\Omega=|\sqrt{m_2^2+\vec{p}^2}-\sqrt{m_1^2+\vec{p}^2}|/2$. The period thus increases for decreasing spacing $\delta m^2 = m_2^2-m_1^2$ and approaches infinity when $\delta m^2 \to 0$. For a general discrete spectrum, if the different eigenvalues $\omega_{\nu,p}$ have conmesurable ratios (a perfectly fine-tuned situation for $\vec{p}\neq0$), then~\refeq{oscillation} is a Fourier series, and $\mathcal{P}_{\mathrm{sur}}(t)$ is periodic, with frequency equal to the greatest common divisor of the $\omega_{\nu,p}$. In the natural case in which some modes do not have commensurable ratios, the survival probability is no longer periodic and the system never goes back to the initial state. However, $\mathcal{P}_{\mathrm{sur}}(t)$ is almost periodic,\footnote{Almost periodic functions were introduced and studied by Harald Bohr~\cite{bohr1925theorie}. They are complex functions $f$ satisfying the following property: for any $\epsilon>0$, the set $T_\eps$ of translation numbers $\tau_\eps$, such that $|f(t+\tau_\eps)-f(t)|<\eps$, is relatively dense in $\mathbb{R}$; that is, a number $L_\eps>0$ exists such that any interval of size $L_\eps$ has at least one element of $T_\eps$.} with $1-\mathcal{P}_{\mathrm{sur}}(t)$ arbitrarily small after a finite time $t$. This result, proven by Bocchieri and Loinger~\cite{PhysRev.107.337}, is the quantum analogue of the Poincar\'e recurrence theorem in classical mechanics. 
%The proof, due to Bocchieri and Loinger~\cite{BocchieriLoinger}, is simple: i) because the state is normalizable for any $t$, a positive number $\Lambda$ can be chosen such that the contribution of modes with $\nu>\Lambda$ can be made arbitrarily small; ii) the remaining contribution is a finite sum of continuous periodic functions and thus an almost-periodic function. 
Given this recurrent behaviour, it makes sense to study time averages $\langle f(t) \rangle_t = \lim_{T\to \infty} \int_0^T dt f(t) / T$. For $h$ strongly peaked\footnote{More precisely, we assume that  $\omega_{\mu,p} \neq \omega_{\nu,q}$ when $\mu^2 \neq \nu^2$ in the momentum region where $h(\vec{p})$ and $h(\vec{q})$ are both non-negligible.} at fixed momentum $\vec{p}$, the time average state of the system (assuming the mass eigenvalues are non-degenerate) is the ``equilibrium'' mixed state
\begin{align}
\rho & : = \langle \ket{\mathcal{A}^{0\tau}_h,t} \bra {\mathcal{A}^{0\tau}_h,t}  \rangle_t  \nn
& =   \frac{1}{\delta(0)} \int \frac{d^3p}{(2\pi)^3} \frac{h(\vec{p})}{\sqrt{ i \tilde{G}_\tau^{(0)}(0,\vec{p})}}  \int \frac{d^3q}{(2\pi)^3} \frac{h(\vec{q})^*}{\sqrt{ i \tilde{G}_\tau^{(0)}(0,\vec{q})}}  \sum_\nu \frac{Z_\nu \tilde{f}_\tau(\omega_{\nu,p})}{N_\nu^4 (2 \omega_{\nu,p})^2}\delta(\vec{p}^2-\vec{q}^2) \ket{\nu,\vec{p}}_0{}_0\bra{\nu,\vec{q}} ,
\end{align}
where the volume factor $\delta(0) := \delta(\vec{p}^2-\vec{p}^2)$ has mass dimension -2.  The normalization is such that $\mathrm{Tr}(\rho)=1$.
The corresponding effective dimension is defined as 
\beq
d^\mathrm{eff}(\rho) = \frac{1}{\mathrm{Tr}(\rho^2)},
\eeq
This is a measure of how many pure states contribute to the mixture, that is, of the number of degrees of freedom that are explored by the time evolution. We find
\begin{align}
d^\mathrm{eff}(\rho) & =\frac{ \left[ \sum_\nu \frac{Z_\nu H_\nu  \tilde{f}_\tau(\omega_{\nu,p})}{\omega_{\nu,p}} \right]^2 } { \sum_\nu \left[ \frac{Z_\nu H_\nu \tilde{f}_\tau(\omega_{\nu,p})}{\omega_{\nu,p}} \   \right]^2    } .
\end{align}
The relevance of these equations is that, for large $d^\mathrm{eff}(\rho)$, the system stays most of the time in states close to the equilibrium state $\rho$. In particular, 
\begin{align}
\langle \mathcal{P}_{\mathrm{sur}}(t) - \frac{1}{d^\mathrm{eff}(\rho)}  \rangle_t & = \langle \mathcal{P}_{\mathrm{sur}}(t) - \mathrm{Tr} (\rho \ket{\mathcal{A}_h^{0\tau}} \bra{\mathcal{A}_h^{0\tau}}) \rangle_t \nn
&\leq \frac{1}{d^\mathrm{eff}(\rho)} ,
\end{align}
where the first identity follows from the calculation of the trace and in the second one we have used a theorem about average expectation values in ref.~\cite{Short:2011pvc}. So, we expect the time evolution of the system for large $d^\mathrm{eff}(\rho)$ to be as follows: The system moves away from the initial state and quickly reaches states close to the equilibrium state $\rho$. In the case of evenly spaced $\omega_{\nu,p}$ the time scale for such equilibration is inversely proportional to the splittings~\cite{PhysRevA.87.032108}. The system then fluctuates around the equilibrium state, with larger fluctuations being more improbable than smaller ones. The average recurrence time $\tau_\eps$, in which the system goes from one state close to the initial state ($1-\mathcal{P}_{\mathrm{sur}}(t)<\eps$) to another state close to the initial state $(1-\mathcal{P}_{\mathrm{sur}}(t+\tau_\eps)<\eps)$, grows exponentially with $d^{\mathrm{eff}}(\rho)$~\cite{PhysRev.111.689}. 
\begin{figure}
  \centering
  \includegraphics[width=0.4\textwidth]{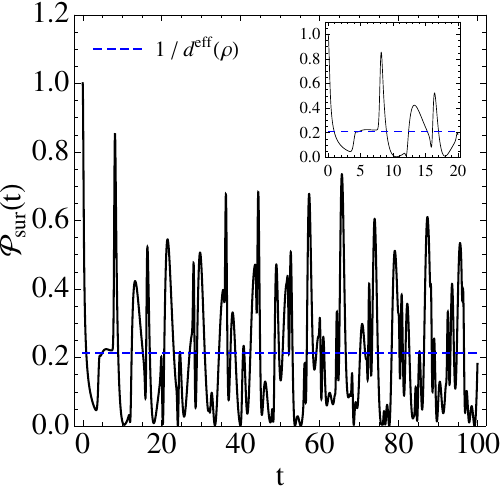}
  \hspace{1cm}
   \includegraphics[width=0.4\textwidth]{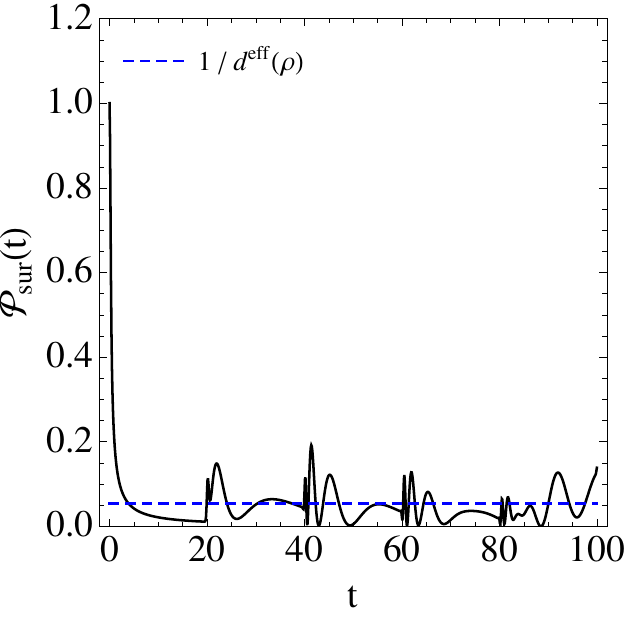}
  \caption{Survival probability of the state created by a free field for $L=2$ (left) and $L=10$ (right). The inserted figure in the left panel is a zoom of the main plot in the range of small $t$. We have considered $g=1$, $\mu_0 = 1$, $M = 10$ and Gaussian smearing with $\tau = 0.1$. Here and in all the plots below we work in the reference frame with $\vec{p}_0=0$.}
  \label{fig:freesurvival_discrete_L1}
\end{figure}
Moreover, for $t\ll 1/\Delta \omega$, with $\omega$ the largest gap between consecutive  $\omega_{\nu,p}$ (for terms in the sum with non-negligible contribution), the sum in $\nu$ can be approximated by an integral over $\mu^2$ with a continuous interpolating function $\sigma_\tau^{(0)}(\mu^2,\vec{p}^2)$. For such times, the survival probability follows closely the one in the theory with the corresponding continuous spectrum, to be studied below. The behaviour of the survival probability in the discrete case is illustrated in ~\ref{fig:freesurvival_discrete_L1}. 

When $d^{\mathrm{eff}}(\rho) \to \infty$, the average survival probability vanishes and the recurrence time is infinite. That is, in this limit the state asymptotically evolves inside a subspace orthogonal to the initial state. This is actually the situation in the continuous case. 

\subsection{Continuous case}
\begin{figure}
  \centering
  \includegraphics[width=0.4\textwidth]{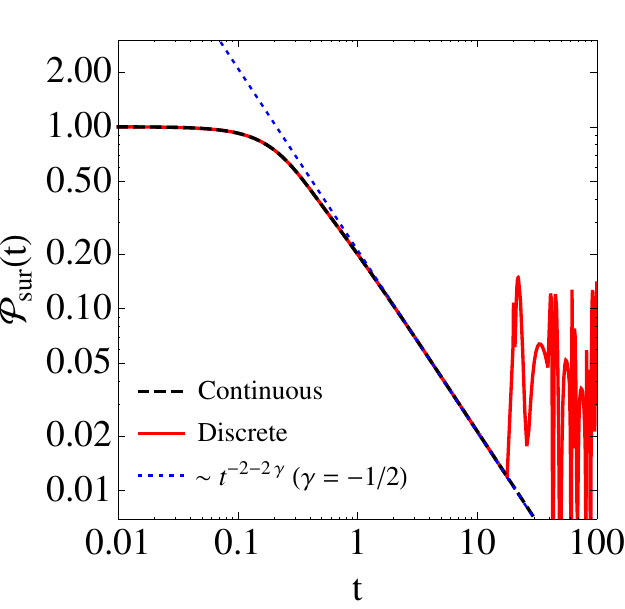}%{freesurvival_continuum.pdf}
  \caption{Survival probability of the state created by a free field in the discrete $(L = 10)$ and continuous case. Comparison with power law for large time behaviour. We have considered $g = 0$, $\mu_0 = 1$ and $M = 10$. We have used Gaussian smearing with $\tau = 0.1$.}
  \label{fig:freesurvival_continuum}
\end{figure}

In the continuous case $\sigma_\tau^{(0)}$ is an absolutely integrable ordinary function in $[0,\infty)$ and the Riemann-Lebesgue lemma  ensures that 
\beq
\lim_{t\to\infty} \mathcal{P}_{\mathrm{sur}}(t) = 0 . \label{RL}
\eeq
So, in this case the initial state decays into orthogonal continuous linear combinations of the one-particle states associated to the free field $A$. The asymptotic decay rate depends on the discontinuities  of $\sigma_\tau^{(0)}$ and its derivatives. Let us assume, in this subsection, that the only discontinuity is the one at the mass gap $\mu_0^2$.  Let $\sigma^{(0)}(\mu^2,\vec{p}^2)/2 \omega_{\mu,p}= \theta(\mu^2-\mu_0^2) \chi(\mu^2)$, with continuous $\chi$ and let $\chi(\mu^2) \simeq C (\mu^2-\mu_0^2)^\gamma/ \kappa^{2(1+\gamma)}$ near $\mu_0^2$, with $\kappa$ some mass scale and $\gamma>-1$ for convergence of the integral. For $t\gg 1/\omega_{\mu_0,p}$ the fast oscillations wash out more strongly the contributions to the integral of regions with larger values of $\mu^2$, so we can approximate (for Gaussian smearing)
\begin{align}
i \tilde{G}_\tau(t,\vec{p}) & \simeq C e^{-i t \omega_{\mu_0,p}} \int_{\mu_0^2}^\infty d\mu^2 e^{-i t \frac{(\mu^2-\mu_0^2)}{ 2 \omega_{\mu_0,p} }} e^{-\omega_{\mu,p}^2 \tau^2}  \left( \frac{\mu^2-\mu_0^2}{\kappa^2} \right)^\gamma \nn
& \simeq C e^{-\omega_{\mu_0,p}^2 \tau^2} e^{-i (\gamma+1) \pi/2} \Gamma(1+\gamma) \kappa^{-2 (1+\gamma)} e^{-i t \omega_{\mu_0,p}} \left( \frac{2\omega_{\mu_0,p}}{t} \right)^{1+\gamma } .
\end{align}
Note that the smearing does not affect the large-time functional form, but it does give rise to a global exponential suppression. We conclude that at large $t$ the decay of the initial state is given by the power law
\beq
\mathcal{P}_{\mathrm{sur}}(t) \propto t^{-2(1+\gamma) }  .\label{freepowerlaw}
\eeq
In figure~\ref{fig:freesurvival_continuum} we show the survival probability in the continuum case and compare it with one in the discrete case and with the large-time approximation in~\refeq{freepowerlaw}, which is excellent for $t> 1/\omega_{\mu_0,p}$. We see that the probability in the discrete and continuous cases are almost indistinguishable for $t < 2L$ (that is, $t<2 \pi / \Delta \omega$) and very different for $t\gtrsim 2 L$. In subsection~\ref{ss:oscillations} we will give a geometric explanation of this abrupt change of behaviour in the discrete case.
%%%%%%%%%%%%%%%%
%%%%%%%%%%%%%%%%

Summarizing, in the free theory the system oscillates between different linear combinations of the free $A$ one-particle states. The oscillations are quasi-periodic in the discrete case, with fluctuations around an equilibrium state, and irreversible in the continuous case. In the latter, the system evolves into a subspace orthogonal to the initial state. We will refer to this irreversible evolution as {\em invisible decay}, as it is unrelated to the visible elementary particles.

%%%%%%%%%%%%%%%%%%%%%%%%%%%%

\section{Time evolution with interaction}
\label{s:interactiontime}

When we turn on the interaction of $A$ with the fields associated to the light particles, that is, when $g\neq 0$, we need to take into account the multi-particle states that are produced by the field $A$. To study the time evolution of the interacting system, which is encoded in the resummed propagator, we follow the same steps as in the free theory. First, we consider the action of the field $A$ on the physical vacuum $\ket{\Omega}$ and evolve the resulting state with the Hamiltonian $H$. Using the completeness relation~\refeq{complete},\begin{align}
\ket{\mathcal{A}_{\vec{p}},t} & = e^{-i t H}  \ket{\mathcal{A}_{\vec{p}}} \nn
& = \int_0^\infty d\mu^2 \rho_1(\mu^2) Z_\mu^{\frac{1}{2}} \frac{1}{2\omega_{\mu,p}} e^{-i t \omega_{\mu,p}} \ket{\mu,\vec{p}} +
\int d\Pi_{X}^{\vec{p}}  \bra{X} A(0) \ket{\Omega} e^{-i t E_X} \ket{X} ,
\end{align}
where  $A(0,\vec{x}) \ket{\Omega} = \int d^3p/(2\pi)^3 e^{-i \vec{x}\cdot \vec{p}} \ket{\mathcal{A}_{\vec{p}}}$ and the second integral is restricted to states with total momentum $\vec{p}$. We have used the vanishing of the one-point function, which implies that $\ket{\mathcal{A}_{\vec{p}}}$ has no vacuum component. 
Note that the two-particle states contribute already at $t=0$, i.e.\ $A(0)$ creates from the physical vacuum a linear combination of one-particle and two-particle states (with no one-particle component in the discrete case). We stress once more that the linear combinations of two-particle states only look like two-particle states when $t\to \infty$. Moreover, the time evolution does not mix the one-particle with the two-particle components of the state (nor with the vacuum). Indeed, let $Q_i$ be the orthogonal projector into the subspace $\mathcal{H}_i$, for $i=0,1,2$ (with $\mathcal{H}_0$ the vacuum one-dimensional space). These projectors commute with the Hamiltonian, which is diagonal in the basis of (orthogonal) generalized eigenvectors. Therefore, $e^{-i H t} Q_i \ket{\mathcal{A}_{\vec{p}}} = Q_i \ket{\mathcal{A}_{\vec{p}},t}  \in \mathcal{H}_i$. 
The overlap of the state at time $t$ with the initial state is
\begin{align}
\braket{\mathcal{A}_{\vec{p}}}{\mathcal{A}_{\vec{q}},t} & = (2\pi)^3 \delta^3(\vec{p}-\vec{q}) \left[\int_0^\infty d\mu^2  e^{-i t \omega_{\mu,p}} \frac{\sigma_1(\mu^2)}{2\omega_{\mu,p}} + \int d\Pi_{X}^{\vec{p}} e^{-i t E_X}  \left| \bra{X} A(0) \ket{\Omega} \right|^2 \right] \nn 
& = (2\pi)^3 \delta^3(\vec{p}-\vec{q}) \left[\int_0^\infty d\mu^2  e^{-i t \omega_{\mu,p}} \frac{\sigma_1(\mu^2)}{2\omega_{\mu,p}} + \int_0^\infty d \mu^2 e^{-i t \omega_{\mu,p}} \frac{\sigma_2(\mu^2)}{2\omega_{\mu,p}}   \right] \nn
& =  (2\pi)^3 \delta^3(\vec{p}-\vec{q}) \int_0^\infty d\mu^2  e^{-i t \omega_{\mu,p}} \frac{\sigma(\mu^2)}{2\omega_{\mu,p}} \nn
& = (2\pi)^3 \delta^3(\vec{p}-\vec{q})  \left[ i \tilde{G}(t,\vec{p}) \theta(t) + \left(i  \tilde{G}(t,\vec{p}) \right)^* \theta(-t) \right].
\end{align}
The first and second identities follow from~\refeq{sigma1} and~\refeq{sigma2} and a change of variables in the first and second integrals, respectively. The fourth identity, with
\beq
\tilde{G}(t,\vec{p}) = \int_{-\infty}^\infty \frac{dE}{2\pi} e^{-i E t} G(E,\vec{p}) ,
\eeq
is obtained just as Eq.~\refeq{overlapprop}, and we have similarly defined $G(p^0,\vec{p}) = G(p^2+i 0^+)$. We see that
\beq
i \tilde{G}(t,\vec{p}) =  \int_0^\infty d\mu^2 e^{-i |t| \omega_{\mu,p}} \frac{\sigma(\mu^2)}{2\omega_{\mu,p}} ,  
\eeq
which agrees with the Fourier transform of the spectral representation of $G$ in Eq.~\refeq{sigma}.

The smearing in time can proceed exactly as in the free case, and Eqs~(\ref{smearA}--\ref{survivalprob}) hold just by removing the ${(0)}$ superindices and using the basis selected in this section. In particular, the survival probability is given by
 \begin{align}
\mathcal{P}_{\mathrm{sur}}(t)  & \simeq \left| \frac{\tilde{G}_\tau(t,\vec{p}_0)}{\tilde{G}_\tau(0,\vec{p}_0)}   \right|^2 \nn
& = \mathcal{N}^{-2} |\tilde{G}_\tau(t,\vec{p}_0)|^2 , \label{survivalprobG}
\end{align}
where
\beq
i \tilde{G}_\tau(t,\vec{p}) =  \int_0^\infty d\mu^2 e^{-i t \omega_{\mu,p}} \frac{\sigma_\tau(\mu^2,\vec{p}^2)}{2\omega_{\mu,p}} ,  \label{Gtau}
\eeq
with
\beq
\sigma_\tau  (\mu^2,\vec{p}^2) =  \tilde{f}_\tau(\omega_{\mu,p}) \sigma(\mu^2) .
 \eeq
We emphasize that $\mathcal{P}_{\mathrm{sur}}(t)$ is the probability of finding the system in the specific initial state $\mathcal{A}_h^\tau$ after a time $t$ has elapsed, and not the probability of finding the system in a linear combination of one-particle states. The latter will be considered in subsection~\ref{ss:oscillations}.
 
The behaviour of the survival probability at very short times is always quadratic, just as in the free theory, with Zeno time $t_Z$ given by~\refeq{Zeno}, this time with the interacting Hamiltonian. Regarding the long-time behaviour, in the interacting theory the time evolution is irreversible in all cases, with the survival probability approaching zero when $t\to \infty$, see Eq.~\refeq{RL}. This follows from the Riemann-Lebesgue theorem, as the spectral function $\sigma_\tau$ is an absolutely integrable ordinary function (due to the absence of real poles in the dressed propagator $G$, the mass gap and the time smearing).  Nevertheless, a closer look reveals a different  behaviour in the discrete and continuous cases.

\subsection{Discrete case}
\label{ss:discrete_interacting}
\begin{figure}
  \centering
\includegraphics[width=0.4\textwidth]{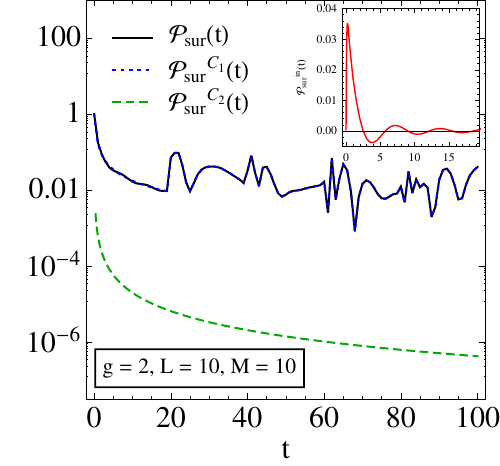}  \hspace{1cm} \includegraphics[width=0.4\textwidth]{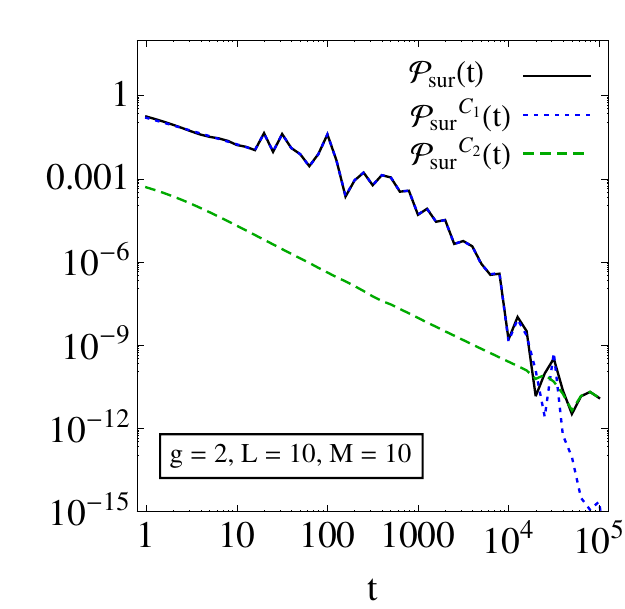} 
\caption{Survival probability in the interacting discrete case up to $t = 100$ (left) and $t = 10^6$ (right). We have displayed the survival probability $\mathcal P_{\rm sur}(t)$, and the contributions of the contours $C_1$ and $C_2$ to the survival probability, $\mathcal P_{\rm sur}^{C_i}(t) \equiv |\tilde G^{C_i}_\tau(t,\vec{p}_0) / \tilde G_\tau(t=0,\vec{p}_0)|^2 \quad i = 1, 2$. The red line in the inserted figure in the left panel is the interference contribution $\mathcal P_{\rm sur}^{\rm in}(t) \equiv 2 \, \textrm{Re}[ \tilde G_\tau^{C_1}(t,\vec{p}_0) \tilde G_\tau^{C_2}(t,\vec{p}_0)^\ast ] / |\tilde G_\tau(t=0,\vec{p}_0)|^2$. We have set $g = 2$, $\mu_0 = 1$, $L = 10$ and $M = 10$ and have used Cauchy smearing with $\tau = 0.1$. Note that the $C_2$ contribution is negligible most of the time, but it becomes dominant at very large $t$. The sharp oscillations of the blue and black curves in the right panel are not accurate, but the result of the interpolation of the numerical calculation at several points. 
\label{fig:polecontributioninteraction}
}
\end{figure}

Let us discuss first the discrete case. As we have already mentioned, the dressed propagator has in this case a series of poles on the second Riemann sheet, which show up in the spectral function as peaks separated by zeros. Let us consider $t\gg \tau$, which allows us to approximate
\beq
i \tilde{G}_\tau(t,\vec{p}) \simeq  \int_{-\infty}^{\infty} \frac{dE}{2\pi} e^{-i E t} \tilde{f}_\tau(E) i G(E,\vec{p}), ~~~ t\gg \tau .  \label{FourierSmearProp}
\eeq 
It will be convenient to consider here Cauchy smearing and to stick to $\vec{p}\neq 0$ to avoid IR subtleties (we can take $\vec{p}\to0$ at the end). 
%%%%%%%%%
\begin{figure}[h]
  \centering
\vspace*{-2cm}  
\includegraphics[width=0.6\textwidth]{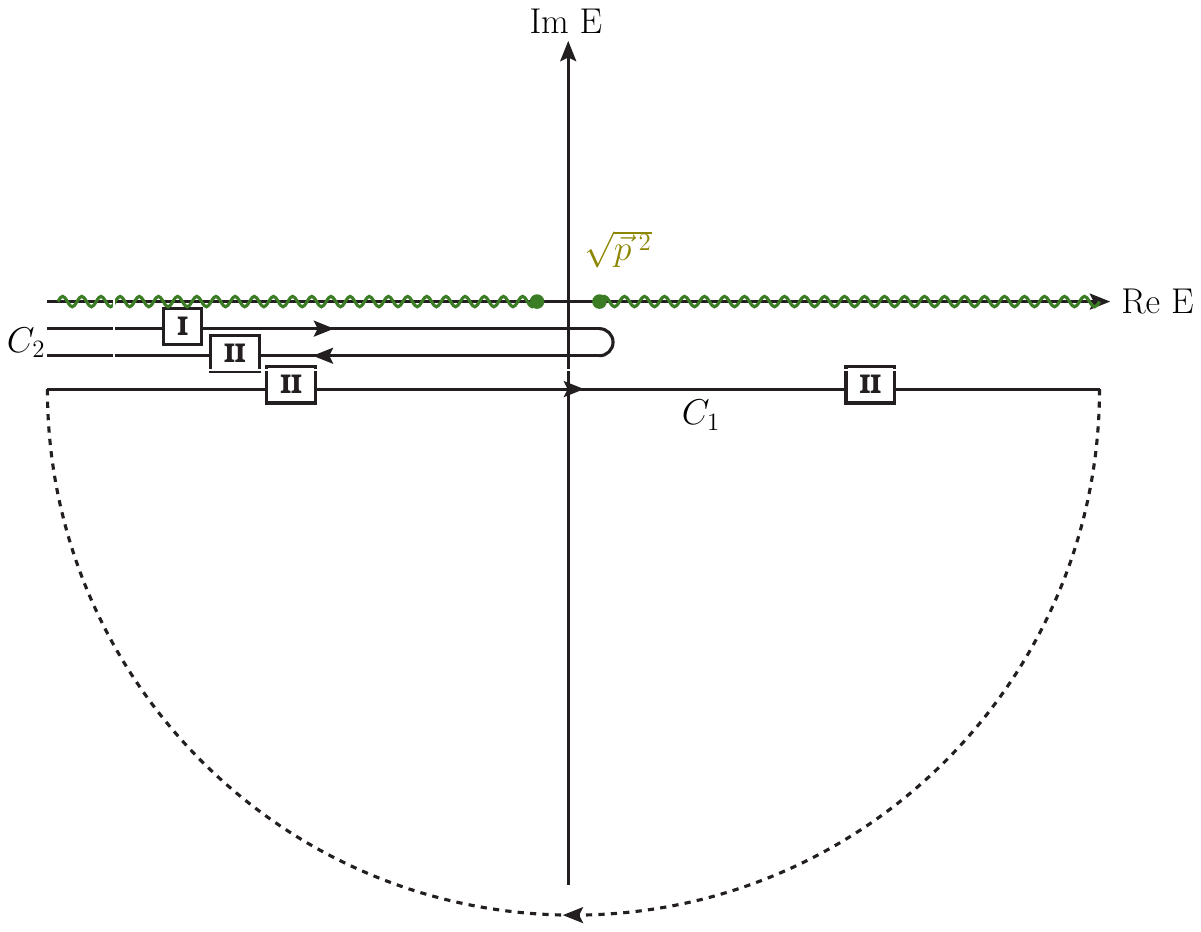}
  \vspace{-3.3cm}
  \caption{Integration contours to evaluate the propagator $\tilde{G}$ as a function of time in the interacting discrete case. The green wavy line represents the branch cut of $G$. The contours $C_1$ and $C_2$ on the complex plane are infinitesimally close to the real axis (their distance to it has been amplified for clarity). The roman numbers I and II in squares designate the Riemann sheet on which $G$ is to be evaluated on the indicated piece of the contours. The dotted semicircle can be used to close the curve $C_1$ on the second Riemann sheet. See the text for more details.}
  \label{fig:contours_discretum}
\end{figure}
%%%%%%%%%
The integral in~\refeq{FourierSmearProp} can then be written as the sum of two contour integrals, as shown in figure~\ref{fig:contours_discretum}:
\beq
i \tilde{G}_\tau = i \tilde{G}_\tau^{C_1} + i \tilde{G}_\tau^{C_2}. \label{C1C2}
\eeq
The second term corresponds to the evaluation of the integrand along the Hankel contour $C_2$, where piece I of the contour lies on the first (physical) Riemann sheet for the propagator, while piece II is on the second Riemann sheet:\footnote{The contour $C_2$ can further be deformed to be parallel to the imaginary axis. The exponent in the exponential $e^{-i E t}$ on this alternative path is real and negative, up to a constant, which is convenient for numerical evaluation.}
\beq
 \tilde{G}_\tau^{C_2}(t,\vec{p}) = \int_{-\infty}^{\sqrt{\vec{p}^2}} \frac{dE}{2\pi} e^{-i E t} \tilde{f}_\tau(E) \left[G((E-i0^+)^2-\vec{p}^2)-G^{\mathrm{II}}((E-i0^+)^2-\vec{p}^2) \right].
\eeq
We observe that
\beq
G(p^2)-G^{\mathrm{II}}(p^2)  = \frac{\Sigma^{\mathrm{II}}(p^2) - \Sigma(p^2)}{(\Pi(p^2) + \Sigma(p^2))(\Pi(p^2)+\Sigma^{\mathrm{II}}(p^2)) } ,
\eeq
and in our model
\beq
\Sigma^{\mathrm{II}}(p^2) = \Sigma(p^2)+2 i \frac{g^2}{16\pi}. 
\eeq
Hence, for weak coupling $\tilde{G}_\tau^{C_2}$ gives a subleading contribution at intermediate times. On the other hand, this contribution is important for the time behaviour at asymptotic times, once the exponential contributions, to be discussed below, are negligible. For Cauchy smearing, the term $\tilde{G}_\tau^{C_1}$ can be directly evaluated with the residue theorem by closing the curve as shown in figure~\ref{fig:contours_discretum}.  The contour $C_1$ lies entirely on the lower half of the second sheet. Therefore, all the poles of the dressed propagator contribute to the integral. In addition, there is a contribution of a pole on the negative imaginary axis in the Cauchy distribution $\tilde{f}_\tau$:
\beq
i \tilde{G}_\tau^{C_1,\mathrm{smearing}}(t,\vec{p})  = \frac{i}{2\tau} G^{\mathrm{II}} (-\frac{1}{\tau^2}+\vec{p}^2) e^{-t/\tau} .
\eeq
This contribution is negligible for not very small time,  $t \gg \tau$, as we were already assuming in this paragraph. The contribution from the propagator poles is
\beq
i \tilde{G}_\tau^{C_1,\mathrm{poles}}(t,\vec{p}) = \sum_n \frac{\mathcal{Z}_n}{2E_n} \frac{1}{1+E_n^2 \tau^2} e^{-i t E_n}, \label{polestime}
\eeq
where $E_n=\omega_n-i\Lambda_n$/2 is the position of the $n$th pole and $\mathcal{Z}_n= 2 E_n \mathrm{Res}_{E_n}\left(G^{\mathrm{II}}(E,\vec{p})\right)$. We have left implicit the $|\vec{p}|$ dependence of the poles and their residues. The widths $\Lambda_n$ are of order $g^2$. When $g\to 0$, Eq.~\refeq{polestime} reduces to Eq.~\refeq{oscillation}, with normalization $H_\nu=1$.  For finite $g$ we have a sum of terms proportional to complex exponentials, with an oscillating and a decaying part. For intermediate times, we expect $\tilde{G}_\tau=  \tilde{G}_\tau^{C_1,\mathrm{poles}}$ to be an excellent approximation. This is corroborated in explicit examples, see figure~\ref{fig:polecontributioninteraction}. In the particular case of only one unstable particle (only one pole), we recover in this approximation the standard Weisskopf-Wigner exponential decay law~\cite{Weisskopf:1930au}. For more than one pole, we need to take into account the interference of the different terms in~\refeq{polestime}. The behaviour is then a power law from $t\simeq \tau$ to $t\simeq 2/L$, followed by an exponential modulated by oscillations. For very long times, the contribution of $\tilde{G}_\tau^{C_2}$ becomes dominant, as shown in the right plot of figure~\ref{fig:polecontributioninteraction}.\footnote{\label{longtimetachyon}Let us mention in passing that for extremely long times, the contribution of the tachyon (which is not physical and we have completely ignored in the discussion) may become visible. But this is irrelevant for all purposes, since the survival probability is already extremely small at those times.}  Even if we have used Cauchy smearing to prove them, all these results at $t\gg\tau$ apply also to other smearings, using 
\beq
i \tilde{G}_\tau^{C_1,\mathrm{poles}}(t,\vec{p}) = \sum_n \frac{\mathcal{Z}_n}{2E_n} \tilde{f}_\tau(E_n) e^{-i t E_n}
\eeq
instead of~\refeq{polestime}.
This fact follows to a good approximation from Jacob-Sachs theorem in~\cite{Jacob:1961zz} when $\tilde{f}_\tau(E)$ is negligible outside a finite interval. It will be used in section~\ref{s:processes}.

%%%%%%%%%
%%%%%%%%%

\subsection{Continuous case}
\begin{figure}[h]
  \centering
  \vspace{-2cm}
  \includegraphics[width=0.6\textwidth]{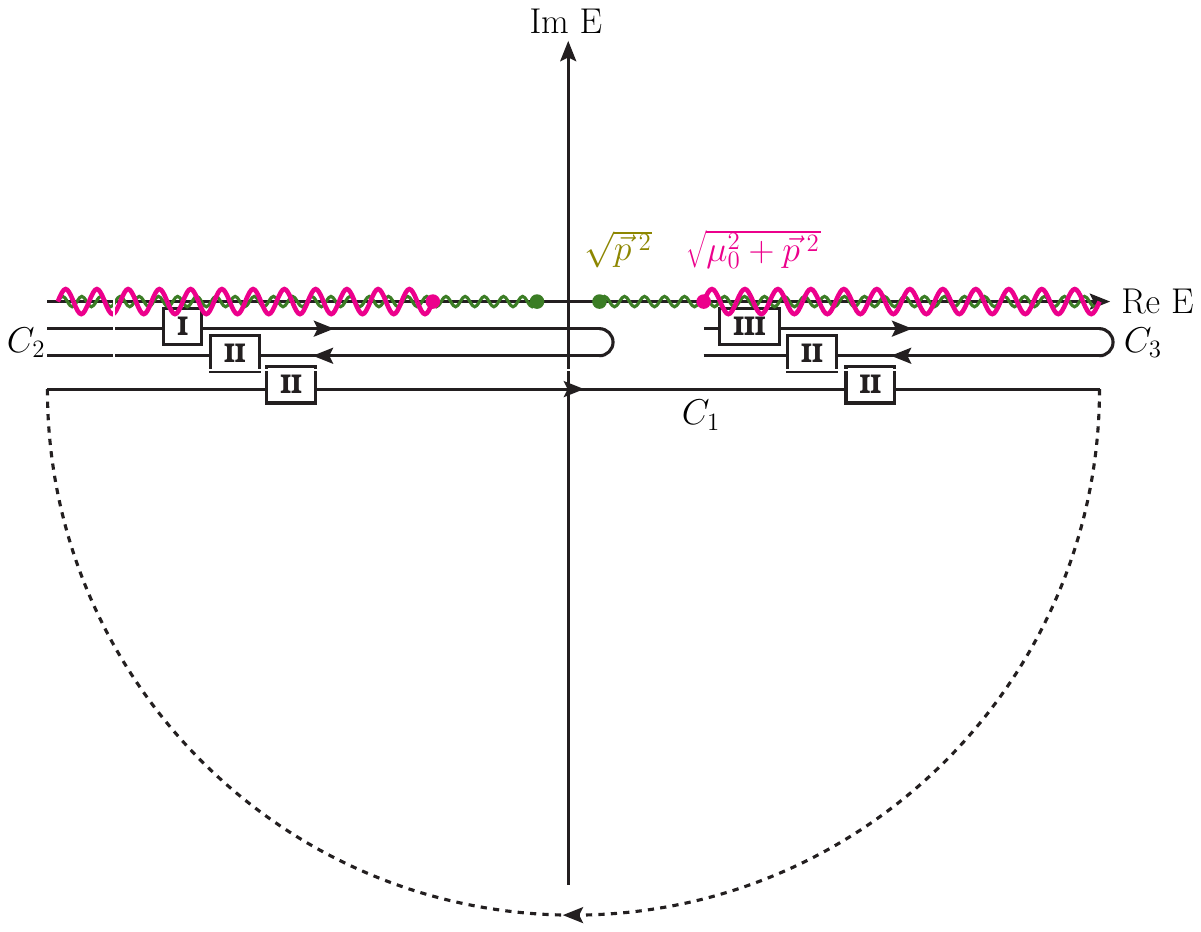}
  \vspace{-3.3cm}
  \caption{Integration contours to evaluate the propagator $\tilde{G}$ as a function of time in the interacting continuum case. The green wavy line represents the branch cut of $G$ associated to $\Sigma$ (branch cut of $G_1$), while the wider red wavy line represents the branch cut of $G$ associated to $\Pi$ (branch cut of $G_2$). The contours $C_1$, $C_2$ and $C_3$ on the complex plane are infinitesimally close to the real axis (their distance to it has been amplified for clarity). The roman numbers I, II and III in squares designate the Riemann sheet on which $G$ is to be evaluated on the indicated piece of the contours. The dotted semicircle can be used to close the curve $C_1$ on the second Riemann sheet. See the text for more details.}
  \label{fig:contours_continuum}
\end{figure}

\begin{figure}
  \centering
  \includegraphics[width=0.4\textwidth]{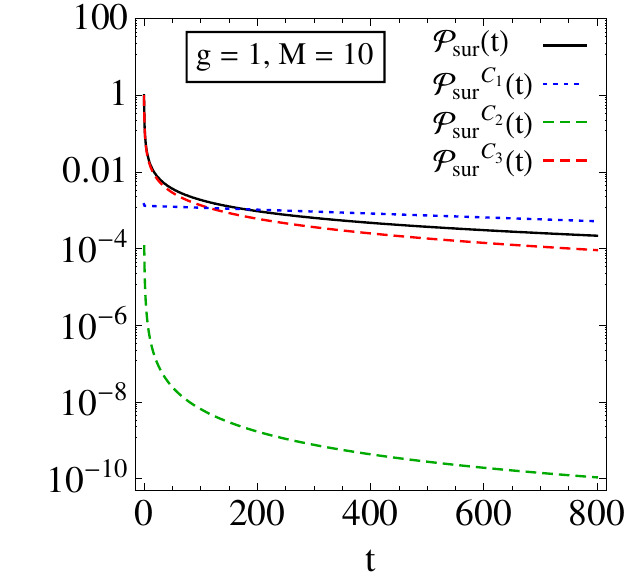} \hspace{0.5cm}   \includegraphics[width=0.4\textwidth]{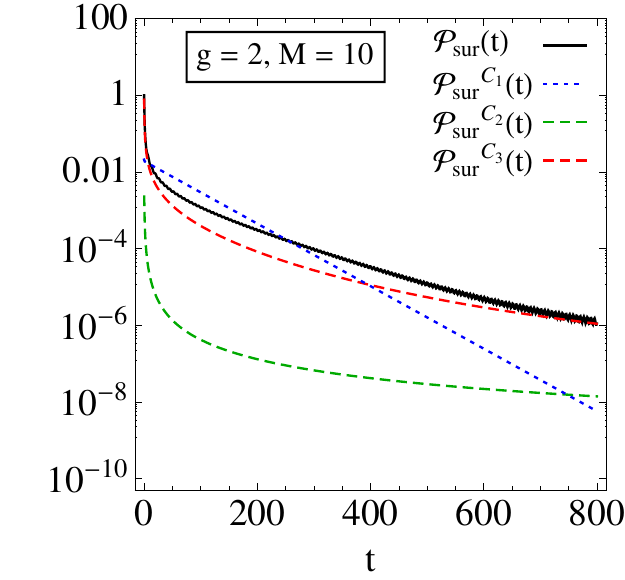} 
  \caption{Survival probability in the interacting continuous case with pole on the second Riemann sheet.  We display also the three contributions $\mathcal P_{\rm sur}^{C_i}(t) \equiv |\tilde G^{C_i}_\tau(t,\vec{p}_0) / \tilde G_\tau(t=0,\vec{p}_0)|^2 \quad i = 1, 2,3$. We have considered $\mu_0 = 1$, $M = 10$, and $g = 1$ (left) and $g = 2$ (right).  We have used Cauchy smearing with $\tau = 0.1$.}
  \label{fig:intsurvival_continuum_pole}
\end{figure}

\begin{figure}
  \centering
  \includegraphics[width=0.4\textwidth]{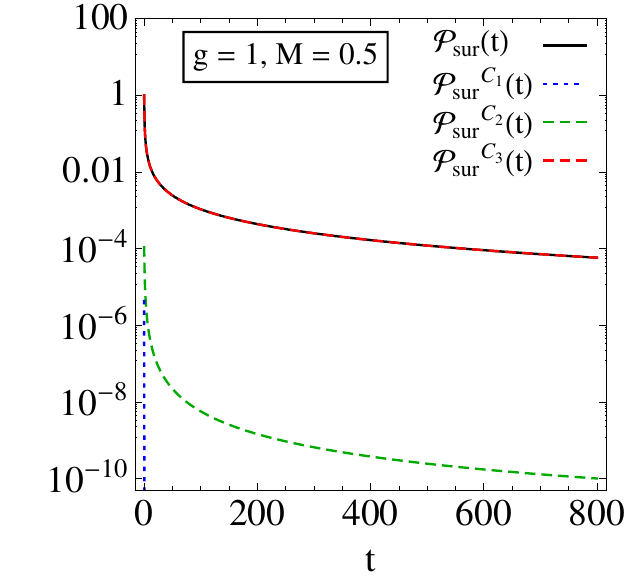} \hspace{0.5cm} \includegraphics[width=0.4\textwidth]{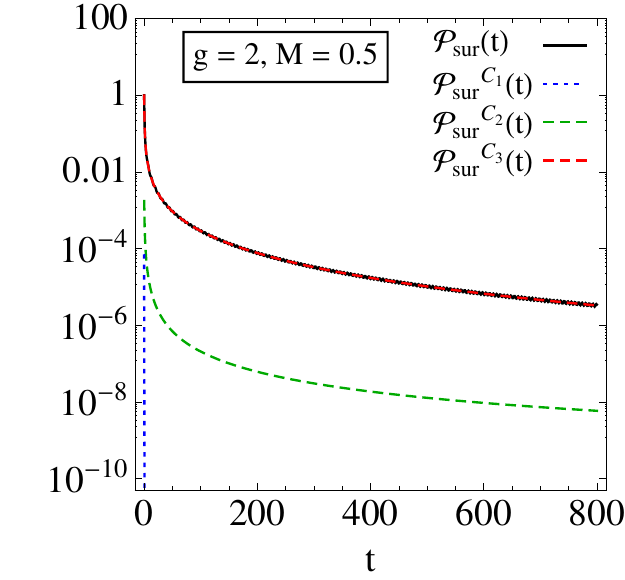}
  \caption{Survival probability in the interacting continuous case without pole on the second Riemann sheet. We display also the three contributions $\mathcal P_{\rm sur}^{C_i}(t) \equiv |\tilde G^{C_i}_\tau(t,\vec{p}_0) / \tilde G_\tau(t=0,\vec{p}_0)|^2 \quad i = 1, 2,3$. We have considered $\mu_0 = 1$, $M = 0.5$, and $g = 1$ (left) and $g = 2$ (right). We have used Cauchy smearing with $\tau = 0.1$; the corresponding pole gives the only contribution to $\mathcal P_{\rm sur}^{C_1}$.}
  \label{fig:intsurvival_continuum_nopole}
\end{figure}

Consider next the continuous case. In this case, the spectral function is smooth, except for the thresholds. Writing $\sigma_\tau=\sigma_{\tau \,1}+\sigma_{\tau \,2}$, with $\sigma_{\tau\, i} = \tilde{f}_\tau \sigma_i$, Eq.~\refeq{Gtau} decomposes into the sum of two integrals, each of them with a single threshold. Then we can approximate the large-time behaviour of  each of them as we did in the free theory, by considering only the region close to the thresholds.  Because in our model $\sigma_2$ has a logarithmic singularity at $\mu^2=0$, we present a generalization of our previous result, using the results of ref.~\cite{WONG1978173}. For $j=1,2$, let $\mu_j:=\mu_0 \delta_{j1}$ and $\sigma_{\tau\, j}(\mu^2,\vec{p}^2)/2 \omega_{\mu,p}= \theta(\mu^2-\mu_j^2) \log^{\beta_j}(\frac{\mu^2-\mu_j^2}{\kappa_j^2}) \chi_j(\mu^2)$, with continuous $\chi_j$, and let $\chi_j(\mu^2) \simeq C_j (\mu^2-\mu_j^2)^{\gamma_j}/ \kappa_j^{2(1+\gamma_j)}$ near $\mu_j^2$, with $\gamma_j>-1$ for convergence of the integrals.
Then, using Gaussian smearing,
\begin{align}
i \tilde{G}_\tau(t,\vec{p})  \simeq   & \sum_{j=0,1} C_j e^{-\omega_{\mu_j,p}^2 \tau^2}e^{-i (\gamma_j+1) \pi/2} \Gamma(1+\gamma_j) \kappa_j^{-2 (1+\gamma_j)} e^{-i t \omega_{\mu_j,p}} \left( \frac{2\omega_{\mu_j,p}}{t} \right)^{1+\gamma_j } \nn
& {\times}  \log^{\beta_j}\left( \frac{2\omega_{\mu_j,p}}{\kappa_j^2 \, t} \right) + \dots \, .
\end{align}
For large-enough time, the term with smaller $\gamma_j$ (larger $\beta_j$ if $\gamma_1=\gamma_2$) will dominate. The dots refer to subleading contributions in the integral of $\sigma_2$, (they can be found in~\cite{WONG1978173}), which may be more important at large time than the leading contribution in the integral of $\sigma_1$ when $\gamma_2<\gamma_1$. A connection with integrals of the propagator on the Riemann surface can be  established:
\beq
\tilde{G}_\tau (t,\vec{p}) = \tilde{G}_\tau^{C_1}(t,\vec{p})+\tilde{G}_\tau^{C_2}(t,\vec{p})+\tilde{G}_\tau^{C_3}(t,\vec{p})
\eeq
where
\beq
i \tilde{G}_\tau^{C_j}(t,\vec{p}) \simeq  \int_{C_j} \frac{dp^0}{2\pi} e^{-i p^0 t} \tilde{f}_\tau(p^0) i G(p^2), ~ t\gg \tau .
\eeq
and $C_j$ are the contours indicated in figure~\ref{fig:contours_continuum}. 
%%%%%%%%%
%%%%%%%%%
As shown in figure~\ref{fig:contours_continuum}, the contour $C_3$ goes from $\omega_{\mu_0,p}$ to $\infty$ on the third Riemann sheet, right below the real axis, and comes from $\infty$ back to $\omega_{\mu_0,p}$ on the second Riemann sheet, also right below the real axis. Here, the propagator on the third Riemann sheet is defined as the one obtained by analytical continuation of the physical propagator when crossing $(\omega_{\mu_0,p},\infty)$, while the propagator on the second Riemann sheet is obtained by analytical continuation across $(\omega_{0,p},\omega_{\mu_0,p})$. The contours $C_1$ and $C_2$ are the ones in figure~\ref{fig:contours_discretum} (using the definition of the second sheet we have just given). 
The integral along $C_1$ can again be evaluated, for Cauchy smearing,  with the residue theorem. In the continuous case we do not have multiple poles on the second sheet, as there were no poles in the free theory. But depending on the functions $\Pi$ and $\Sigma$ there may be at least a simple pole on the second sheet, at a finite distance from the real axis~\cite{Delgado:2008gj}.  Such a pole  gives an exponentially decaying contribution,
\beq
 i \tilde{G}_\tau^{C_1,\mathrm{pole}} = \frac{Z_\mathrm{pole}}{2 E_\mathrm{pole}} \frac{1}{1+E_\mathrm{pole}^2\tau^2} e^{-i t E_\mathrm{pole}},
\eeq
where $E_\mathrm{pole}=\omega_\mathrm{pole}-i \Gamma_\mathrm{pole}/2$ is the position of the pole and $Z_\mathrm{pole} = \mathrm{Res}_{E_\mathrm{pole}}\left(G^{\mathrm{II}}(E,\vec{p})\right)$, with the dependence on $|\vec{p}|$ implicit in the pole position and residue. The residue, and hence the exponential contribution, is suppressed for weak coupling. In the flat holographic model, for small $g$ and vanishing $\vec{p}$ we find 
\begin{align}
& \frac{E_{\mathrm{pole}}}{\mu_0} \simeq 1 + \frac{(\pi^2-4c^2) g^4}{512\pi^4 \mu_0^2} -i \frac{c g^4}{128 \pi^3 \mu_0^2}  ,
\nn
& Z_\mathrm{pole} \simeq \frac{2c + i \pi}{8\pi^2} g^2,
\end{align}
for $c=\log ({M}/{\mu_0})>0$, and no pole otherwise.
Despite this suppression, as shown in figure~\ref{fig:intsurvival_continuum_pole}, the contribution of the pole is significant at intermediate times. We also see that the contribution of $\tilde{G}_\tau^{C_3}$, which is associated to the continuum branch cut,  is crucial in getting the correct time dependence of the survival probability. This contribution gives a power law and thus becomes dominant at some point. Note that the survival probability is not equal to the sum of the $\mathcal P_{\rm sur}^{C_i}$, showing the relevance of the interference of the different contributions to the propagator. The contribution of $\tilde{G}_\tau^{C_2}$ (also a power law) can be mostly ignored most of the time, but it can become dominant at very late times (not shown in the plots). We can also notice in  figure~\ref{fig:intsurvival_continuum_pole} small oscillations in the total survival probability, arising from the interference of the different contributions. In figure~\ref{fig:intsurvival_continuum_nopole} we show the survival probability and the different contributions in an example in which the propagator has no pole on the second Riemann sheet ($c<0$). In this case, the only contribution to $\tilde{G}_\tau^{C_1}$ comes from the smearing and it is apparent that it is indeed negligible for $t>\tau$. Therefore, the survival probability is very well described by $\tilde{G}_\tau^{C_3}$ alone, except for the small oscillations from its interference with $\tilde{G}_\tau^{C_2}$, and at very late times, when it is already very small and the contribution of $\tilde{G}_\tau^{C_2}$ becomes dominant. Again, the same results hold also for other smearings, as the Jacob-Sachs theorem~\cite{Jacob:1961zz} also applies to the continuum case. 
%%%%%%%%%
%%%%%%%%%
%%%%%%%%%
%%%%%%%%%

Summarizing, in the continuous case we can write\footnote{The remark in~\cref{longtimetachyon} also applies here.}
\beq
i \tilde{G}_\tau \simeq  i \tilde{G}_\tau^{C_1,\mathrm{pole}} + i \tilde{G}_\tau^{C_2} + i \tilde{G}_\tau^{C_3} , ~~~ t \gg \tau.
\eeq
We have neglected the contribution from the smearing, $ \tilde{G}_\tau^{C_1,\mathrm{smearing}}$, as it is suppressed  by $e^{-t /\tau}$ and thus very small when $t \gg \tau$.
   
%%%%%%%%%
\begin{figure}
  \centering
  \includegraphics[width=0.4\textwidth]{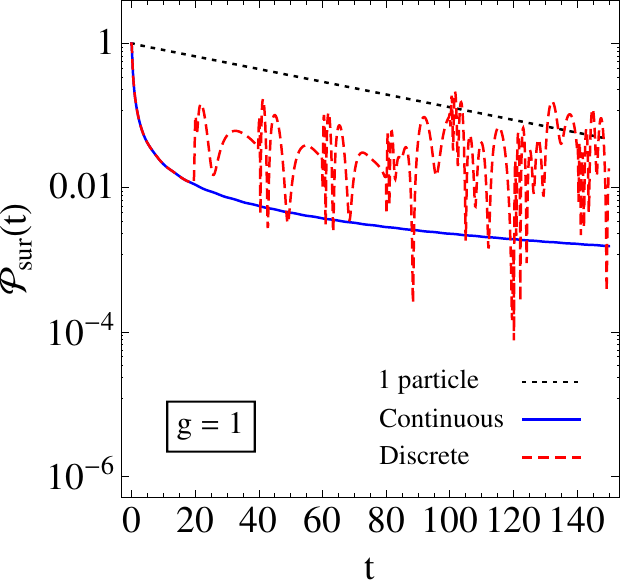} \hspace{1cm}    \includegraphics[width=0.4\textwidth]{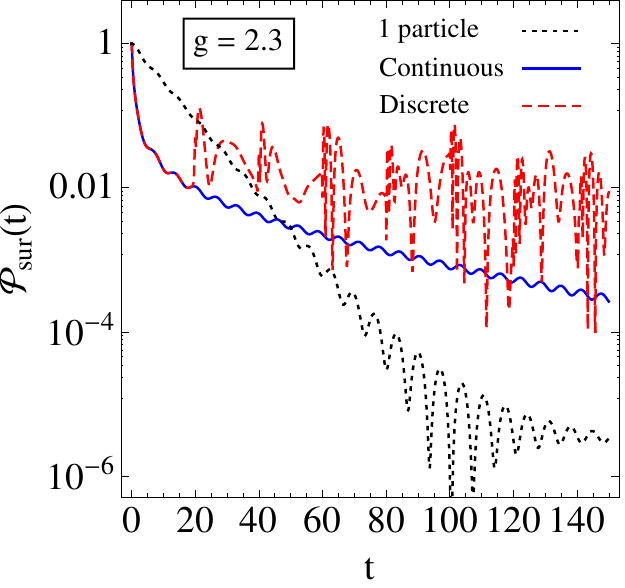} 
  \caption{Survival probability in the interacting discrete and continuous cases. For comparison, we display also the standard result for a single particle of mass $\mu_0=1$. We have considered $\mu_0 = 1$, $L = 10$, $M = 10$, and $g = 1$ (left) and $g = 2.3$ (right). We have used Gaussian smearing with $\tau = 0.1$.}
  \label{fig:intsurvival_discretecontinuum}
\end{figure}
%%%%%%%%%

It is interesting to compare the results in the continuum case with the continuum limit of the discrete case. In the discrete case, the contour $C_2$ is far from the poles. Therefore, when the spacings between modes become small, $\tilde{G}_\tau^{C_2}$ approaches smoothly the result of the same integral in the continuum case. When $t \ll 1/\Delta \omega$, with $\Delta \omega$ the largest gap between consecutive $\omega_n$, the widths $\Lambda_n$ approach zero, except for the one associated with the lowest $\omega_n$, which as we have observed in some cases stays finite.  Hence, we can approximate $\tilde{G}_\tau^{C_1,\mathrm{poles}}$ by an integral over $\mu^2$ with a continuous interpolating function $\bar{\sigma}_{\tau \, 1}(\mu^2,\vec{p}^2)$ (with support in $[\mu_0,\infty)$ to reproduce the position of the poles) plus the possible contribution of the pole that remains isolated in the continuum limit, which corresponds to $\tilde{G}_\tau^{C_1,\mathrm{pole}}$. To simplify the discussion, consider the case without pole ($c<0$). It is clear then that $\bar{\sigma}_{\tau \, 1}(\mu^2,\vec{p}^2)$ approaches $\sigma_{\tau \, 1}(\mu^2,\vec{p}^2)$ in the continuum limit. Away from the strict limit, the contribution of the poles is well approximated by the contribution of the branch cut in $G_1$ for the continuous case when $t \ll 1/\Delta \omega$, while it is completely different at later times.
In figure~\ref{fig:intsurvival_discretecontinuum}, we compare the survival probability for a discrete case with the one for the corresponding continuum case (obtained in the limit $L\to \infty$ of the model with the same parameters). We see again that they are almost indistinguishable for $t<2L \simeq 2 \pi/\Delta\omega$ and very different for $t \gtrsim 2L$. This holds independently of the size of the coupling $g$. The oscillations in the continuous case have period $\Delta t = 2\pi / \mu_0$ if $|\vec{p}| = 0$ and arise as mentioned above from the interference with the contribution $\tilde{G}_\tau^{C_2}$. They are more apparent in the right plot because their amplitude is larger for larger coupling, as $\tilde{G}_\tau^{C_2}$ is of order $g^2$.

\subsection{Visible decay}
\label{ss:particledecay}
The description of time evolution in terms of the asymptotic states, which we have used above, is precise, but arguably not very intuitive. In this sort of ``exact'' description, a muon, for instance, is at any time a particular multi-particle state formed by an electron, a muon neutrino and an electron antineutrino. However, for practical purposes it is usually more convenient to think of the muon as an independent degree of freedom: a particle of charge -1 with mass 105.7~$\mathrm{MeV}$ and mean lifetime 2.2~$\mathrm{\mu s}$. In this familiar picture, based on perturbation theory, a muon with well-defined momentum is a one-particle eigenstate of the free Hamiltonian. To gain more intuition about time evolution in our problem, we next apply such a ``perturbative'' description to our model. For this, we need to consider the ``free'' states $\ket{\mathcal{A}^0}\sim A(0)\ket{0}$ and $\ket{\varphi \bar{\varphi}}_0 \sim \varphi^\dagger(0) \varphi(0) \ket{0}$, where $\ket{0}$ is the Fock vacuum. We would then like to compute the probabilities of finding, in a measurement at time $t$, that the system is in the one-particle state $\ket{\mathcal{A}^0}$ or in any free two-particle state $\ket{\varphi \bar{\varphi}}_0$, given that it was in the state $\ket{\mathcal{A}^0}$ at time 0. The time evolution is still to be calculated with the complete Hamiltonian. The interest of these probabilities is that they allow us to distinguish how much of the depletion of the initial state is due, at each instant, to decay into the elementary particles, which we call {\em visible decay\/}. 

These probabilities involving free states are unfortunately more difficult to calculate than the ones involving asymptotic states. The reason is that energy is not conserved in the vertices of the corresponding diagrams, which in turn happens because there appear integrals over finite or semi-infinite time intervals.\footnote{This is similar to what happens in the in-in formalism.} Then it is highly non-trivial, if possible at all, to resum the diagrams contributing to the required modified propagator, $\bra{0} A(0,\vec{p})  e^{-i t H} A(0,\vec{q}) \ket{0}$. For this reason we will approximate 
$\ket{\mathcal{A}^0}$ by $\ket{\mathcal{A}}$ and use the usual propagator $\bra{\Omega} A(0,\vec{p})  e^{-i t H} A(0,\vec{q}) \ket{\Omega}$ as an approximation to the survival probability of  $\ket{\mathcal{A}^0}$ (which will then be the same as the survival probability of $\ket{\mathcal{A}}$ considered so far) and also as an approximation to the propagator that appears in the perturbative evaluation of the decay probability. This approximation is only good up to corrections of order $g^2/(16\pi^2)$. The instantaneous probability of decay into a pair of elementary particles precisely at time $t$, which we will call {\em visible decay rate}, is then approximated by
\beq
\frac{d\mathcal{P}_{\bar{\varphi}\varphi}}{dt}(t_1) \simeq \frac{g^2}{8\pi} \frac{1}{|\tilde{G}_\tau(0,\vec{0})|} \left| \tilde{G}_\tau(t_1,\vec{p}_0) \right|^2 .\label{decaylightinstant}
\eeq
This is the probability density associated to a measurement of the time delay in a displaced vertex, up to details to be discussed in section~\ref{s:processes}. Within our approximation, it is proportional to the survival probability.
The probability of visible decay, that is, the probability of finding at time $t$ that the initial state has decayed into elementary particles is well approximated by the accumulated probability
\beq
\mathcal{P}_{\mathrm{vis}}(t)  \simeq \int_0^t d t_1 \frac{d\mathcal{P}_{\bar{\varphi}\varphi}}{dt}(t_1) . \label{decaylightaccumulated}
\eeq
This simple expression, which justifies the notation for the instantaneous probability, is the result of (i) choosing $\ket{\mathcal{A}^0}$ as the initial state, so that the visible decay probability vanishes at $t=0$, in agreement with the perturbative intuition, and (ii) neglecting the interference between the decay amplitudes at different values of $t_1$. A more precise expression that takes into account this interference is
\beq
\mathcal{P}_{vis}(t) = \frac{g^2}{16\pi^2} \frac{1}{|\tilde{G}_\tau(0,\vec{0})|} \int_0^\infty d E_f \left| \int_0^t d t_1 e^{i t_1 E_f} \tilde{G}_\tau(t_1,\vec{0}) \right|^2 .\label{decaylight}
\eeq
The outer integral accounts for the final-state phase space, with $E_f$ the total (free) energy of each momentum configuration. But using this expression in a explicit calculation is more complicated than using~\refeq{decaylightaccumulated}, as we need to compute numerically three iterated integrals (the two integrals explicit in \refeq{decaylight} and the Fourier transform \refeq{FourierSmearProp}) instead of two. The convenient approximation~\refeq{decaylightaccumulated} is obtained from~\refeq{decaylight} by extending  the integration in $E_f$ to the interval $(-\infty,\infty)$. Indeed, working in the center-of-mass frame,
\begin{align}
\mathcal{P}_{\mathrm{vis}}(t) & \simeq \frac{g^2}{16\pi^2} \frac{1}{|\tilde{G}_\tau(0,\vec{0})|}  \int_{-\infty}^\infty d E_f 
\int_0^t d t_1 \int_0^t d t_2  e^{i (t_1-t_2) E_f}  \tilde{G}_\tau(t_1,\vec{0})  \tilde{G}_\tau(t_2,\vec{0})^*  \nn
& = \frac{g^2}{16\pi^2} \frac{1}{|\tilde{G}_\tau(0,\vec{0})|}  \int_0^t d t_1 \int_0^t d t_2  2\pi \delta(t_1-t_2) \tilde{G}_\tau(t_1,\vec{0})  \tilde{G}_\tau(t_2,\vec{0})^* \nn
& =  \frac{g^2}{8\pi} \frac{1}{|\tilde{G}_\tau(0,\vec{0})|}  \int_0^t d t_1 \left| G_\tau(t_1,\vec{0})  \right|^2 .
\end{align}
We have checked that the approximation~\refeq{decaylightaccumulated} is very good, for not too large $g$, when $t \gtrsim 1/\mu_0$. So, it will be used in the plots.
%%%%%%%%%
\begin{figure}
  \centering
  \includegraphics[width=0.4\textwidth]{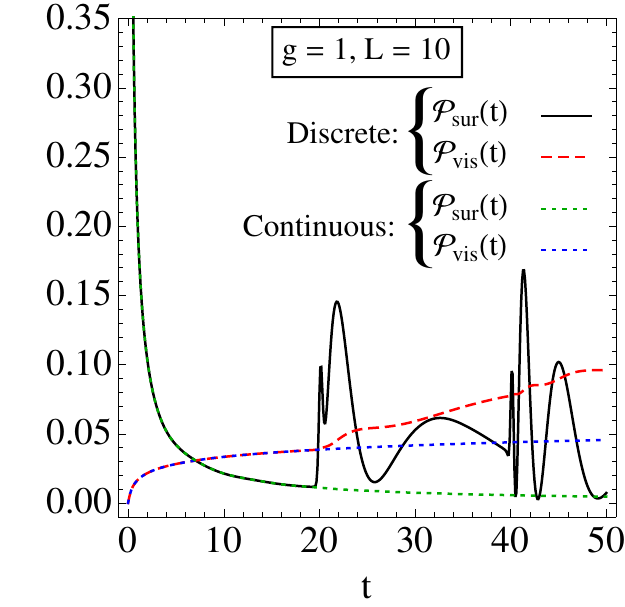}
  \hspace{1cm} \includegraphics[width=0.38\textwidth]{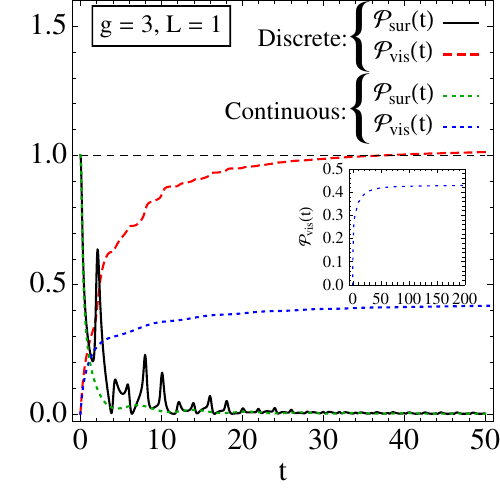}
  \caption{Survival probability and visible decay probability in the interacting discrete and continuous cases. The inserted figure in the right panel is the visible decay rate in the interacting continuous case up to $t = 200$, a range in which an approximately constant behaviour can be observed.  We have considered $\mu_0 = 1$, $M = 10$, and $(g=1, L = 10)$ (left) and $(g=3, L = 1)$ (right). We have used Gaussian smearing with $\tau = 0.1$.}
  \label{fig:visibledecay_discrete}
\end{figure}
%%%%%%%%%
In figure~\ref{fig:visibledecay_discrete} we plot the probabilities of survival and of visible decay in the continuous and discrete cases for different values of $L$ and $g$, using the approximation~\refeq{decaylightaccumulated}. The slope of the visible decay is proportional to the survival probability, as implied by~\refeq{decaylightaccumulated}. Therefore, the visible decay probability and the visible decay rate in the discrete case are well approximated by the ones in the continuous case for $t<2L$, as can be appreciated in the plots. Note that the sum of both probabilities is smaller than 1 at all times. In the discrete case, $\mathcal{P}_{\mathrm{vis}}(t) \to 1$ as $t\to\infty$, as can be seen in the right panel (in the plot it actually gets a bit larger than 1, but this is an error due to our approximations). On the other hand,  when $t\gtrsim 2L$ the probability of visible decay is significantly smaller in the continuum case. As it can be appreciated in the inserted figure in the right panel, it asymptotes to a value smaller than 1. Since the survival probability approaches 0 when $t\to\infty$, this indicates the presence of extra asymptotic states in the continuous case. Note that the reason for a smaller visible decay in the continuum when $t \to \infty$ is not the asymptotic behaviour, dominated by a similar $\tilde{G}^{C_2}$ in both cases, but the smaller visible decay rate at intermediate times.

\subsection{Oscillations and invisible decay}
\label{ss:oscillations}
In the free theory, the complement of the event of survival of the initial state is its oscillation into free one-particle states orthogonal to it. In the interacting theory, we can keep using the free basis and distinguish four exclusive events in the A1 approximation: the measured final state can be i) $\ket{\mathcal{A}_{\vec{p}}^{0}}$, ii) an arbitrary $\ket{\varphi \bar{\varphi}}_0$ free two-particle state, iii) an arbitrary orthogonal combination of free $A$ one-particle states, and iv) the Fock vacuum.\footnote{For simplicity we neglect here the effects of time and spatial smearing.}  The last possibility has finite probability when $g\neq 0$ because $\bra{0} U(t) A(0) \ket{0} \neq \bra{\Omega} A(0) \ket{\Omega}=0$, but it will be small if $g$ has perturbative values, and we neglect it in the following. The first and second possibilities have been discussed above. In this subsection, we study the third one: oscillations.

Even if a general analysis is in principle possible, we restrict ourselves here to the holographic theory in section~\ref{s:holography}. This will be computationally convenient and will provide an intuitive picture of the oscillations. For this, we first consider the propagation of the field from the UV boundary to an arbitrary point in the bulk. The corresponding dressed five-dimensional propagator has been introduced in section~\ref{s:holography}.
We can define the state with definite three-momentum created by $B(z)$ acting on the physical vacuum as
\beq
\ket{\mathcal{B}_{\vec{p}}(z)} = \int d^3 x e^{i \vec{x}\cdot\vec{p}} B(0,\vec{x},z) \ket{\Omega} . 
\eeq
Note that these states are not completely orthogonal for different $z$.\footnote{This is related to the impossibility of localizing particles in quantum field theory.}
For $t\geq 0$, we can write
\begin{align}
\bra{\mathcal{B}_{\vec{p}}(z^\prime)}e^{-i t H} \ket{\mathcal{B}_{\vec{q}}(z)}  & = (2\pi)^3 \delta^3(\vec{p}-\vec{q}) i \tilde{G}(t,\vec{p};z,z^\prime)   \nn
& := (2\pi)^3 \delta^3(\vec{p}-\vec{q}) \int_{-\infty}^\infty \frac{dE}{2\pi} e^{-i E t} i G(E^2-\vec{p}^2+i 0^+; z,z^\prime). \label{G0zB}
\end{align} 
 After proper smearing in time and momentum, the probability of finding the system in the state $\ket{\mathcal{B}_h^\tau(z)}$ at time $t$, given that it was in the state $\ket{\mathcal{A}_h^\tau}=\ket{\mathcal{B}_h^\tau(0)}$ at time 0, is
\beq
\mathcal{P}_{\mathrm{sur}}(t,z)  \simeq \frac{\left| \tilde{G}_\tau(t,\vec{p}_0,0,z)\right|^2}{\left|\tilde{G}_\tau(0,\vec{p}_0) \tilde{G}_\tau(0,\vec{p}_0;0,z)   \right|} . \label{probz}
\eeq
%%%%%%%%%
\begin{figure}
  \centering
  \includegraphics[width=0.4\textwidth]{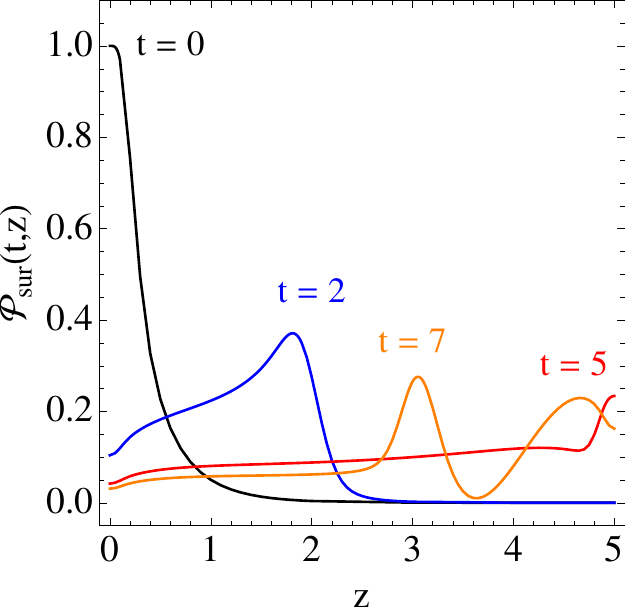}
  \caption{Survival probability $\mathcal{P}_{\rm sur}(t,z)$ in the interacting discrete case at different times, cf. Eq.~(\ref{probz}). We have considered $g = 1$, $\mu_0 = 1$, $L = 5$ and $M = 10$. We have used Gaussian smearing with $\tau = 0.1$.}
  \label{fig:profile_discrete}
\end{figure}
%%%%%%%%%
%%%%%%%%%
\begin{figure}
  \centering
  \includegraphics[width=0.4\textwidth]{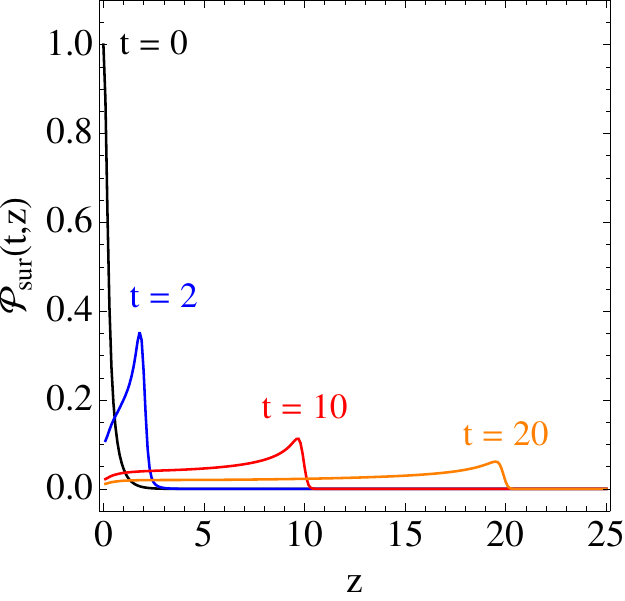} 
  \caption{Survival probability $\mathcal{P}_{\rm sur}(t,z)$ in the interacting continuous case at different times, cf. Eq.~(\ref{probz}). We have considered $g = 1$, $\mu_0 = 1$ and $M = 10$. We have used Gaussian smearing with $\tau = 0.1$.}
  \label{fig:profile_continuum}
\end{figure}
%%%%%%%%%
We plot this probability at different times for the discrete and continuous cases in figs.~\ref{fig:profile_discrete} and~\ref{fig:profile_continuum}, respectively. We see that the state spreads and propagates inside the bulk, much as in the quantum mechanical evolution of a wave packet in one dimension. The amplitude of the survival probability (equal to $\mathcal{P}_{\rm sur}(t,0)$) is given by the overlap of such ``wave packet'' with the UV boundary. This explains why it becomes smaller and smaller initially, even when $g=0$. But in the discrete case, corresponding to a compact extra dimension, the ``wave packet'' finds a wall at $z=L$ and rebounds against it. At  time $t=2L$ it starts reaching the UV brane, which explains geometrically the resurgence of the survival probability at precisely that time. The subsequent bounces at the UV and IR boundaries, and the resulting interference of left-moving and right-moving waves account for the oscillations of the survival probability in the discrete case. In the continuous case, on the other hand, there is no bounce and the state is eventually diluted deep inside the bulk, giving rise to irreversible invisible decay.\footnote{This happens also in other geometries with a continuum. Indeed, for any geometry and bulk potential, the equation of motion for the field can be written as a Schr\"odinger equation. The case with a continuum corresponds to the situation in which the potential asymptotes in the far IR to a constant value, which gives the mass gap~\cite{Falkowski:2008fz}. For energies above this mass gap, nothing stops the field solution from moving further and further towards the IR (maybe after some bounces if there were previous higher barriers).} Note that for $g\neq 0$ the visible decay rate follows the same pattern, since decay into elementary particles localized on the UV boundary can only occur at $z=0$. In the continuous case, the visible decay rate thus becomes smaller and smaller, in such a way that the probability of visible decay has an upper bound smaller than 1. This is the counterpart in time space of the branching ratio smaller than 1 unparticle decay into elementary particles. In the holographic description, the hidden asymptotic states can be understood as infinitely diluted wave packets lost in the far IR region.

One might think that $\int_0^L dz \mathcal{P}_{\mathrm{sur}}(t,z)=1$ in the free theory, as it seems to be the probability of finding the system in any of the states $\ket{\mathcal{B}_{\vec{p}}(z)}$, but this is not true. The reason is that the events of finding $\ket{\mathcal{B}_{\vec{p}}(z)}$  or $\ket{\mathcal{B}_{\vec{p}}(z^\prime)}$ are not exclusive for $z^\prime \neq z$, due to the non-orthogonality of these states. 
For this reason it is more convenient to work instead with the following quantities: 
\begin{align}
& \mathcal{P}_{\mathrm{1P}}(t) = \frac{1}{i\tilde{G}_\tau(0,\vec{p_0})} \int_0^\infty d\mu^2 \sigma^{(0)}(\mu^2) \frac{2\omega_{\mu,p_0}}{\tilde{f}_\tau(\omega_{\mu,p_0})} \left| \int_0^L dz f_\mu(z) \tilde{G}_\tau(t,\vec{p}_0;0,z)\right|^2 , \\
& \mathcal{P}_{\mathrm{osc}}(t) = \mathcal{P}_{\mathrm{1P}}(t) - \mathcal{P}_{\mathrm{sur}}(t). 
\end{align}
To explain the interest of these probabilites, let us consider the free theory for a moment. 
In the free theory, the states $\ket{\mathcal{B}_{\vec{p}}(z)}$ are related to the one-particle states by
\beq
\ket{\mathcal{B}_{\vec{p}}(z)} = \int_0^\infty d\mu^2 \sigma^{(0)}(\mu^2) f_\mu(z) \frac{Z_\mu^{-\frac{1}{2}}}{2\omega_{\mu p}} \ket{\mu,\vec{p}}_0 ~~(\mbox{free}). \label{Bz_onepart}
\eeq
The inverse relation,
\beq
\ket{\mu, \vec{p}}_0 = 2\omega_{\mu p} Z_\mu^{\frac{1}{2}} \int_0^L dz f_\mu(z) \ket{\mathcal{B}_{\vec{p}}(z)} ~~(\mbox{free}),
\eeq
shows that the states $\ket{\mathcal{B}_{\vec{p}}(z)}$ form a complete (non-orthogonal) set for free $A$ one-particle states.  Using the first line of~\refeq{G0zB}, together with~\refeq{Bz_onepart} and the orthogonality of the profiles, it is easy to see that in the free theory $\mathcal{P}_{\mathrm{1P}}(t) = 1$ at any $t$. Indeed, it is just the probability of evolving into an arbitrary one-particle state. We propose to also use $\mathcal{P}_{\mathrm{1P}}(t)$ in the interacting theory to approximate the probability of remaining at a given time in any free-one-particle state and $\mathcal{P}_{\mathrm{osc}}(t)$ to approximate the probability of oscillations into free $A$ one-particle states different from the initial one. According to this, we should have at any time
\beq
\mathcal{P}_{\mathrm{sur}}(t)+\mathcal{P}_\mathrm{osc}(t) +\mathcal{P}_{\mathrm{vis}}(t) \simeq 1. \label{unitarityt}
\eeq
This is the time-space version of the optical theorem. 
The numerical calculation of $\mathcal{P}_{\mathrm{1P}}$ involves several iterated integrals. Even if we have managed to evaluate it in some examples, let us mention in passing that there is a simpler alternative:
\begin{align}
& \tilde{\mathcal{P}}_{\mathrm{1P}}(t) = \frac{\int_0^L dz \left| \tilde{G}_\tau(t,\vec{p}_0,0,z)\right|^2}{\int_0^L dz \left| \tilde{G}_\tau(0,\vec{p}_0,0,z)\right|^2} , \label{tildeP1P}\\
& \tilde{\mathcal{P}}_{\mathrm{osc}}(t) =  \tilde{\mathcal{P}}_{\mathrm{1P}}(t) - \mathcal{P}_{\mathrm{sur}}(t).
\end{align}
Because the numerator of \refeq{tildeP1P} is independent of $t$ in the free theory,
\beq
\int_0^L dz \left| \tilde{G}^{(0)}_\tau(t,\vec{p}_0,0,z)\right|^2 = \int_0^\infty d\mu^2 \sigma^{(0)}(\mu^2) \left(\frac{\tilde{f}_\tau(\omega_{\mu,p})}{2\omega_{\mu,p}}\right)^2,
\eeq
we also have $\tilde{\mathcal{P}}_{\mathrm{1P}}(t)=1$ in this case.
Furthermore, at any time, $\tilde{\mathcal{P}}_{\mathrm{1P}}(t) =0$ if and only if $\mathcal{P}_{\mathrm{1P}}(t)=0$. 

%%%%%%%%%
\begin{figure}
  \centering
  \includegraphics[width=0.4\textwidth]{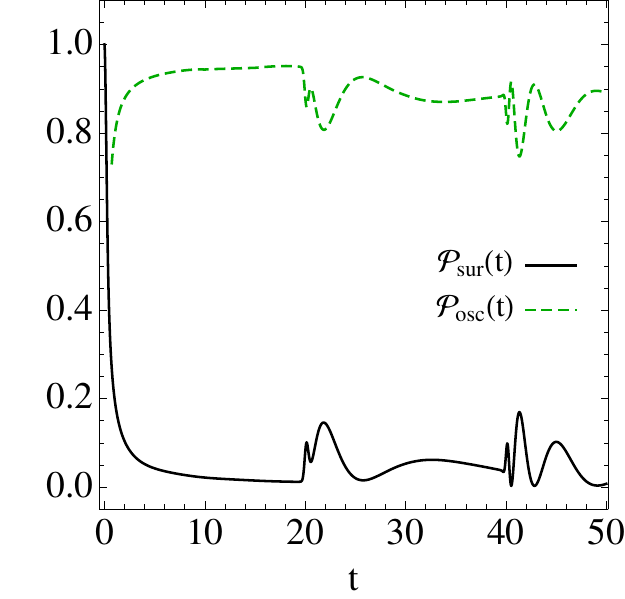} \hspace{1cm} \includegraphics[width=0.4\textwidth]{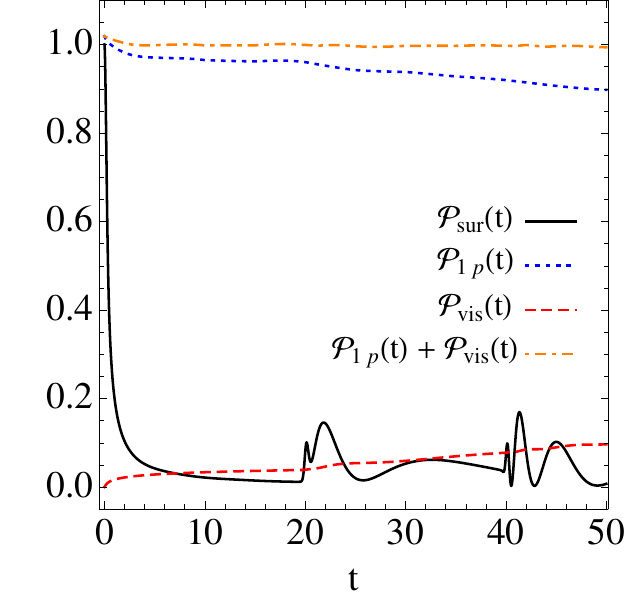} 
  \caption{Left panel: Oscillation probability in the interacting discrete case, compared with survival probability. Right panel: We display $\mathcal P_{\rm sur}(t)$, $\mathcal P_{\rm 1p}(t)$ and $\mathcal P_{\rm vis}(t)$, as well as the summation $\mathcal P_{\rm 1p}(t) + \mathcal P_{\rm vis}(t)$. We have considered $g = 1$, $\mu_0 = 1$, $L = 10$ and $M = 10$. We have used  Gaussian smearing with $\tau = 0.1$.  \label{fig:oscillation_discrete} }
\end{figure}
%%%%%%%%%
In the left panel of figure~\ref{fig:oscillation_discrete} we display the oscillation and survival probabilities in the discrete case. It is clear that the initial state mostly evolves (for small coupling and not very late times) into orthogonal one-particle states. However, for non-vanishing coupling these are not the only possibilities. This is better appreciated in the right panel of figure~\ref{fig:oscillation_discrete}, in which we plot  
${\mathcal{P}}_{\mathrm{1P}}(t)$, $\mathcal{P}_{\mathrm{vis}}(t)$ and their sum (together with  $\mathcal{P}_{\mathrm{sur}}(t)$). We see that ${\mathcal{P}}_\mathrm{1P}(t)$ decreases monotonically, but the sum gives 1 at all times to a good approximation. Even if not shown in this figure,  our results above imply that $\lim_{t\to \infty}{\mathcal{P}}_\mathrm{1P}(t) = 0$. 
%%%%%%%%%
\begin{figure}
  \centering
  \includegraphics[width=0.4\textwidth]{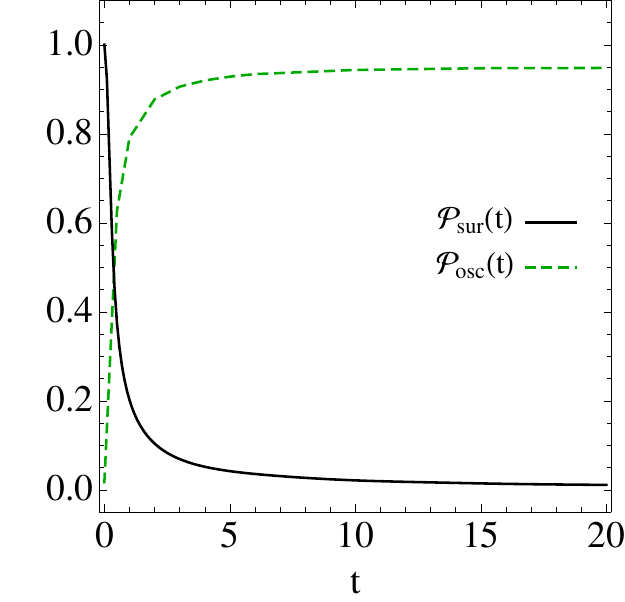} \hspace{1cm} \includegraphics[width=0.4\textwidth]{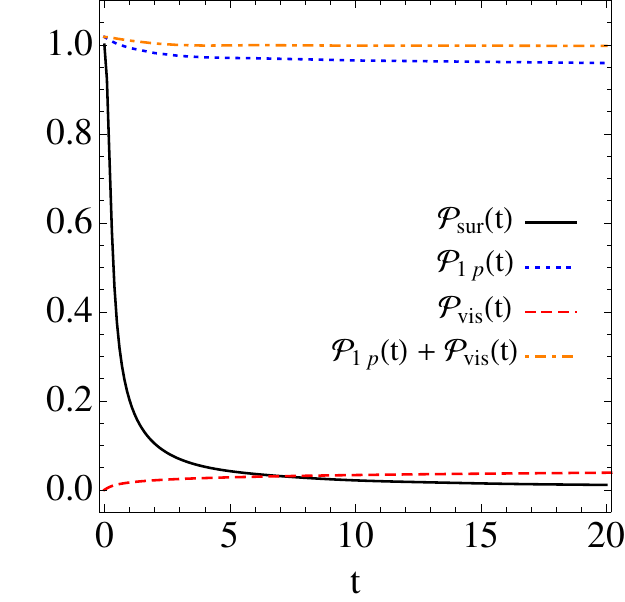} 
  \caption{Left panel: Oscillation probability in the interacting continuous case, compared with survival probability. Right panel: We display $\mathcal P_{\rm sur}(t)$, $\mathcal P_{\rm 1p}(t)$ and $\mathcal P_{\rm vis}(t)$, as well as the summation $\mathcal P_{\rm 1p}(t) + \mathcal P_{\rm vis}(t)$.  We have considered $g = 1$, $\mu_0 = 1$ and $M = 10$.  We have used Gaussian smearing with $\tau = 0.1$.}   \label{fig:oscillation_continuum}
\end{figure}
%%%%%%%%%
The same quantities are plotted for the continuous case in figure~\ref{fig:oscillation_continuum}. We see that the unitarity relation~\refeq{unitarityt} is also fulfilled approximately. But in this case, ${\mathcal{P}}_\mathrm{1P}(t)$ is bounded from below and approaches a strictly positive number as $t\to \infty$. Because $\mathcal{P}_\mathrm{sur}(t)$ approaches 0 in this limit, this implies that $\lim_{t\to \infty} \mathcal{P}_\mathrm{osc}(t)>0$. This is a direct manifestation of the existence of hidden one-particle asymptotic states deep inside the bulk, also in the presence of interactions with elementary particles much lighter than the mass gap. So, we see that in the interacting continuous case there are two decay channels with non-vanishing branching ratios: visible decay and invisible decay. They are in one-to-one correspondence with the visible cross section~\refeq{stotalcrosssection} and the excess~\refeq{excess}, respectively, in the energy representation. 

%%%%%%%%%

%%%%%%%%%%%%%%%%%%%%%%

\section{Physical processes}
\label{s:processes}
In sections~\ref{s:freetime} and~\ref{s:interactiontime} we have studied in detail the time evolution of a state created at some instant by the field $A$.  The physical way of (approximately) creating such an unstable state is by scattering of stable particles. So, the complete scattering process of stable particles must be considered in order to make predictions for observable quantities that can be compared to experiments. In this section, we discuss these physical processes and their relation to the previous analysis of the propagator. As external stable particles we can use the light particles $\varphi$ themselves, or, in extended models, any other particles that couple to $A$. In particular, we could consider the possibility of additional probe particles $\psi$ with coupling much smaller than $g$, such that their contributions to the self-energy can be neglected. 

To start with, it should be clear that the narrow-width approximation is not valid in general in our model. Indeed, this approximation not only requires that the different resonances be narrow with respect to their widths, but also with respect to the energy-momentum precision.  Otherwise, the effects of interference of the different modes completely change the cross sections, as we show below. The resolution in energy and momentum is constrained by limitations in the measuring devices and also by the space-time localisation of the initial and final states, which is actually required to study the space-time dependence of the process. Resolving the individual modes is of course difficult for compressed spectra and impossible in the continuum case. In these cases the narrow-width approximation cannot be used and it does not make sense to consider separately the production and decay of individual modes. This is the problem we see in the remarks on unparticle decay in~\cite{Stephanov:2007ry}, upon which we have already commented in the introduction.

We can distinguish two types of scenarios, according to the localization in space and time of the incoming and outgoing particles in a given experimental setup.\footnote{This discussion is largely based on the one for neutrino oscillations in ref.~\cite{Akhmedov:2010ua}.} This is determined by the preparation of the initial state and the measurements on the final state, and can be approximately described in terms of production and detection space-time regions. In the first scenario, these regions strongly overlap, so one can speak of a single interaction region. In the second one, their overlap can be neglected, so the production and detection regions are separated. Intermediate or more complicated situations are also possible, but we will not consider them in detail. In all cases, we can use the S-matrix formalism with normalizable asymptotic states. The transition amplitude for  $\varphi \bar{\varphi} \to \varphi \bar{\varphi}$ scattering\footnote{Consistently with the $A1$ approximation, we do not consider the possibility of soft or collinear emission of additional massless elementary particles.} is given by the LSZ formula: 
\begin{align}
\bra{g^\prime} S \ket{g} &= \int \left(\prod_{i,j=1}^2 d^4 x_i  d^4y_j g_i(x_i) [g^\prime_j(y_j)]^* \Box_{x_i}  \Box_{y_j} \right) \bra{\Omega} T \varphi^\dagger(y_1) \varphi(y_2) \varphi^\dagger(x_1) \varphi(x_2) \ket{\Omega} \nn
& =  \int \left(\prod_{i,j=1}^2 \frac{d^3p_i}{(2\pi)^3 2\omega_{0,p_i}} \frac{d^3q_j}{(2\pi)^3 2\omega_{0,q_j}} \tilde{g}_i(\vec{p}_i) [\tilde{g}^\prime_j(\vec{q}_j)]^* \right) \mathcal{M}(\vec{q}_1,\vec{q}_2;\vec{p}_1,\vec{p}_2).  \label{LSZ}
\end{align}
Here, $\mathcal{M}$ is the on-shell scattering amplitude for in and out states with well-defined momenta, given by amputated diagrams and with the energy-momentum-conservation delta included, while $g_i$ and $g_j^\prime$ are the wave packets of the incoming and outgoing particles, respectively, with
\beq
g_i^{(\prime)}(x_i) = \int \frac{d^3 p}{(2\pi)^3 2\omega_{0,p}} \tilde{g}_i^{(\prime)}(\vec{p}) e^{-i (\omega_{0,p} x_i^0 - \vec{p} \cdot \vec{x})}.
\eeq
They satisfy the massless Klein-Gordon equation. For the asymptotic formalism to be valid, the two $g_i(x_i)$ and the two $g_j^\prime(y_j)$ must have no spatial overlap when $x^0_i \to -\infty$ and $y^0_j \to \infty$, respectively. On the other hand, there should be  overlap in some space-time region(s) for a non-trivial scattering. We also assume that the functions $\tilde{g}_i$ and $\tilde{g}_j^\prime$ have disjoint support, to exclude disconnected contributions to the amplitude. %Finally, the fields $\varphi$ are canonically normalized ($Z_\varphi=1$). 
The probability of finding $\bar{\varphi}$ and $\varphi$ in a final state $\ket{g^\prime}$ when scattering  $\bar{\varphi}$ and $\varphi$ in a state $\ket{g}$ is  $\mathcal{P}_{g^\prime g}=|\bra{g^\prime} S \ket{g} |^2$.

%%%%%%%%%
\begin{figure}
  \centering
  \includegraphics[width=0.45\textwidth]{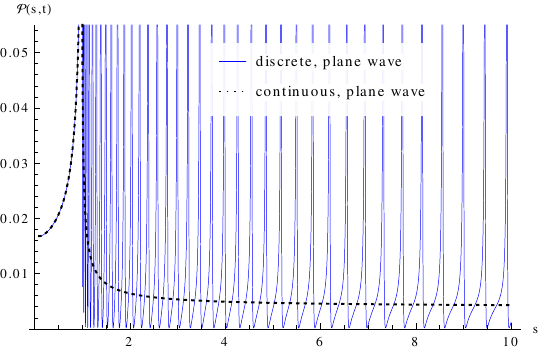} \hspace{1cm} \includegraphics[width=0.45\textwidth]{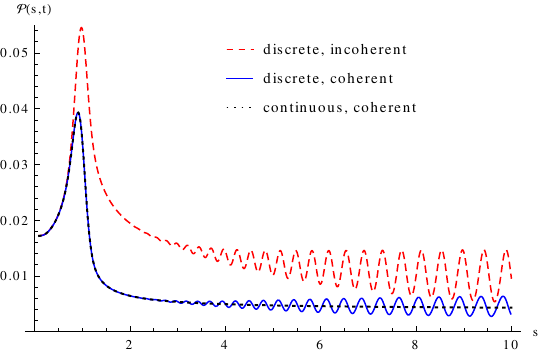} 
  \caption{Probability $\mathcal{P}(s,t)$ in discrete (blue solid lines) and continuous cases (black dotted lines) of $\bar{\varphi}\varphi \to \bar{\varphi}\varphi$ scattering as a function of the invariant mass $s$ for fixed squared momentum transfer $t=-0.1 \mu_0^2$. Left panel: plane waves.  Right panel: wave packets with Gaussian overlap function of width $0.1 \mu_0$ (see eq.~\refeq{overlapfunction}) and incoherent Gaussian smearing of the same width (in discrete case only). We have set  $g = 3$, $\mu_0 = 1$,  $L=40$ and $M = 1$. }   \label{fig:crosssection}
\end{figure}
%%%%%%%%%
The scenario with a single interaction region corresponds to the case in which all $g_1$, $g_2$, $g^\prime_1$ and $g^\prime_2$ have a strong overlap in a common space-time domain.  Then, all the cross diagrams in $M$ will contribute to the amplitude of the physical process. The $s$-channel contribution to the amplitude at leading order (with resummed $A$ propagator) is
\begin{align}
\bra{g^\prime} S \ket{g}^{(s)} = & - i g^2   \int \left(\prod_{i,j=1}^2  \frac{d^3p_i}{(2\pi)^3 2\omega_{0,p_i}} \frac{d^3q_j}{(2\pi)^3 2\omega_{0,q_j}}  \tilde{g}_i(\vec{p}_i) [\tilde{g}^\prime_j(\vec{q}_j)]^* \right) 
 G\left(\omega_{0,p_1}+\omega_{0,p_2}, \vec{p}_1+\vec{p}_2 \right)  \nn
& \mbox{} \times (2\pi)^4 \delta^{(3)}(\vec{p}_1+\vec{p}_2-\vec{q}_1-\vec{q}_2) \delta(\omega_{0,p_1}+\omega_{0,p_2}-\omega_{0,q_1}-\omega_{0,q_2}) . \label{schannel}
\end{align}
The $t$-channel contribution is the same, but with $\vec{p}_2 \to -\vec{q}_2$ in the arguments of $G$. 
Let us consider in particular the standard situation with $\tilde{g}_i$ and $\tilde{g}_j^\prime$ peaked at $\vec{\mathbf{p}}_i$ and $\vec{\mathbf{q}}_j$, respectively. Then, (peak) energy and momentum are approximately conserved: a non-vanishing amplitude requires $\vec{\mathbf{p}}_1+\vec{\mathbf{p}}_2 \simeq \vec{\mathbf{q}}_1+\vec{\mathbf{q}}_2$, $\omega_{0,\mathbf{p}_1} + \omega_{0,\mathbf{p}_2} \simeq \omega_{0,\mathbf{q}_1} + \omega_{0,\mathbf{q}_2}$. Nevertheless, in the case of compressed spectra, the finite spreads around the peak values can be very important. They must be taken into account whenever the energy-momentum uncertainties are larger than the mass spacing~\cite{Perez-Victoria:2008pbb,deBlas:2012qp}. Indeed, as shown on the left panel of figure~\ref{fig:crosssection}, the cross section oscillates strongly  with the invariant mass $s=(\omega_{0,\mathbf{p}_1}+\omega_{0,\mathbf{p}_2})^2 - (\vec{\mathbf{p}}_1+\vec{\mathbf{p}}_2)^2$ for discrete compressed spectra in the plane-wave limit. Here we have taken width mixing into account~\cite{Cacciapaglia:2009ic}. The effect on the cross section of a finite resolution in the wave functions is twofold. On the one hand, the oscillations are averaged out. On the other, the destructive interference between contributions with different momenta and energies gives rise to a suppression of the cross section. As a result of these two effects, for large enough uncertainties the resulting cross section for a compressed spectrum precisely mimics the one for the corresponding continuous spectrum. This is illustrated on the right panel of figure~\ref{fig:crosssection} (black dotted and blue solid lines). We also plot there the probability resulting from an incoherent smearing of the plane-wave probability in the discrete case (red dashed line). In this case, the oscillations are only averaged, with no suppression. The significant difference of the coherent and incoherent smearings can be understood from our previous analysis of time evolution. For incoherent smearing, the amplitude is calculated in the standard way with plane waves and the only asymptotic states that can appear in the final state are formed by the stable particles $\varphi$ and $\bar{\varphi}$.  For the coherent smearing in~\refeq{LSZ}, on the other hand, the wave packets are localized within a finite time interval. Then, another channel is possible, with a final state formed by quasi-stable oscillation one-particle $A$ states, which are effectively stable within that time interval. The branching ratio into visible particles is decreased due to this competing channel. Moreover, for energy uncertainty larger than the mass spacing, only the times smaller than the inverse mass spacing, for which the probabilities have continuum behaviour, are probed.

For a strictly continuous spectrum the cross section varies slowly with $s$, and the size of the energy-momentum uncertainties has a very small impact. One way of understanding this is that no matter how small the uncertainties are, there are always infinite continuum modes that interfere among themselves. Therefore, we can approximate
\beq
\bra{g^\prime} S \ket{g}^{(s)} \simeq  - g^2 i G\left(\omega_{0,\mathbf{p}_1}+\omega_{0,\mathbf{p}_2}, \vec{\mathbf{p}}_1+\vec{\mathbf{p}}_2 \right)  (2\pi)^4 \delta^{(4)}(\mathbf{p}_1+\mathbf{p}_2-\mathbf{q}_1-\mathbf{q}_2)~~~\mathrm{(continuum)}. \label{aproxamplitudecont}
\eeq
It is somewhat paradoxical that the continuous case can be well described in terms of idealized initial and final states with well defined momenta and energies, just as in the case in which the individual modes near the relevant invariant mass are resolved, while this is a bad approximation in the intermediate case of unresolved discrete spectrum. The explanation is that the limit of external particles with well-defined momenta does not commute with the continuum limit of the spectrum of the field $A$. When the actual resolution is larger than the mass spacing, the correct approximation procedure is to perform the continuum limit first, so that the employed effective resolution is always kept larger than the mass spacing and smaller than the real one.\footnote{This fact has been used in the past, for instance in~\cite{Giudice:1998ck} in the context of large extra dimensions, which feature compressed spectra of Kaluza-Klein gravitons coupled to four-dimensional SM fields. A graviton that disappears into a large extra dimension precisely corresponds to the process we have described in our holographic model when $L$ is finite but large.} Note finally in figure~\ref{fig:crosssection} that the invariant-mass distribution in the continuous case is quite peaked at the mass gap in our model. This feature is due to the square root divergence of $G^0(s)$ at that point (independently of the existence of a pole on the second Riemann sheet). Even if the shape is different, with poor resolution it may be difficult to distinguish this peak from the one of an ordinary resonance. 

Let us consider next the scenario with distinct production and detection regions. This corresponds to the case in which $g_1$ and $g_2$ overlap in some space-time region $R_P$ and $g_1^\prime$ and $g_2^\prime$ overlap in another space-time region $R_D$ with $R_P \cap R_D \approx \,$\O. Then, the $t$-channel diagram can be neglected, and the amplitude is given by~\refeq{schannel}. Let $T$ and $l$ be the typical distance along the time and spatial directions, respectively, between points in $R_P$ and points in $R_D$. A space-time analysis of the process is only possible if these distances are macroscopic and much larger than the corresponding sizes of $R_P$ and $R_D$. This occurs, for instance, in setups with a far detector or when the measurements identify a displaced vertex at a distance $l$. In this situation, each mode $\mu$ propagates nearly on-shell with classical velocity $v_\mu=p/\mathrm{Re}\, \omega_{\mu,p}$. A sizable contribution of a mode $\mu$ in the propagator to the amplitude requires $l \approx v_\mu T$. This condition, and a similar one in the transverse direction, will only be met by part of the modes, giving rise to some degree of decoherence. This phenomenon is well known in the context of neutrino oscillations~\cite{Nussinov:1976uw}. To be more explicit and make more apparent the relation with the discussion in sections~\ref{s:freetime} and~\ref{s:interactiontime}, we follow the developments in~\cite{Jacob:1961zz} and \cite{Beuthe:2001rc}. First, we choose
\begin{align}
&\tilde{g}_i(\vec{p}) = \bar{g}_i(\vec{p}) e^{i\omega_{0,p} T_P - i \vec{p}\cdot \vec{X}_P} ,\nn
&\tilde{g}^\prime_j(\vec{q}) = \bar{g}^\prime_j(\vec{q}) e^{i\omega_{0,q} T_D - i \vec{q}\cdot \vec{X}_D} ,
\end{align}
with  $\bar{g}_i$ and $\bar{g}_j^\prime$ peaked at $\vec{\mathbf{p}}_i$ and $\vec{\mathbf{q}}_j$, respectively, and satisfying the approximate symmetry $\bar{g}_i(\vec{\mathbf{p}}_i+\vec{k}) \simeq \bar{g}_i(\vec{\mathbf{p}}_i-\vec{k})$, $\bar{g}^\prime_j(\vec{\mathbf{q}}_j+\vec{k}) \simeq \bar{g}^\prime_j(\vec{\mathbf{q}}_j-\vec{k})$, for any $\vec{k}$. This gives rise to overlap regions $R_P$ and $R_D$ centered at $X_P:=(T_P,\vec{X}_P)$ and $X_D:=(T_D,\vec{X}_D)$, respectively. So, $T=T_D-T_P$, $l=|\vec{X}_D-\vec{X}_P|$. We can then rewrite~\refeq{schannel} as
\beq
\bra{g^\prime} S \ket{g} = -g^2 \int \frac{d^4 k}{(2\pi)^4} \psi(k) i G(k^2)  e^{-i k (X_D-X_P)}, \label{covariantamp}
\eeq
where we have defined the overlap function
\begin{align}
\psi(k) =  & \int \left(\prod_{i,j=1}^2  \frac{d^3p_i}{(2\pi)^3 2\omega_{0,p_i} } \frac{d^3q_j}{(2\pi)^3 2\omega_{0,q_j}}  \bar{g}_i(\vec{p}_i) [\bar{g}^\prime_j(\vec{q}_j)]^* \right)  \nn
& \mbox{} \times (2\pi)^4 \delta^{(3)}(\vec{p}_1+\vec{p}_2-\vec{q}_1-\vec{q}_2) \delta(\omega_{0,p_1}+\omega_{0,p_2}-\omega_{0,q_1}-\omega_{0,q_2})  \nn
& \mbox{} \times (2\pi)^4  \delta^{(3)}(\vec{k}-\vec{p}_1-\vec{p}_2) \delta(k^0-\omega_{0,p_1}-\omega_{0,p_2}).  \label{overlapfunction}
\end{align}
Choosing to integrate first on $k^0$ in~\refeq{covariantamp}, we can write
\beq
\bra{g^\prime} S \ket{g} =  -g^2 \int \frac{d^3 k}{(2\pi)^3} J(T,\vec{k})  e^{i \vec{k} \cdot (\vec{X}_D-\vec{X}_P)},  \label{spacialint}
\eeq
with 
\beq
J(T,\vec{k}) = \int_{-\infty}^\infty \frac{d E}{2\pi} i G(E,\vec{k})  \psi(E,\vec{k}) e^{-i E T} \label{temporalint}.
\eeq
The right hand side of~\refeq{temporalint} is just like \refeq{FourierSmearProp}, with the particular smearing function $\tilde{f}_\tau(E)=\psi(E,\vec{k})$. The function $\psi$ has two main differences with respect to the smearing functions we have used before. First, $\psi(E,\vec{k})$ necessarily vanishes when $E<0$. This does not entail a significant numerical difference because the region with $E<0$ is far from the poles (and the extra branch cut, in the continuous case). Second, $\psi(E,\vec{k})$ will be peaked at $E=\omega_{0,\mathbf{p}_1}+\omega_{0,\mathbf{p}_2}$. The remaining integral over $\vec{k}$ is akin to~\refeq{hintegral}, with a particular wave packet $h$. In previous sections we have chosen for simplicity a very strongly peaked $h$. In a real process, however, the time and spatial uncertainties are correlated, and we cannot take the limit of well-defined initial (and final) momentum without loosing the localization in time. 
%Therefore, the oscillations we have studied will also be partially averaged out by a momentum uncertainty, since $\omega_n$ in \refeq{polestime} depends on $\vec{p}$. This adds to the similar effect of the time uncertainty. A consequence of all this is that, also in this scenario, a highly compressed spectrum will behave like the continuum, for $T$ smaller than the inverse mass spacing. As we have emphasized above, the behaviour is very different for longer times. 
The main effect of the integral over $\vec{k}$ is to introduce a dependence of the amplitude on $l$, correlated with the T dependence. All the features of the time evolution we have studied in section~\ref{s:interactiontime}, including oscillations and non-exponential behaviour, will also show up in the dependence on $l$.
In many experimental observables the time elapsed between production and detection is not measured, so to extract the dependence on $l$ we should integrate the probability $\mathcal{P}_{g^\prime g}(T,l)$ over $T$. 
Assuming Gaussian wave packets, an approximate analytic expression to the integrated probability $\mathcal{P}_{\mathrm{int}}(l) = \int_{-\infty}^\infty  dT \, \mathcal{P}_{g^\prime g}(T,l)$ can be found at large enough $l$ in the discrete case by using the pole approximation to the integral~\refeq{temporalint}, a stationary-phase approximation in~\refeq{spacialint} and Laplace's method for the $T$ integral (see~\cite{Beuthe:2001rc} for details of the calculation, including the conditions under which the approximations are valid). The result is
\beq
\mathcal{P}_{\mathrm{int}}(l) \simeq g^4 N_C \frac{C_{\vec{\mathbf{p}}} (\vec{X}_D-\vec{X}_P)}{l^2} \sum_{mn} \mathcal{Z}_m \mathcal{Z}^*_n  \exp \left[- i \frac{\omega_m^2-\omega_n^2}{2\mathbf{p}} l -\frac{l}{L_{mn}^{\mathrm{decay}}} 
- 2 \pi^2  \left(\frac{l}{L^{\mathrm{coh}}_{mn}}\right)^2 \right] . \label{Lprob}
\eeq
The following definitions have been used. First, $\vec{\mathbf{p}} = \vec{\mathbf{p}}_1 + \vec{\mathbf{p}}_2$ is the mean total momentum of the initial state (approximately equal to the one of the final state) and  $\mathbf{p} = \sqrt{\vec{\mathbf{p}}^2}$, its modulus. The poles of the propagator on the second Riemann sheet $G^{\mathrm{II}}(E^2)$ are located at $E_n=\omega_n - i \Gamma_n/2$ and have residues $\mathcal{Z}_n/(2E_n)$, as in subsection~\ref{ss:discrete_interacting}. It is assumed that $\omega_m^2-\omega_n^2 \ll \mathbf{E}^2$, with $\mathbf{E}$ the mean energy of the initial (and final) state, for all the modes $m$, $n$ that give a sizable contribution to the sum in~\refeq{Lprob}. The width $\sigma_p$ is approximately the smallest of the production and detection momentum widths, and we have also assumed that $\omega_m^2-\omega_n^2 \ll \sigma_p \mathbf{p}$ for the relevant modes. The function
\beq
C_{\vec{\mathbf{p}}}(\vec{x}) = e^{-\frac{(\vec{\mathbf{p}} \times \vec{x})^2}{4 \vec{x}^2 \sigma_p^2}}
\eeq
represents a cone of axis $\vec{\mathbf{p}}$ and angle $\arcsin (\sigma_p/\mathbf{p})$, with the corresponding normalization constant given by
\beq
N_C^{-1} = \int d\Omega C_{\vec{p}}(\vec{x}). 
\eeq
The length
\beq
L_{mn}^{\mathrm{decay}} = \frac{2 \mathbf{p} }{\omega_m \Gamma_m + \omega_n \Gamma_n}
\eeq
is an average decay length for the modes $m$ and $n$, associated to decay into elementary particles. And finally,
\beq
L_{mn}^{\mathrm{coh}} = \frac{4 \mathbf{p}^2}{\sqrt{2} \sigma_p^\prime (\omega_m^2-\omega_n^2)} 
\eeq
is the coherence length between modes $m$ and $n$. The corresponding exponential suppression for $n\neq m$ arises from the decoherence effect due to the separation of the ``propagating wave packets" of those modes, as a result of their different velocities, and also from the longitudinal dispersion of such wave packets, which, when large enough, averages to zero the interference terms in the time integral. Here, $\sigma^\prime_p \leq \sigma_p$ is an effective width in the overlap function $\psi$ (see \cite{Beuthe:2001rc}  for its precise definition), associated to energy uncertainties in the production and detection process. Clearly, the first term in the exponent of ~\refeq{Lprob} represents oscillations as a function of distance, while the second one represents a distance-dependent decay into elementary particles. These mimic accurately the behaviour we have studied above as a function of time (at intermediate times). In fact, ignoring the geometric factors and assuming infinite coherence length together with the assumptions above, $\refeq{Lprob}$ can be simply reproduced from the time-dependent visible decay rate for well-defined momentum $\vec{\mathbf{p}}$ (with time smearing) by considering classical on-shell propagation of each mode $n$, which allows to substitute $t$ by $(\omega_n(\mathbf{p})/\mathbf{p}) l$. 
Eq.~\refeq{Lprob} applies, as written, to the discrete case only. Nevertheless, the time-integrated probability in the continuous case can be obtained as the continuum limit of this equation, which does not violate the assumptions above. Therefore, it can be approximated by the discrete formula for $l$ smaller than the inverse mass spacing. We can also use the trick of classical propagation to directly convert our results for the continuous case in section~\ref{s:interactiontime} into a distance-dependent probability.

The probability of occurrence of the $\bar{\varphi} \varphi \to \bar{\varphi} \varphi$ process is proporcional to the visible decay rate $\mathcal{P}_{\mathrm{\bar{\varphi}\varphi}}$. This is in a sense a disappearance experiment, in which the survival probability of the initial state is measured. The same applies to processes in which $\varphi$ or $\bar{\varphi}$ particles are initially present in the detector. None of these processes allows to probe directly the oscillations into orthogonal one-particle states, except for their collective effect in reducing the survival probability. In order to probe oscillations directly, a non-local coupling of $A$ to some visible field would be necessary.  Such couplings would be induced in the effective theory if these visible fields (partially) belonged to the hidden sector and were not completely integrated out. In a strongly-coupled extra sector with a higher-dimensional holographic dual, these fields would propagate in the bulk. An example of this is provided by the gauge fields in the 5D Unhiggs scenario~\cite{Falkowski:2008yr}.

%%%%%%%%%%%%%%%%

\section{Summary}
\label{s:summary}

The excellent agreement of the Standard Model (SM) with the experimental measurements strongly suggests that any new degrees of freedom should be either heavier than the energies we have probed so far or feebly interacting with the known particles. Beyond this, little is known about the nature of physics beyond the SM. In this work, we have studied some aspects of hypothetical new degrees of freedom with a gapped continuous distribution of masses. This type of new physics might elude to some extent the conclusion of the first sentence of this summary, due to the fact that it is more elusive to direct searches. First, it does not produce sharp peaks in the differential cross sections, so the signals cannot be easily separated from the background. Second, its contribution to cross sections with SM particles in the final states is typically suppressed. We have shown here that the origin of this suppression is a reduced branching ratio due to additional invisible  final states in the scattering of SM particles. These two phenomenological features of continuous spectra are present as well for new physics with tightly compressed spectra, as a result of energy-momentum uncertainties in the initial and final states of the processes.

The presence of invisible decay channels in these scenarios looks surprising when the decay channel into lighter visible particles is open. Indeed, in the familiar case of one unstable particle, such as a Z boson, the unstable particle eventually decays, so in the far future of the scattering process all the final states are made of visible particles. The same applies to the case of several unstable particles with significant mass differences, such that their resonances can be resolved. However, this intuitive explanation of the completeness of the stable visible states relies on the narrow-width approximation, which obviously fails for continuous or very compressed spectra. In these cases, the strong interference effects between different modes with the same decay products completely changes the picture and opens the way to invisible final states. 

Our main purpose in this work has been to provide a clear understanding of the nature of such states from the point of view of the non-local effective theory that describes the interactions of the extra sector with the SM fields. To achieve this, we have studied in detail the time evolution of quantum states in a toy model that captures the most important features of such an effective theory. The crucial feature of this model is a form factor in the quadratic term of a mediator field, which acts as an interpolating field for asymptotic states in the extra sector. We have studied the two cases in which these asymptotic states form a continuous and a discrete spectrum. In a perturbative description, the time evolution can be outlined as follows. 

First, the scattering of visible particles gives rise to a state that is a superposition of one-particle states with different energy. Then, this initial state evolves non-trivially. After a finite time has elapsed, the system can be found, in a measurement, either in the same initial state  ({\em survival}), in an orthogonal state formed by linear combinations of the extra modes created by the mediator ({\em oscillation}), or in a multi-elementary-particle state ({\em visible decay}). In the continuous case, the survival probability decreases following a power law, in contrast with the classical exponential decrease of a single unstable particle. The same is true in the discrete case, up to a time of the order of the inverse mass spacing. Before that, the discrete and continuous cases are almost indistinguishable. Afterwards, on the other hand, a partial revival of the initial state takes place in the discrete case, and the survival probability oscillates around a decaying exponential. The survival probability always goes to zero when $t \to \infty$.

As regards the visible decay, its rate is at the beginning of the evolution smaller in the continuum than for one particle with the same coupling. The reason is the existence of infinitely-many orthogonal oscillation states, which enhances the probability of oscillation as the result of the larger phase space. This situation persists during a certain time interval, which is longer for smaller coupling. As a result, at the end of this period the probability of visible decay is significantly suppressed in the continuous case. Even if at later times it increases faster than for exponential decay, this is not sufficient to compensate for the previous deficit, and the visible decay probability ends up being asymptotically smaller than one. 
This shows that in the continuous case the system can survive decay and be found in the far future in a stable subspace of oscillation states. These are the invisible final states in scattering processes.

In the discrete case, the visible decay rate is suppressed initially, just as in the continuous case. But after a time of the order of the inverse mass spacing it increases and then oscillates. This eventually compensates the initial deficit, so the final states are formed exclusively by visible particles. In practice, however, for very compressed spectra the decay may occur outside the detector, much as in the case of ordinary long-lived particles. In this sense, the continuum limit is smooth. The apparent discontinuity is the result of the non-commutation of the continuum limit and the idealized $t\to \infty$ limit in scattering theory.

These results have been derived without  any assumptions beyond the ones incorporated in the effective Lagrangian, so they are very general and valid for any ultraviolet completion. Nevertheless, we have also made use of a particular completion that provides an alternative point of view: a model in five dimensions, which is a toy model for the holographic dual of a strongly-coupled extra sector.
%\footnote{If the non-local effective action is obtained by exactly integrating out the extra degrees of freedom and all the sources that can be used to probe them are kept, the effective theory keeps all the information of the complete theory and is actually an equivalent theory written in terms of different degrees of freedom. This is the case for our holographic model if the bulk degrees of freedom can only be probed via the coupling at the UV boundary.} 
It gives rise to discrete and continuous spectra when the extra dimension is compact and non-compact, respectively.\footnote{Some of the main ideas we have emphasized here, like oscillation into different linear combinations of Kaluza-Klein modes and the consequence of a reduced visible cross section, were already studied and used in~Ref.~\cite{Dienes:1999gw}, where a compact model with a large extra dimension and a bulk axion was proposed and analyzed.} The mass spacing is conveniently controlled by the compactification radius and the local coupling to the elementary sector is realized by a coupling of the bulk field to four-dimensional fields localized on the ultraviolet boundary. The oscillation states correspond to states inside the bulk of the extra dimension and time evolution can be described as motion in the extra dimension: An initial state created on the UV boundary moves further and further inside the bulk as time goes by. Its smaller overlap with the ultraviolet boundary translates into a smaller visible decay rate, due to the localization of the elementary fields. When the extra dimension is non-compact, if no visible decay has been observed before, the state is eventually diluted deep inside the bulk. Such infinitely diluted states form the hidden final states in this completion. When the extra dimension is compact, on the other hand, the state bounces off the infrared boundary and at some point reaches again the ultraviolet boundary, leading to a resurgence of the visible decay rate. 

\section{Discussion and outlook}
\label{s:discussion}

We conclude with a discussion of variations of the basic scenario we have considered so far and of possible phenomenological implications. 

The simple effective theory we have studied could be complicated in several ways, besides the obvious one of adding more mediators and more visible fields. 
One possibility is that the lightest elementary fields to which the mediator couples linearly have a mass that is larger than half the mass gap. Then, part of the modes created by the mediator will be stable. In the continuous conformal cases studied in~\cite{Delgado:2008gj}, these modes include a discrete part associated to a real pole on the physical sheet below the mass gap. In this situation, the stable part of the spectrum, which clearly forms possible final states, will behave as the free theory in section~\ref{s:freetime}, while the unstable modes above the production threshold will behave much as the interacting theory in section~\ref{s:interactiontime}. These two sides of the spectrum are not mixed during the time evolution, as the exact Hamiltonian is diagonal in that basis. 

An important ingredient of the theory we have considered is the locality of the interactions of the mediators with the visible particles. Non-local interactions may arise in the context of strongly-coupled setups when some visible fields are a mixture of elementary and composite fields (partial compositeness). For instance, in composite-Higgs models, including the UnHiggs case~\cite{Stancato:2008mp,Falkowski:2008yr,Falkowski:2009uy}, the SM gauge fields are partially composite and have non-local couplings to the Higgs field. Such fields could be used to probe more efficiently the hidden states but in general they do not have dramatic consequences for their decay. The exception is the case in which the mediators couple linearly to light partially-composite particles and can decay into them. Depending on the model, this may even preclude the existence of stable hidden states in the continuum. Note that the presence of non-local interactions in the effective theory depends on which combinations of degrees of freedom are integrated out and which are kept as visible fields. 

We have neglected in our analysis the impact of higher-point functions of the extra sector, which would manifest in the effective theory as non-local self-interactions of the mediator field. They are suppressed with respect to the two-point functions when the extra sector is a large N theory, but not for a generic weakly-coupled or strongly-coupled theory. In holographic theories, they are associated with bulk couplings, which are present in most models in extra dimensions. Furthermore, self-interactions of the mediator field are generated even in our simple effective theory by loops of the visible fields, although these contributions are suppressed by higher powers of the coupling and go beyond the approximation we have used. The most important effect of higher-point functions is that they give rise to final states with higher multiplicity of visible particles.\footnote{Let us stress that these multi-particle states cannot account for the excess in the imaginary part of the amplitudes in the continuous case. First, the corresponding cross sections depend on the self-coupling and cannot be equal for generic values of this coupling to terms that are insensitive to it. Second, the phase space for these processes does not give rise to the same momentum dependence as in the imaginary part. These final sates do match other (higher-loop) contributions to the imaginary part of the amplitudes, and are necessary for the unitarity of the effective theory with self-interactions.} This could lead to striking hidden-valley signals if the mass gap is small~\cite{Strassler:2006im,Strassler:2008bv}. Otherwise, higher multiplicities will have a significant kinematical suppression. The other effect, which is more relevant for our study, and is related to the previous one by unitarity, is that the higher-point functions will allow for loops of the mediator in the propagator, which should also be resummed. The new thresholds will give rise to new branch points. All this will change of course our explicit results, but not our qualitative conclusions. Indeed, the resulting spectral function will have the same general form, up to new discontinuities that indicate the decay of heavy mediator modes into lighter ones. The lighter modes will behave exactly as in the model we have studied. Therefore, in the continuum case with self-interactions the invisible final states can be formed not only by one-particle oscillation states, as in our model, but also by multi-particle oscillation states. 

We have had implicitly in mind throughout the paper strongly-coupled extra sectors, as their distinction from a visible sector to which they couple weakly is better motivated and the effective point of view is more useful. But let us point out that the same sort of effective theory could result from a weakly-coupled completion. In that context, our findings are nothing but the effective-theory description of very familiar phenomena. Consider for instance an extra sector formed by a field $\Psi$ of mass $\mu_0/2$, which is free except for an interaction $\lambda \Psi \Psi A$ with a mediator $A$. This is an example with a continuous spectrum with mass gap $\mu_0$, in which our continuous ``one-particle'' states are actually  $\Psi \Psi$ two-particle states. These are produced via the mediator at a small spatial region. Afterwards,  the wave packet of each $\Psi$ separates more and more from the other one. At the beginning, the overlap of the two wave functions is sizable and they can interact with the mediator---the interaction is local and involves two $\Psi$ fields---and thus produce visible particles. But after a while, the overlap of the two wave functions is negligible, so they cannot interact any longer with the mediator and will not give rise to visible particles any more. Instead, the two-particle state will continue its evolution inside a stable subspace of configurations. In such a model, the simultaneous existence of visible and $\Psi \Psi$ decay channels comes as no surprise. Note also that this model with a free extra sector has unsuppressed higher-point functions, which give rise to self-interactions of $A$ with the only suppression of a factor of $\lambda$ for each point.
We could even think of compactifying the three spatial dimensions, which would give rise to a discrete mass spectrum and to the eventual annihilation of all $\Psi \Psi$ pairs (assuming a contrived model in which the elementary fields could still propagate in a non-compact space). We find interesting the fact that, in these respects, the same role of ordinary space in weakly-coupled completions is played in strongly-coupled completions by the extra (radial) dimension of their holographic duals. This is not completely unexpected, given the infrared/ultraviolet connection in holographic dualities. 

It may be helpful to illustrate the ideas in this work by translating to this language some features of a realistic process in the SM. Consider for instance $e^+ e^-$ collisions at energies up to 3.5~GeV and let us treat QCD as the extra sector of the theory.\footnote{We thank the referee for suggesting this example.} The corresponding effective theory is obtained by exactly integrating out the quarks and gluons (which of course cannot be done explicitly, but we can be guided by experimental data~\cite{ParticleDataGroup:2022pth}). For the analysis of $e^+ e^-$ annihilation, we can use the photon field as the mediator. The resummation of vacuum polarization corrections gives rise to a dressed photon propagator that, besides the pole at the vanishing momentum, has branch cuts related to the different final states, including the hadronic ones.\footnote{This relation allows in particular, via dispersion relations, for the evaluation of the hadronic vacuum polarization contributions to the muon anomalous magnetic moment from experimentally measured $e^+ e^- \to \mbox{hadrons}$ cross sections.}  What we have called the free (effective) theory corresponds to turning off the {\em explicit} electroweak couplings {\em in the effective theory}. Note that the resulting effective theory is actually not free, as there are non-trivial higher-point functions for the photon/mediator field. At any rate, in this ``free theory'', the spectral density of the photon field has an isolated point at $\mu^2=0$, corresponding to the physical photon $\gamma$, and a continuum starting at the $\pi^0 \gamma$ threshold, with further discontinuities due to other final states, such as $\pi^+ \pi^-$, and peaks associated to the resonances ($\rho$, $\omega$, $\phi$, etc.). Therefore, we are here in the continuous case, except for the isolated photon mode. At energies above $\sim$ 2~GeV, the final states are best understood inclusively, as a continuum of hadrons containing $u$, $d$ and $s$ quarks, and the cross sections agree well with perturbative QCD calculations.  In the ``free theory", the charged pions are stable, but the neutral pion can decay into $\gamma\gamma$, which in this picture is only possible because the physical photon is partially composite. When the electroweak couplings are turned on, almost all of these states completely decay into visible particles, that is, into leptons and the physical photon. This occurs via additional interactions between the composite and elementary sectors (the weak interactions of the quarks), which do not involve the mediator. Such interactions are not present in our toy model. However, there are also $p\bar{p}$ final states. Because the protons and antiprotons are stable, this continuous set of final states would be missed if we only measured leptons and photons, so the visible cross section is slightly smaller than in the absence of this channel. These states (and others involving protons and/or antiprotons) form the invisible states that we have emphasized in this paper. Of course, in this case they are very easy to detect, since the elementary components of these ``invisible'' states couple directly to a visible particle, the physical photon, and furthermore, unlike in the case of unparticle stuff beyond the SM, our world is largely made of one of these components: the proton.

%%%%%%%%%%%%%%%%

The phenomenology of the new-physics scenario with continuous or very compressed spectra and stable or quasi-stable invisible states is in many regards similar to the one of long-lived particles, with signals such as missing energy and displaced vertices. The main difference for displaced vertices is that the distance dependence will not follow an exponential but a power law  (up to a distance of the order of the inverse spacing), with an exponent that depends mostly on the form of the spectrum near the mass gap and is fairly insensitive to the value of the coupling. The implications for experiments depend on the values of the different parameters. We leave the corresponding analysis for the future. Here let us just mention that one could find larger event rates than expected at large distances, which could be targeted by far detectors. This is specially so in the discrete case, due to the resurgence of the decay rate.

In a different vein, the stability or quasi-stability of the hidden states for continuous and very compressed spectra, respectively,  suggests that this scenario could give rise to dark matter, or rather, {\em dark stuff}. The interest of this possibility is twofold. First, due to the suppression of cross sections stressed above, the limits from direct searches would be less stringent than for ordinary particles. Second, there is no need of a $Z_2$ symmetry because (quasi) stability does not require the absence of linear couplings in our scenario.  Continuum dark matter has actually been proposed recently in~\cite{Csaki:2021gfm} (see also~\cite{Csaki:2021xpy,Csaki:2022lnq} for related work and~\cite{Fichet:2022ixi,Fichet:2022xol} for alternatives). In that work, the strong suppression of the cross section in direct searches was emphasized, although its origin is different from the one we have pointed out here. A $Z_2$ symmetry was imposed to ensure stability. The novelty of the idea we propose here relies on the fact that no such discrete symmetry is necessary for dark stuff, in principle.\footnote{This proposal actually reduces to the case of standard dark matter for weakly-coupled completions of the extra sector. For instance, the simple model with hidden $\Psi$ fields used as an example above does have a $Z_2$ symmetry $\Psi \to - \Psi$, which forbids linear couplings to the elementary fields and thus the decay of the isolated $\Psi$ particles. A similar completion of the models in~\cite{Csaki:2021gfm} would have a $Z_4$ symmetry, forbidding $\Psi \Psi$ annihilation.} We will explore the viability of this intriguing possibility in future work.

%%%%%%%%%%%%%%%%%%%%% 

\acknowledgments{We thank Adam Falkowski and Jos\'e Santiago for discussions that were the early seeds of this project.  We also thank
Sylvain Fichet, Carlos Garc\'ia Canal and Rogerio Rosenfeld for useful discussions and collaboration on related topics. 
The work of EM is supported by the project PID2020-114767GB-I00 and by the Ram\'on y Cajal Program under Grant RYC-2016-20678 funded by MCIN/AEI/10.13039/501100011033 and by ``FSE Investing in your future", by the FEDER/Junta de Andaluc\'{\i}a - Consejer\'{\i}a de Economı\'{\i}a y Conocimiento 2014-2020 Operational Programme under Grant A-FQM-178-UGR18, by Junta de Andaluc\'{\i}a under Grant FQM-225, and by the ``Pr\'orrogas de Contratos Ram\'on y Cajal'' Program of the University of Granada.
The work of MPV is 
partially supported by grants PID2019-106087GB-C22 and PID2022-139466NB-C22 funded by
MCIN/AEI/10.13039/ 501100011033 and by ERDF A way of making Europe, and also by grant FQM 101 funded by the Junta de Andaluc\'ia. 
The work of MQ is supported by the Departament d'Empresa i Coneixement, Generalitat de Catalunya, Grant No.~2021 SGR 00649, and by the Ministerio de Econom\'{\i}a y Competitividad, Grant No.~PID2020-115845GB-I00. IFAE is partially funded by the CERCA program of the Generalitat de Catalunya.}

 %%%%%%%%%%%%%%%%

\bibliographystyle{ieeetr}
\bibliography{undecay}{}

\end{document}